\documentclass[useAMS,usenatbib,minimal]{mn2e}

\usepackage{graphicx,amsmath,amssymb}
\usepackage{natbib}
\usepackage{enumerate}
\usepackage{url}
\usepackage{color}
\usepackage{mathrsfs}
\newcommand {\reac}[6] {$\rm\,{}^{#2}\kern-0.8pt{#1}\,({#3}\,,{#4})
\,{}^{#6}\kern-0.8pt{#5}\,$}
\title[Molecular chemistry in the CSEs of TP-AGB stars]
{Connecting the evolution of thermally pulsing asymptotic giant branch stars
to the chemistry in their circumstellar envelopes -- I. 
The case of  hydrogen cyanide}
\author[P. Marigo et al.]
{Paola Marigo$^{1}$\thanks{E-mail
paola.marigo@unipd.it},
Emanuele Ripamonti$^{1}$, 
Ambra Nanni$^{1}$, 
Alessandro Bressan$^{2}$, \newauthor
and L\'eo Girardi$^{3}$ \\
$^{1}$Department of Physics and Astronomy G.\ Galilei, University of Padova
        Vicolo dell'Osservatorio 3, I-35122 Padova, Italy\\
$^{2}$Astrophysics Sector, SISSA, Via Bonomea 265, I-34136 Trieste, Italy \\
$^{3}$Astronomical Observatory of Padova -- INAF,
            Vicolo dell'Osservatorio 5, I-35122 Padova, Italy}
\begin{document}

\date{Accepted 2015 xxx 15. Received 2015 April xx; in original form 2015 May xx}
 
\pagerange{\pageref{firstpage}--\pageref{lastpage}} \pubyear{2011}

\maketitle

\label{firstpage}

\begin{abstract}
We investigate the formation of hydrogen cyanide (HCN) in the inner circumstellar
envelopes of thermally pulsing asymptotic giant branch (TP-AGB) stars.
A dynamic model for periodically shocked atmospheres, which includes an extended 
chemo-kinetic network, is for the first time coupled to
detailed evolutionary tracks for the TP-AGB phase computed with the \texttt{COLIBRI} code.
We carried out a calibration of the main shock parameters 
(the shock formation radius $r_{\rm s,0}$, and the effective adiabatic index 
$\gamma_{\rm ad}^{\rm eff}$) using the circumstellar 
HCN abundances recently measured for a populous 
sample of pulsating TP-AGB stars. 
Our models recover the range of the observed HCN concentrations 
as a function of the mass-loss rates, and successfully reproduce 
 the systematic increase of HCN moving   
along the M-S-C chemical sequence of TP-AGB stars, that traces the increase of the  
surface C/O ratio.
The chemical calibration brings along two important implications
for the physical properties of the pulsation-induced shocks: i) the first shock  
should emerge very close to the photosphere $(r_{\rm s,0} \simeq 1\, R)$, and ii) 
shocks are  expected to have a dominant isothermal character 
$(\gamma_{\rm ad}^{\rm eff}\simeq 1)$ in the denser region close to the star 
(within $\sim 3-4\, R$), implying that
radiative processes should be quite efficient.
Our analysis also suggests that the HCN concentrations in the inner circumstellar envelopes 
are  critically affected by the H-H$_2$ chemistry during the post-shock relaxation stages. 
Given the notable sensitiveness of the results to stellar parameters, 
this paper shows that such chemo-dynamic analyses may indeed provide a significant 
contribution to the broader goal of attaining a  
comprehensive calibration of the TP-AGB evolutionary phase.

\end{abstract}


\section{Introduction}
Circumstellar envelopes (CSE) around TP-AGB stars are 
active sites of complex physical processes.  They involve the development 
of shock waves driven by long-period pulsations,  
the abundant formation of many different molecules, 
the condensation and growth of dust grains that, absorbing 
momentum from the radiation field and transferring it to the gas via collisions,
are believed to be the key drivers of stellar winds at higher luminosities on the AGB
\citep[][]{Olofsson_03}.

Indeed, the  characterization of the chemistry across the CSEs of TP-AGB stars 
has had  an impressive progress in recent years thanks to the advent of radio and 
sub-millimeter telescopes at very high resolution and signal-to-noise ratio 
(e.g., Herschel, ALMA).
On observational grounds progress is proceeding mainly on two fronts.

On one hand,  the inventory of
molecular species in resolved CSEs is lengthening with an increasing level 
of detail through studies devoted to specific objects.
A striking case is provided by the carbon star IRC+10216,
where more than 70 molecular species have been detected
\citep{Agundez_etal14, Agundez_etal11, Neufeld_etal11a, Decin_etal10b,
Decin_etal10c, Cernicharo_etal10a, Cernicharo_etal10b,
MauronHuggins_10, Pulliam_etal10}, while the dust-condensation zone is
being resolved and probed thoroughly \citep{Fonfria_etal14,
Cernicharo_etal13, Cernicharo_etal11, Decin_etal10b}.  Other examples
of well-studied stars which several chemical analyses have focused on
are the oxygen star IK Tau \citep{DeBeck_etal13, Vlemmings_etal12,
Decin_etal10a, Decin_etal10d, Kim_etal10, Duari_etal99}, and the
S-type star $\chi$ Cig \citep{Schoier_etal11, Justtanont_etal10,
Cotton_etal10, DuariHatchell_00}.

On the other hand chemical studies, targeted to  larger samples of AGB stars, 
have been undertaken just recently. 
Thanks to these works the available statistics are being importantly increased, allowing us 
 to get the first insights on the dependence of the CSE chemistry 
on relevant stellar parameters,  such as the spectral type (M, S, C), the pulsation
period, and the rate of mass loss.
Information of this kind is now available for the circumstellar abundances 
of SiO \citep{Ramstedt_etal09,  Schoier_etal06, GonzalesDelgado_etal03}, 
SiS \citep{Schoier_etal07}, and HCN \citep{Schoier_etal13}.
As for H$_2$O, we have measures  for few  M-type stars \citep{Maercker_etal09}; 
for carbon stars precise estimates are in most cases still missing, though 
the presence of H$_2$O in significant concentrations is confirmed to be a   
widespread feature \citep{Neufeld_etal11a, Neufeld_etal11b}.

On theoretical grounds,  most of the efforts are directed to
analyze the chemical and physical characteristics of individual, 
resolved CSEs around AGB stars.
Again, the case of the carbon star IRC+10216 is definitely the most outstanding example, 
with a large number of theoretical studies dedicated to it
\citep[e.g.,][]{DeBeck_etal12, Cherchneff_12, Cherchneff_11, 
Agundez_etal08, AgundezCernicharo_06, Cau_02, WillacyCherchneff_98, 
WillacyCherchneff_97, CherchneffGlassgold_93,  Cherchneff_etal92}.

All these works need to specify the input values for basic stellar
parameters, -- such as the stellar mass $M$, the pulsation period $P$, the
effective temperature $T_{\rm eff}$, the photospheric radius $R$, 
the density $\rho$ and the temperature $T$ at the shock formation radius, 
the photospheric C/O ratio, the
mass-loss rate $\dot M$, etc., --  which are usually derived or constrained from
observations.  

It is worth emphasizing that in all of these works these parameters refer to a particular object
(e.g. the carbon star IRC+10216, or the M star IK Tau),  at a particular
stage of the AGB evolution. 
Clearly, the results of any ad-hoc
chemo-dynamic model, able to reproduce (within a given degree of
accuracy) a set of measured CSE abundances for that star, represent an important
test for the validation of the same model. At the same time, however, the
specificity of the input assumptions does not allow to investigate the dependence 
of the CSE chemistry on the stellar parameters as they vary during the TP-AGB evolution.

A first important step in the direction of extending the exploration of the parameters
was performed by \citet{Cherchneff_06}, who carried out an extensive analysis
of the non-equilibrium chemistry in the inner pulsation-shocked CSE at varying  
the photospheric C/O ratio, which was increased from  C/O=0.75
to C/O=1.1, to represent stars of types M, S, and C.
\citet{Cherchneff_06} explored the impact on the CSE 
chemistry caused by a single quantity, i.e. the photospheric carbon abundance,
while all other stellar parameters were kept fixed and 
equal to the values that are believed to best represent the case of the M star TX Cam, i.e.
$T_{\rm eff}=2600$ K, $R=280\, R_{\odot}$, $P=557$ days, $M=0.65$\, ${\rm M}_{\odot}$.
 
The purpose of the present study is to make a further leap forward, for the first time
connecting a chemo-dynamic model for CSEs to the 
stellar evolution during the TP-AGB.   
In this way we will be able to follow the changes of the CSE chemistry 
as a star evolves along the TP-AGB varying its luminosity, radius, effective
temperature, photospheric density, pulsation period,  
surface chemical composition  due to convective mixing
processes (third dredge-up) and hot-bottom burning 
(for initial mass $M_{\rm i}\ga 4\, {\rm M}_{\odot}$), 
while its mass decreases due to stellar winds.
In a broader context, coupling the TP-AGB evolution of a star with the 
chemo-physical properties of its CSE is a timely step that may importantly 
contribute to the far-reaching goal of attaining a comprehensive 
calibration of the TP-AGB phase \citep{Marigo_14}.

In this work we focus on the inner, warmer and denser,  
regions of the CSE, where 
the shocks caused by stellar pulsations are thought to critically affect
the dynamics and the chemistry. As a first application
 we will investigate the formation of the hydrogen cyanide molecule.

The structure of the paper is as follows.
The basic characteristics of the TP-AGB stellar models are briefly recalled
in Sect.~\ref{sect_tpagb}. The adopted dynamic description together 
with the chemistry network are described in 
Sect.~\ref{sect_dyn}. Then, Sect.~\ref{sect_hcn} is dedicated to 
present the results of the CSE chemo-dynamic integrations applied to 
a selected grid of TP-AGB evolutionary tracks,  focusing on the abundances of HCN.
We analyse the dependence on the main stellar and shock parameters,  
and discuss  their calibration 
on the base of the measured HCN abundances 
for a populous sample of AGB stars that includes all chemical types 
(M, S, and C).
The paper is closed with Sect.~\ref{sect_calibration}, which outlines 
the major implications arising from the calibration, mainly 
in terms of the dynamic properties of the inner CSEs of AGB stars.

\section{TP-AGB evolution}
\label{sect_tpagb}
The evolution  during the entire TP-AGB phase, from the first thermal pulse up
to the complete ejection of the envelope by stellar winds, is computed with the
\texttt{COLIBRI} code \citep[][with updates as in \citet{Rosenfield_etal14} 
and \citet{Kalirai_etal14}]{Marigo_etal13}, to which the reader should refer for all the
details.
Suffice to recall here the input physics and basic features of \texttt{COLIBRI},
that are relevant for the present work. 
The physical conditions at first thermal pulse are extracted from the 
\texttt{PARSEC} sets of evolutionary tracks \citep{Bressan_etal12}.
For the purposes of this work we adopt a solar-like 
initial chemical composition, with abundances 
(in mass fraction)  of 
helium $Y_{\rm i}=0.279$ and metals $Z_{\rm i}=0.017$.
The initial chemical mixtures for metals is scaled-solar according to
the \citet{Caffau_etal11}.
This assumption is suitable for comparing model predictions with
observations of Galactic AGB stars (see Sect.~\ref{sect_hcn}).

\texttt{COLIBRI} is designed to optimize the ratio between computational issues 
(mainly computing time
requirements, and numerical  instabilities due to the development of thermal pulses) 
and physical accuracy. Specifically, the kernel of \texttt{COLIBRI} is 
a detailed {\em deep} envelope model, that includes the atmosphere and     
the convective mantle,  and extends downward in mass to encompass the hydrogen
burning shell. Across this region numerical integrations of the four stellar structure
equations are carried out at each time step, together with the on-the-fly computation
of the equation of state and gas opacities, which ensures full consistency with the varying 
chemical composition.

This latter is  a unique feature of \texttt{COLIBRI} which incorporates
the \texttt{\AE SOPUS} code \citep{MarigoAringer_09} as a subroutine to compute 
the concentrations of more than 800 species (300 atoms and 500 molecules) for temperatures
$1\,500 {\rm K} \la T\la 20\,000 {\rm K}$ under the assumption 
of instantaneous chemical equilibrium, as well as the corresponding Rosseland
mean opacities due to many continuum absorptions, atomic and molecular line transitions,  
and scattering  processes. 
\begin{figure}
\centering
\resizebox{0.7\hsize}{!}{\includegraphics{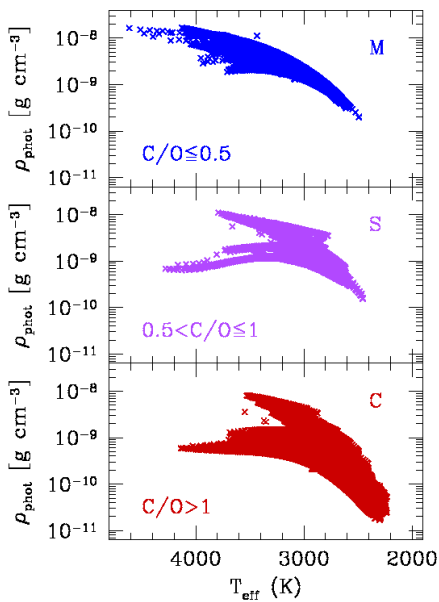}}
\caption{Photospheric gas density (at optical depth $\tau = 2/3$) as a
function of the effective temperature for roughly seventy 
 complete TP-AGB evolutionary tracks, spanning the whole range of relevant 
initial stellar masses, from $0.6\, {\rm M}_{\odot}$ to $6.0\, {\rm M}_{\odot}$.
The initial chemical composition corresponds to a metallicity $Z_{\rm i}=0.017$.
Models are grouped according to their surface C/O ratios to represent the chemical
classes of M stars (C/O$\,\le 0.5$), MS-S-SC stars ($0.5 <\,$C/O$\,\le 1.0$), and C stars (C/O$\,> 1.0$).}
\label{fig_atmoeq}
\end{figure}

Figure \ref{fig_atmoeq} displays a large number of atmosphere models, in terms
of effective temperature $T_{\rm eff}$ and gas density at the photosphere 
$\rho_{\rm phot}$,  
computed during the TP-AGB evolution of stars with initial metallicity $Z_{\rm i}=0.017$, 
and covering  the range of initial masses from  $0.6\, {\rm M}_{\odot}$ to $6.0\, {\rm M}_{\odot}$.
In general, models occupy a relatively wide region, with 
carbon stars extending  to lower effective temperatures and lower gas
densities compared to the oxygen-rich models.
We emphasize that both $T_{\rm eff}$ and $\rho_{\rm phot}$ 
are key input parameters to the  chemo-dynamic integrations
of the circumstellar envelopes, described in Sect.~\ref{sect_dyn}.
\label{sect_chemeq}
\begin{figure}
\centering
\begin{minipage}{0.35\textwidth}
\resizebox{\hsize}{!}{\includegraphics{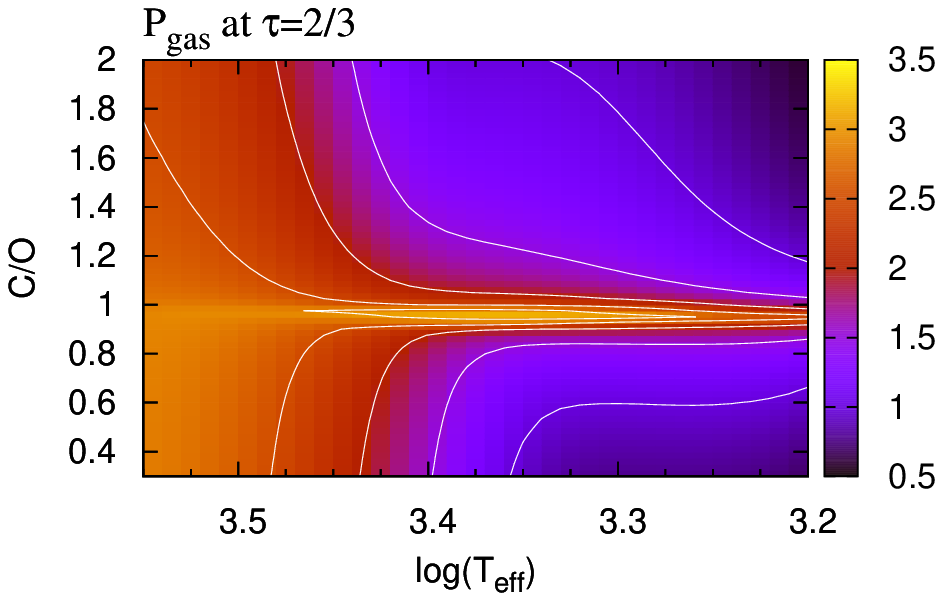}}
\end{minipage}
\hfill
\begin{minipage}{0.35\textwidth}
\resizebox{\hsize}{!}{\includegraphics{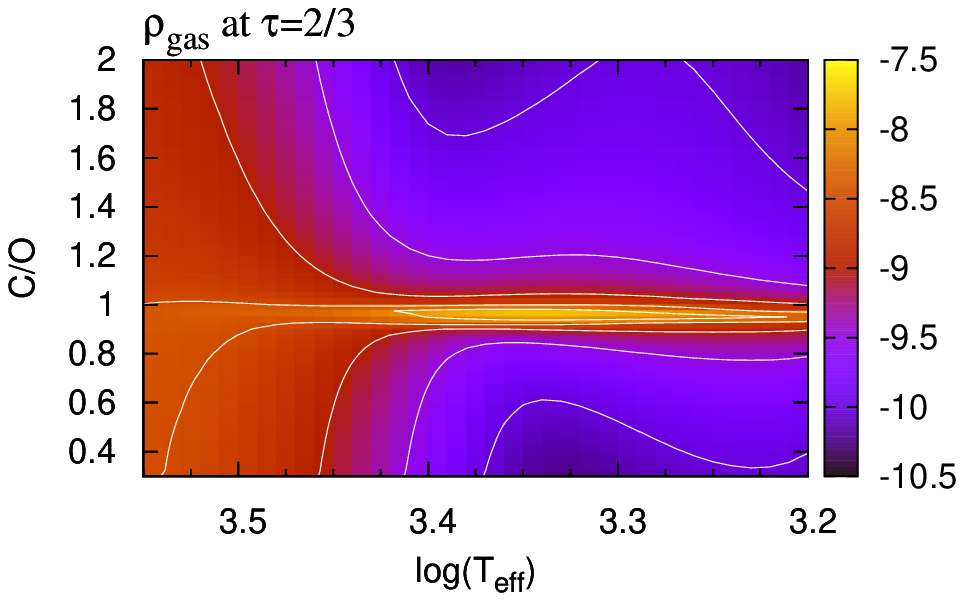}}
\end{minipage}
\hfill
\begin{minipage}{0.35\textwidth}
\resizebox{\hsize}{!}{\includegraphics{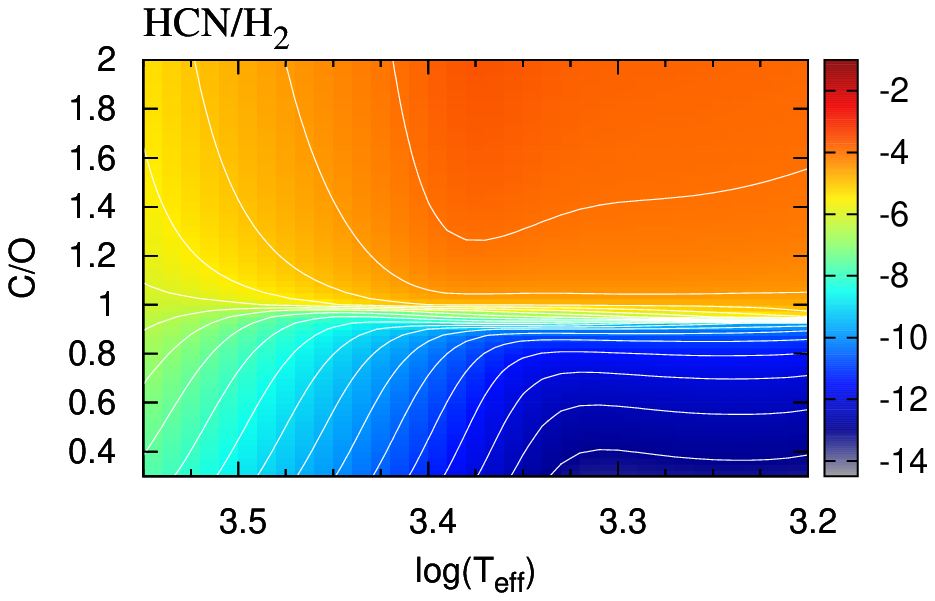}}
\end{minipage}
\caption{Bi-dimensional map of the 
logarithms of  gas pressure, mass density, and  equilibrium abundance of HCN relative to molecular hydrogen, 
as a function of the effective temperature and surface C/O ratio. 
The C/O ratio  is increased from
0.3 to 2.0, in steps of 0.0025, while all other elements 
are kept constant to their values at the first thermal pulse.
Results are obtained with the equilibrium chemistry and opacity routines  
of the \texttt{\AE SOPUS} code \citep{MarigoAringer_09}, applied to a static stellar atmosphere with total stellar mass $M=2\, {\rm M}_{\odot}$, 
initial metallicity $Z_{\rm i}=0.017$, and luminosity $L=10^4\, {\rm L}_{\odot}$. }
\label{fig_eqHCN}
\end{figure}

\begin{figure*}
\centering
\resizebox{0.6\hsize}{!}{\includegraphics{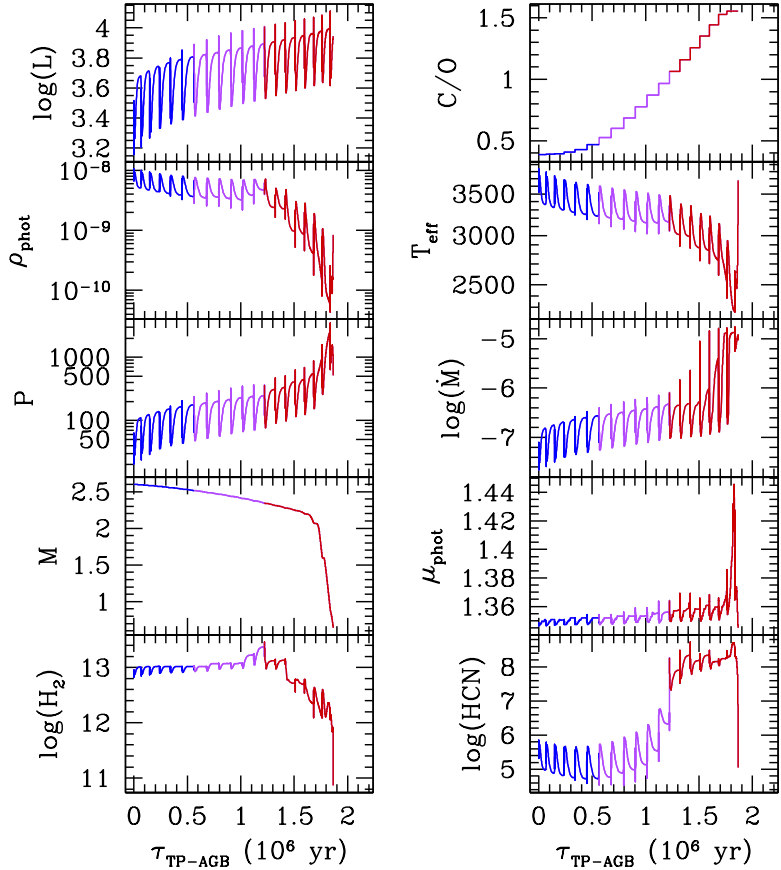}}
\caption{Evolution of several key quantities during the whole TP-AGB evolution 
of a model with initial mass $M_{\rm i} = 2.6\, {\rm M}_{\odot}$ and metallicity 
$Z_{\rm i}=0.017$. From top-left to bottom-right the eight panels show:
the logarithm of the  luminosity [${\rm L}_{\odot}$], the surface C/O ratio, the photospheric
density [g cm$^{-3}$], the effective temperature [K], 
the pulsation period [days] in the fundamental mode, the logarithm of the 
mass-loss rate [${\rm M}_{\odot}$ yr$^{-1}$], the stellar mass [${\rm M}_{\odot}$], 
the photospheric mean molecular weight [amu], the logarithm of the number density
 [cm$^{-3}$] of molecular hydrogen H$_{2}$, and of hydrogen cyanide HCN 
under the assumption of instantaneous chemical equilibrium.
All curves are colour-coded as a function of the C/O ratio, namely: 
C/O$\,\le 0.5$ in blue,  $0.5<$C/O$\,\le 1.0$ in purple, and C/O$\,>1.0$ in red.}
\label{fig_2.6z0.017tpagb}
\end{figure*}

Due the severe uncertainties that still affect our knowledge of   
stellar convection and stellar winds, both processes are treated in \texttt{COLIBRI} 
with a parametrized description, mainly in terms of the efficiencies of
the third dredge-up and mass loss.

It is common
practice to adopt some analytical formula for the mass loss rate as a
function of stellar parameters in stellar evolution models. Doing so, we
can then directly compare the modelled HCN abundance as a function of
mass loss with observations. However, this practice generates an
inconsistency in the present chemo-dynamic models of the wind of TP-AGB
stars, as we do not consider dust formation in the HCN formation
zone and yet assume dust forms at a specific radius to generate a mass
loss through radiation pressure on dust grains.

Mass loss is described  assuming 
that it is caused by two main mechanisms, 
dominating at different stages. Early on the AGB, when stellar winds 
are likely related  to magneto-acoustic waves operating below the stellar chromosphere, 
we adopt the semi-empirical relation 
by \citet{SchroderCuntz_05}, modified according to \citet{Rosenfield_etal14}.
Later on the AGB, when 
the star enters the dust-driven wind regime, we adopt  an exponential form
$\dot M \propto \exp({M^{a} R^{b}})$,  
as a function of stellar mass and radius \citep[see for more details][]
{Bedijn_88, Girardi_etal10, Rosenfield_etal14}, which
yields results similar to those obtained with the mass-loss law of \citet{VassiliadisWood_93}.
The relation was calibrated on a sample
of Galactic long-period variables with measured mass-loss rates, pulsation
periods, masses, effective temperatures, and radii. 
At any time on the TP-AGB,  the actual mass-loss rate is taken as the maximum among 
the two regimes.

The onset of the third dredge-up is evaluated with the aid of envelope integrations at the stage
of the post-flash luminosity peak, while its efficiency is computed with analytic fits
to the results of full stellar models provided by \citet{Karakas_etal02}, as a function of 
current stellar mass and metallicity.  For the metallicity $Z=0.017$ considered here, 
stellar models able to become carbon stars have initial masses 
$M_{\rm i} \ga 2.0 M_{\odot}$.

Hot-bottom burning (HBB) in the most massive AGB models 
($M_{\rm i} \ga 4\, {\rm M}_{\odot}$) is fully taken into account, 
with respect to both energetics and nucleosynthesis, adopting a complete nuclear
network (including the pp chains, CNO, NeNa, and MgAl cycles) 
coupled to a diffusive description of stellar convection.
 
In particular, with \texttt{COLIBRI} we are able to follow in detail 
the evolution of the surface C/O ratio, which is known to have a 
dramatic impact on molecular chemistry, opacity, and 
effective temperature every time it crosses the critical region around unity 
\citep[e.g.,][]{Marigo_02, Marigo_etal03, Cherchneff_06, MarigoAringer_09}. 
In turn, the surface C/O ratio plays a paramount role in determining the
chemistry of dust in the CSEs of AGB stars \citep[e.g.,][]{FerrarottiGail_06, 
Nanni_etal13}.

The high sensitivity of the atmospheric parameters to the surface C/O ratio  is
illustrated in Fig.~\ref{fig_eqHCN}.
The  gas pressure and the mass density
at the photosphere (at optical depth $\tau=2/3$) attain maximum values 
in the neighborhood of C/O$\,\simeq 1$, and then
decrease on both sides, corresponding  to oxygen-rich (C/O$\,<1$) and 
carbon-rich models (C/O$\,>1$).
The decrease is more pronounced in the models with lower effective temperatures.
As for HCN,   the sharp change in its abundance (relative to that of H$_2$) around 
C/O$\,\simeq 1$ is even more dramatic, and the equilibrium concentration of this
carbon-bearing molecule drops by orders of magnitude in the region 
characterised by C/O$\,<1$, while it rises notably in the opposite 
region defined by C/O$\,>1$.

\subsection{Chemical definition of M, S, and C classes}
TP-AGB stars are observationally classified in three groups (M, S, and C), mainly on the base of
the molecular absorption features that characterize their spectra.
While M stars  exhibit prominent molecular bands due to O-bearing molecules
(e.g., VO, TiO, H$_2$O), C stars  are dominated by 
strong bands of C-bearing molecules (e.g. C$_2$, CN, HCN, C$_2$).
S stars are  distinguished by strong ZrO
bands in addition to TiO bands typical of  M giant spectra.
In the past, attempts were made to interpret the  phenomenological criteria of spectral classification
in terms of  stellar parameters: photospheric C/O ratio, effective temperature, gravity, abundances
of S-process elements \citep[e.g.,][]{Keenan_54, Keenan_etal76}.

It is generally believed that the C/O ratio is the key-factor that drives the spectral dichotomy
between M stars (C/O$\,<1$) and C stars (C/O$\,>1$).  In this scheme 
 S stars would be  transition objects with C/O ratios within a very narrow range
(C/O$\approx 1$).
From a theoretical point of view, the increase of the C/O ratio is explained by the recurrent
surface enrichment of primary carbon due to the third dredge-up at thermal pulses, which
is also generally associated with the  increase in the concentrations of heavy elements
produced by slow-neutron captures,  among which Zr \citep{Herwig_05}.

Following the predictions of new model atmospheres,  
\citet{vanEck_etal11} have recently pointed out that the interval of C/O
associated to S stars would be much wider than considered so far, that
is $0.5<$C/O$\,\le 1.0$.
This is essentially because
the characteristic ZrO bands observed in the spectra of S stars
require that enough free oxygen, not locked in carbon monoxide 
and other O-bearing molecules, is available.

In the light of the above considerations, in this work we adopt 
the following operative definition to  assign  a given TP-AGB 
stellar model to one for the three chemical classes:
\begin{itemize}
\item {\em M stars}: C/O$\,\le 0.5$\,,
\item {\em S stars}:\,\,  $0.5 <\,$C/O$\,\le 1.0$\,,
\item {\em C stars}:\, C/O$\,> 1.0$\,.
\end{itemize}

With this definition the C/O range adopted 
to characterize the S-star group is meant to gather     
the  sub-classes of MS, S and SC stars \citep{SmithLambert_90}.
In Fig.~\ref{fig_2.6z0.017tpagb} we show the evolution of other relevant quantities during
the whole TP-AGB phase experienced by a stellar model with initial mass 
$M_{\rm i} = 2.6\, {\rm M}_{\odot}$ and metallicity $Z_{\rm i}=0.017$. Following the
occurrence of third dredge-up events, 
this model is expected to experience
a complete chemical evolution regulated by the photospheric C/O ratio,  
starting from the M class,  transiting through the S class, 
and eventually reaching the C class  where major mass loss takes place.
At the transition to the C-star domain, most of the variables 
exhibit a notable change in their mean trends, which results into a decrease of the effective 
temperature  and of the photospheric density, an increase of 
the pulsation period, and an intensification of the mass-loss rate.
The bottom panels present the photospheric concentrations of H$_2$ and
HCN, derived with the \texttt{\AE SOPUS} code under the assumption of
instantaneous chemical equilibrium. As already discussed above, 
the molecule of hydrogen cyanide is critically affected by the evolution of the C/O ratio.
Further modulations in the HCN and H$_2$ abundances are driven by the quasi-periodic 
occurrence of thermal pulses, clearly recognizable in the oscillating trends 
on timescales of the inter-pulse period (see also Sect.~\ref{ssect_tpcycle}). 

\section {Chemo-dynamic evolution of the inner circumstellar envelope}
\label{sect_dyn}
Several theoretical studies clearly pointed out that
the strong pulsations experienced by AGB stars cause 
periodic shock waves to emerge from sub-photospheric regions,
whence they propagate outward through the stellar atmosphere and the circumstellar
envelope 
\citep{Wood79, WillsonHill_79, FoxWood_85, BertschingerChevalier_85, Bowen_88}.
To model the shock dynamics and the non-equilibrium
chemistry across the circumstellar region that extends from the photosphere at $r=1\,R$
to  $r= 5\,R$, which we refer to as ``inner wind'',
we adopt the basic scheme developed by the works of
\citet[][]{Cherchneff_etal92} and
\citet[]{WillacyCherchneff_98}.

Below we will recall the basic ingredients  and introduce the main  modifications 
with respect to the original prescriptions. 
We will adopt the standard notation that 
marks with the subscripts 1 and 2 the gas conditions in front of
(pre-shock) and behind the shock (post-shock), respectively. 
The functional forms for time-averaged radial temperature and density profiles 
of the pre-shock gas are the sames as in \citet{Cherchneff_etal92}, and we refer
to that paper for many technical details. 

\subsection{The time-averaged pre-shock temperature}
\label{ssect_t}
The radial stratification of the  pre-shock gas temperature, $\langle T_1(r)\rangle$, 
is expressed as a power law

\begin{equation}
\label{eq_tr}
\langle T_1(r) \rangle= T_0 \left(\frac{r}{r_0}\right)^{-\alpha}\, ,
\end{equation}
where $r$ is the radial distance from the stellar centre, 
$T_0$ is a reference gas temperature attained at a radius $r_0$, 
and $\alpha$ is an exponent that typically varies from 0.4 to 0.8. 

\begin{figure}
\centering
\resizebox{0.8\hsize}{!}{\includegraphics{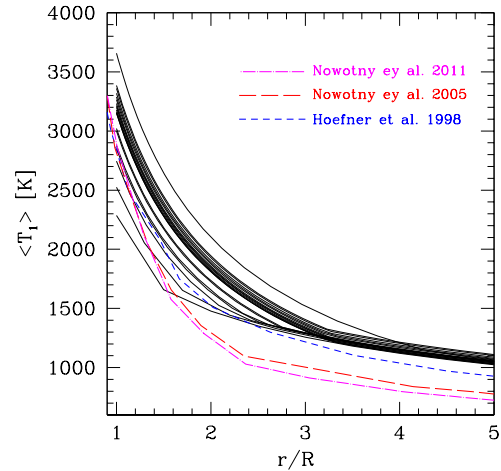}}
\caption{Radial profile of the average pre-shock temperature of the gas, across 
a distance from  $r=1\,R$ up to $r=5\,R$. 
Each solid line corresponds to the temperature structure  at the quiescent
luminosity just preceding each thermal pulse during the whole TP-AGB evolution of
a $(M_{\rm i}=2.6\, {\rm M}_{\odot},\, Z_{\rm i}=0.017)$ 
model computed with the \texttt{COLIBRI} code. For comparison we over-plot 
three temperature profiles
taken from detailed hydrodynamic models, averaged over the pulsation phases 
\citep{Nowotny_etal11, Nowotny_etal05, Hoefner_etal98}.}
\label{fig_tpreshock}
\end{figure}

Following the results from more recent detailed models of pulsating atmospheres 
\citep[][see Fig.~\ref{fig_tpreshock}]{Nowotny_etal11, 
Nowotny_etal05, Hoefner_etal98}, we notice that 
the phase-averaged gas temperature profile presents a change of the average slope $\alpha$ 
in the region where dust condensation starts to be efficient, 
 and a reasonable description is obtained with the combinations
\[
  [r_0; T_0; \alpha]=\begin{cases}
             [R; T_{\rm eff}; 0.8]\,\,\,\,\,\,\,\,\,\,\,\,\,\,\,\,\,\,\,\,\,\,\,\,\,\,\,
             \mathrm{for}\,\, R \le r <r_{\rm cond} \\
              [r_{\rm cond}; T_1(r_{\rm cond});  0.4] \,\,\,\,\mathrm{for}\,\, r  \ge r_{\rm cond}\, ,
            \end{cases}
\]
where $T_{\rm eff}$ is the effective temperature, $R$ is the photospheric radius, 
$r_{\rm cond}$ is the dust condensation temperature at which 
 the pre-shock gas temperature attains the value $T_1(r_{\rm cond})$.
The condensation radius $r_{\rm cond}$ is  estimated from the dust 
temperature distribution:
\begin{equation}
\label{eq_tdust}
T_{\rm dust}(r) = T_{\rm eff} W(r)^{\textstyle \frac{2}{4+p}}\,, 
\end{equation}
with the condition that $T_{\rm dust}(r_{\rm cond}) = T_{\rm cond}$.
The condensation temperature $T_{\rm cond}$ has 
typical values in the range $1000 \la T_{\rm cond} \la 1500$ K for the relevant species.
In Eq.~(\ref{eq_tdust}) $W(r) = 0.5[1 - \sqrt{1 - (r/R)^2}]$ is the geometrical 
dilution factor of the radiation field intensity, 
and $p$ is the spectral index of the power law that 
approximates the wavelength dependence of the dust absorption coefficient
$Q_{\rm A} \propto a\, \lambda^{p}$ (with $a$ being the grain size). 
The exponent $p$, in general, depends on the particular grain composition.

For carbon-rich models (C/O$>1$)
we take $p = 1$, which is a reasonable assumption for 
amorphous carbon grains in the relevant spectral region between 1-30 $\mu$, and 
$T_{\rm cond}=1500\,{\rm K}$ \citep{Jager_etal98, Cherchneff_etal91}.
These values for $p$ and   $T_{\rm cond}$ lead
to $r_{\rm cond}$ rather close to the star 
($r_{\rm cond}/R \la 2-4$) for carbon stars with intense mass loss, in line
with detailed dust models and observational indications 
\citep[e.g.,][]{FerrarottiGail_06, Nanni_etal13, Karovicova_etal13}.

For oxygen-rich models (C/O$<1$) the situation is more complex and uncertain,
since $p$ is shown to change significantly,  i.e. from 2 to -1 when passing from iron-rich 
olivine grains to forsterite 
\citep{Andersen_07, Jager_etal03, Dorschner_etal95}.
We were not able to identify a suitable unique prescription in the form 
of Eq.~(\ref{eq_tdust}) that performs well for all TP-AGB models.
Empirically, we find that the relation $r_{\rm cond}/R = 0.24\, \exp(1.87/M) + 0.027 M +2.68$ 
(where $M$ is current stellar mass)  leads to density and temperature profiles
suitable for HCN chemistry in O-rich models.
We get  typical values $r_{\rm cond}/R \la 3-4$ in line with interferometric  observations
\citep{Wittkowski_etal07, Karovicova_etal11, Norris_etal12}, and theoretical studies
\citep{Bladh_etal13, Bladh_etal15}.
We plan to improve the above prescription on a more physical base in follow-up studies.

We also remark that the issue of dust formation is beyond
the purpose in this initial work,  and the related quantities ($T_{\rm cond}$,
$r_{\rm cond}$) are relevant here only for the definition of the 
boundary conditions of the dynamic model (see Sect.~\ref{sect_boundc}).
 
Figure~\ref{fig_tpreshock} presents the predicted temperature gradients across the CSE of 
a stellar model during the TP-AGB phase, each line corresponding to the stage 
just prior the occurrence of a thermal pulse.
The temporal sequence is characterized by the decrease of the effective temperature, 
that coincides with  the gas temperature at $r=1R$. The flattening of the slope in the
$\langle T_1(r)\rangle$ relations for $r>r_{\rm cond}$
is particularly evident for the coolest tracks, which refer to the latest stages
on the TP-AGB characterized by strong mass loss. 
For comparison, we plot a few curves that refer to the results of detailed dynamic
models for pulsating C-rich atmospheres, all corresponding to an effective temperature
$T_{\rm eff} =2600\,{\rm K}$ \citep{Nowotny_etal11, Nowotny_etal05, Hoefner_etal98}.
Indeed, the average trends are well 
recovered by our models, though they tend to be shifted at somewhat higher $T$.
In this respect, one should also note that the input stellar parameters are not the same, 
e.g. most of our pre-flash TP-AGB stages correspond 
to effective temperatures $T_{\rm eff}  > 2600\,{\rm K}$. 

\subsection{The time-averaged pre-shock density}
\label{ssect_rho}
The description of the time-averaged pre-shock gas density is that
presented  by \citet{Cherchneff_etal92} following the formalism 
first introduced by \citet{WillsonBowen_86} on the base of 
dynamic models for periodically
shocked atmospheres.  Three wind regions are
considered: 1) the static region, which is controlled but a static scale
height, 2) the pulsation-shocked region, which is controlled by an
extended scale height due to shock activity, and 3) the wind region
where the wind fully develops once dust has formed.
The equation of momentum conservation is integrated over radial
distance by assuming spherical symmetry. 

The pre-shock density $\langle\rho_1(r)\rangle$ is then described by an 
exponential decay:

\begin{equation}
\label{eq_rhor}
 \langle\rho_1(r)\rangle = \rho_0 \times \exp\left [ - \int_{r_0}^{r}  \frac{1}{H(r')}\, dr'\right ]
\end{equation}
in which the reference initial density  $\rho_0$ corresponds to the distance $r=r_0$.

The characteristic scale-height, $H(r)$, as well as the reference point at $r= r_0$,
depend on the local mechanical conditions across  the pulsating atmosphere.  
Let us denote with $r_{\rm s,0}$ the radius at which the first shock develops.
\begin{figure}
\centering
\resizebox{0.8\hsize}{!}{\includegraphics{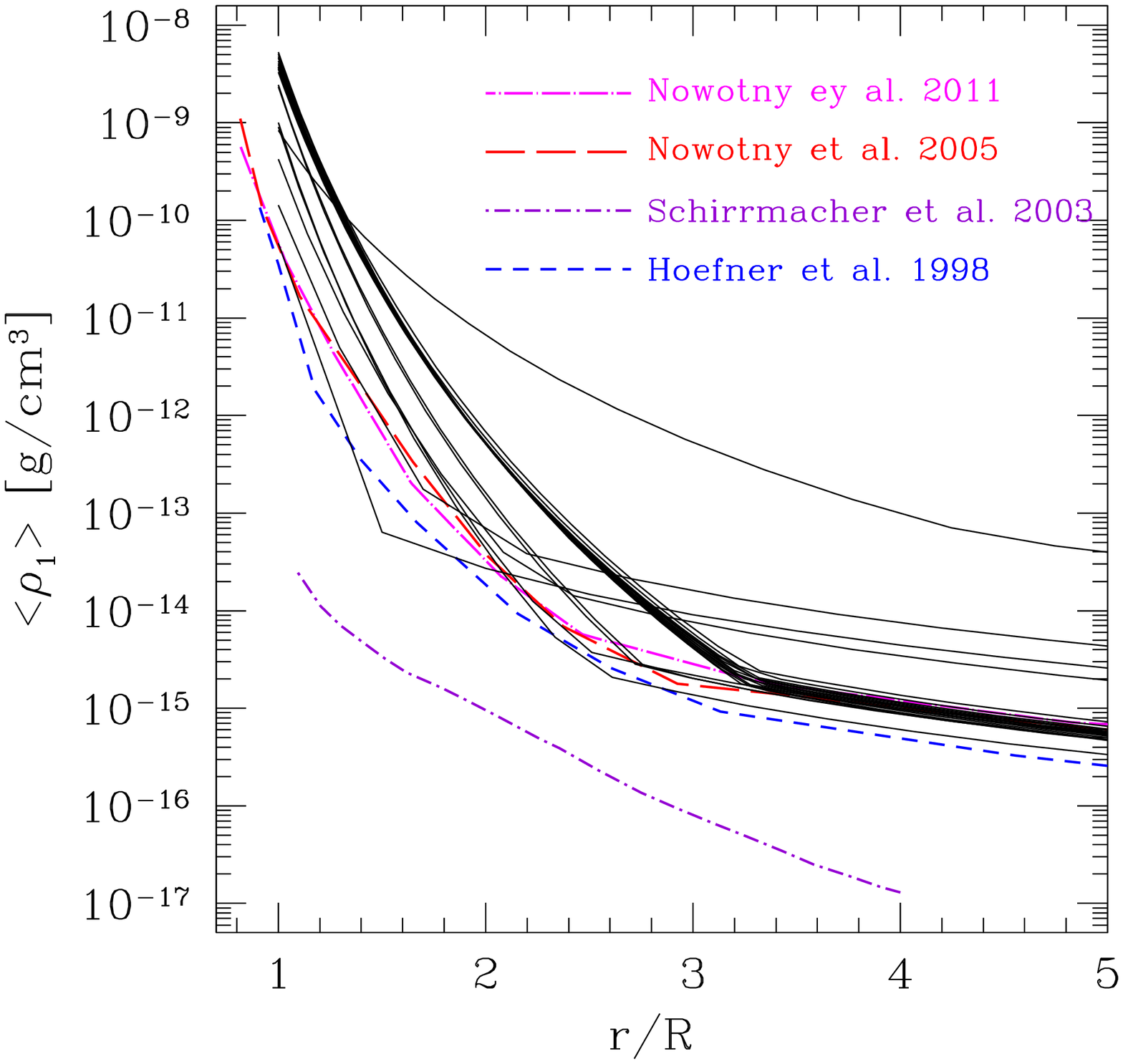}}
\caption{The same as in Fig,~\ref{fig_tpreshock}, but for the pre-shock density.
We also plot the  phase-averaged density structures from detailed
dynamic atmospheres  
\citep{Nowotny_etal11, Nowotny_etal05, Schirrmacher_etal03, 
Hoefner_etal98}. }
\label{fig_rhopreshock}
\end{figure}

\paragraph*{The static region.}
Across the unperturbed region from the photosphere $(r=R)$ 
to $r=r_{\rm s,0}$, 
the condition of hydrostatic equilibrium holds and the 
decay of the density follows a static scale-height:
\begin{equation}
H_0(r) = \frac{k T_1(r)}{\mu_1 g(r)}\, ,
\label{eq_H0}
\end{equation}
where $k$ is the Boltzmann constant, $g(r)=(G M)/{r^2}$ 
is the local gravitational acceleration exerted by a 
star with total mass $M$, $G$ is the gravitational constant, 
and $\mu_1$ is the mean molecular weight of the unperturbed gas.

\paragraph*{The pulsation-shocked region.}
Across the shock-extended zone $(r=r_{\rm s,0})$ 
up to the dust condensation distance $(r=r_{\rm c})$
a dynamic scale-height $H(r)$ is adopted:  

\begin{equation}
H_{\rm dyn}(r)  = \frac{H_0(r)}{1 -\gamma_{\rm shock}^2}\,.
\label{eq_Hdyn}
\end{equation}
Here, $\gamma_{\rm shock}$ is a parameter that describes the shock strength
and it is defined as   
\begin{equation}
\gamma_{\rm shock} = \frac{\Delta \upsilon}{\upsilon_{\rm e}}\,,
\label{eq_gamshock}
\end{equation}
 that is the ratio between 
the shock amplitude $\Delta \upsilon $ and the escape velocity 
$\upsilon_{\rm e} = \sqrt{2 G M / r}$.

The shock amplitude $\Delta \upsilon$ is the infall  velocity of the 
pre-shock gas measured in the shock rest-frame. 
Following the formalism developed by \citet{BertschingerChevalier_85}, 
the shock velocity $\upsilon_{\rm shock}$ (measured in the stellar rest-frame) is 
computed with
\begin{equation}
\upsilon_{\rm shock} = \delta \Delta \upsilon\,,
\label{eq_vsho}
\end{equation}
where the parameter $\delta$ is obtained from the solution of the
fluid equations over the shock deceleration stages as a function 
of the shock parameters (see Sect.~\ref{ssect_pulshock}). 
For the our best set of models we find typical values of $\delta \simeq 0.4$.

We note that, as long as $\gamma_{\rm shock} < 1$ (see Fig.~\ref{fig_dv0}),
$H_{\rm dyn}(r) > H_0(r)$, which implies that in the shocked region the average density
is higher than otherwise predicted under the assumption of hydrostatic equilibrium.
This fact is a key point for the non-equilibrium molecular chemistry that is active in 
that region.  

Following \citet{Cherchneff_etal92} and \citet{Cherchneff_12}
we assume that, once the first shock forms at $r=r_{\rm s,0}$, the initial amplitude
$\Delta \upsilon_{0} =\Delta \upsilon(r=r_{\rm s,0}) $
is damped as the shock propagates outward according to the scaling
law 
\begin{equation}
\Delta \upsilon(r) =   \Delta \upsilon_0 \left(\frac{r}{r_{\rm s,0}}\right)^{-0.5}
\label{eq_dvr}
\end{equation}
which derives from a linear proportionality of $\Delta \upsilon$ with the escape velocity $\upsilon_{\rm e}$,
as suggested by earlier dynamic studies \citep{WillsonBowen_86}.
The damping of the shock strength is confirmed by more recent detailed simulations
of pulsating atmospheres \citep[e.g.,][]{Schirrmacher_etal03}.

\paragraph*{The wind region.}
We assume that at the dust condensation radius $r_{\rm cond}$ the stellar wind starts
to develop.  The driver is thought to be the radiative
acceleration of dust particles with subsequent collisional transfer of
momentum from the grains to the surrounding gas \citep{GustafssonHofner_04}.

The density profile is derived from the continuity equation for a spherically symmetric,
stationary wind:
\begin{equation}
\label{eq_rhow}
\langle \rho_1(r)\rangle  =\rho_{\rm wind}(r) = \frac{\dot M}{4 \pi r^2 v(r)}\, 
\end{equation}
where  $v(r)$ is the gas expansion velocity and $\dot M$ is the constant mass-loss rate. 

Figure~\ref {fig_rhopreshock} illustrates a temporal 
sequence of pre-shock density profiles, referring to the same TP-AGB evolution
as in Fig.~\ref{fig_tpreshock}. Calculations were carried out 
with $r_{\rm s,0}=1\,R$, so that the CSE does not include a 
static region and the pulsation-shocked zone starts already from the photosphere.
Each curve shows a clear change in the average slope, which becomes flatter at 
the condensation radius whence the stationary wind regime is assumed to start. 

For comparison, the results from detailed models of pulsating atmospheres are
also shown. A nice agreement is found in most cases.
In particular, the radial decline of the density  predicted by the stationary wind condition
compares surprisingly well  with the phase-averaged slopes
of the detailed shocked-atmosphere models, just over 
the distances where dust is expected to be efficiently
forming.

\subsection{Boundary conditions for the density and the initial shock amplitude}
\label{sect_boundc}
The parameter $\gamma_{\rm shock}$ is a critical quantity since 
it determines the dynamic scale height of the density across the shock-extended region 
(see Eq.~\ref{eq_Hdyn}).
\citet{Cherchneff_etal92}  adopted $\gamma_{\rm shock}=0.89$ 
in their  dynamic model for the carbon-rich star  IRC+10216, following considerations
about the location of the shock formation radius $r_{\rm s,0}$.

Since then, to our knowledge, all works in the literature based on the \citet{Cherchneff_etal92}
formalism always kept the same  value,  $\gamma_{\rm shock}=0.89$, even if they  
analyse  a different object -- such as the S star $\chi$ Cig, or the oxygen-rich 
stars TX Cam and IK Tau --, or they adopt different prescriptions for 
other parameters -- such as the shock formation radius $r_{\rm s,0}$, the 
initial shock strength $\Delta \upsilon_{0}$, the pulsation period, etc. --
\citep{Cherchneff_12, Agundez_etal12, Cherchneff_11, Cherchneff_06, AgundezCernicharo_06, 
Cau_02, DuariHatchell_00, Duari_etal99, WillacyCherchneff_98}.

\begin{figure}
\centering
\resizebox{0.7\hsize}{!}{\includegraphics{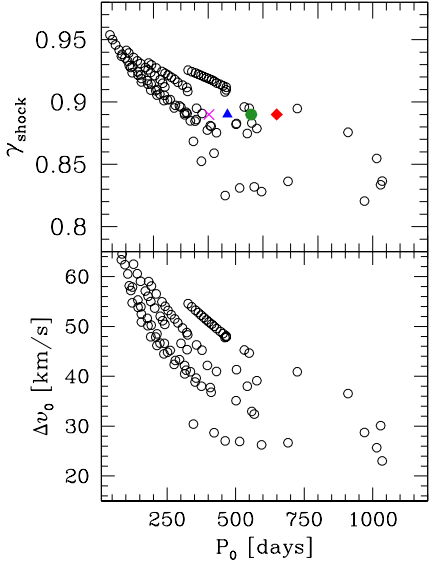}}
\caption{Evolution of the dynamic parameters $\gamma_{\rm shock}$ (top panel) 
and maximum shock amplitude  $\Delta \upsilon_{0}$ (bottom panel) 
as a function of the pulsation period in the 
fundamental mode. Predictions for the whole TP-AGB evolution of a few selected models
of different initial mass ($M_{\rm i} = 1-4\, {\rm M}_{\odot}$; empty circles) are shown.
Note the significant dispersion for $\gamma_{\rm shock}$, extending well above and 
below the constant value of 0.89  assumed 
in several existing models (referenced in the text), dealing with different stars, such as 
IRC+10216 (square), TX Cam (circle), IK Tau (triangle), $\chi$ Cyg (cross).}
\label{fig_dv0}
\end{figure}

In this work we relax the assumption of a constant, predetermined 
$\gamma_{\rm shock}$.
Rather, we consider $\gamma_{\rm shock}$ as a free parameter 
to be constrained, for each integration of the CSE,
with the aid of a suitable boundary condition.
The most natural condition is to impose the matching of the density laws in the 
pulsation-shocked and wind regions at the dust condensation radius 
$r_{\rm cond}$, that is

\begin{equation}
\label{eq_rhocond}
\rho_{\rm dyn}(r_{\rm cond}) = \rho_{\rm wind}(r_{\rm cond}) 
= \frac{\dot M}{4 \pi\, r_{\rm cond}^2\, \upsilon_{{\rm s}, T}(r_{\rm cond})}\,, 
\end{equation}
where 
\begin{equation}
\label{eq_cs}
\upsilon_{{\rm s},T}(r_{\rm cond}) = \sqrt{k T_{\rm cond} /\mu}
\end{equation}
is the isothermal
sound velocity at the dust condensation radius with temperature $T_{\rm cond}$.

With Eq.~(\ref{eq_rhocond}) we  assume that the sonic point coincides with
the condensation radius $r_{\rm cond}$, which is a reasonable condition as it is
generally thought that radiation on dust grains 
is the main process driving supersonic outflows in AGB stars
\citep[][]{LamersCassinelli_99}.

In summary, the boundary condition for the density, given by Eq.~(\ref{eq_rhocond}), 
translates into a function of the parameter $\gamma_{\rm shock}$, which
is eventually found with a standard numerical iterative technique.
Once $\gamma_{\rm shock}$ is known, we can obtain the corresponding 
value for the initial shock amplitude 
$\Delta \upsilon_0 = \gamma_{\rm shock} \times \upsilon_{\rm e}(r_{\rm s, 0})$, 
where the escape velocity is evaluated
at the shock formation radius.
In this way we consistently couple two key parameters of the
dynamic model,  $\gamma_{\rm shock}$ and $\Delta \upsilon_0$ which 
are often kept as independent input quantities in the works cited above.

Figure~\ref{fig_dv0} shows the evolution
of $\Delta\upsilon_0$ (assuming that $r_{\rm s,0}=R$) for a representative sample of models with different initial
stellar masses over the entire TP-AGB evolution.  The initial shock amplitude  $\Delta\upsilon_0$ is
found to progressively decrease during the TP-AGB evolution. 
This finding is in line with the general trends of detailed
calculations of periodically shocked atmospheres \citep{Bowen_88}, in which
$\Delta\upsilon_0$ is predicted to reduce with increasing pulsation period (see his table 6) 
and decreasing effective temperature (see his table 9), a trend that also happens as the 
TP-AGB evolution proceeds. 

\begin{figure}
\centering
\resizebox{0.8\hsize}{!}{\includegraphics{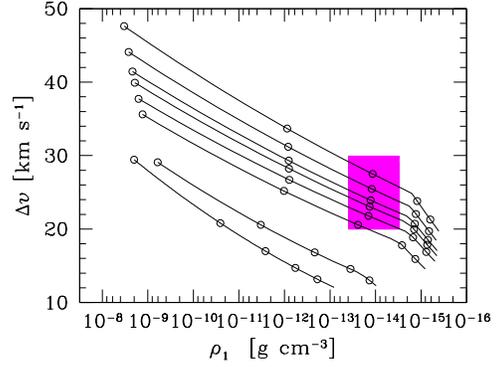}}
\caption{Evolution of the shock amplitude $\Delta\upsilon$ as a function 
of the quiescent pre-shock density $\langle \rho_1(r)\rangle$ as the shock propagates 
across the CSE, during the whole TP-AGB evolution of a star with initial mass 
$M_{\rm i}=1.4$ and metallicity $Z_{\rm i}=0.014$. Each track corresponds to the stellar
parameters just preceding a thermal pulse (the temporal sequence 
of TPs proceeds from top  to bottom). 
The hatched region in magenta defines 
the suitable region of the $\Delta\upsilon -  \langle \rho_1(r)\rangle$ plane where both 
normalised peak fluxes of the permitted and forbidden Fe II emission lines 
fit the observed data for a sample of Mira stars, following the analysis of 
\citet{Richter_etal03}, and \citet{RichterWood_01}. The empty circles mark 
the positions through the CSE from $r/R=1$ to $r/R=5$ in steps $\Delta(r/R)=1$.}
\label{fig_rhodv}
\end{figure}

Once  $\Delta\upsilon_0$ at $r=r_{\rm s, 0}$  is singled out for each combination 
of the stellar parameters, the weakening of the shock strength $\Delta\upsilon(r)$ 
at increasing $r$ is described according to Eq.~(\ref{eq_dvr}).
Figure~\ref{fig_rhodv} shows the predicted damping  of $\Delta\upsilon(r)$  as
a function of the pre-shock density across the CSE of a low-mass TP-AGB stellar model 
with initial mass $M_{\rm i} = 1-4\, {\rm M}_{\odot}$
and metallicity $Z_{\rm i}=0.017$. Interestingly, at distances around $r/R\simeq 3$ the tracks
enter the region of the $(\rho_1, \Delta\upsilon)$ space that seems to provide the suitable
conditions for the  formation of the observed Fe II
and [Fe II] peak line fluxes and line ratios in a sample of 
Mira stars \citep{Richter_etal03, RichterWood_01}.

\begin{figure}
\centering
\resizebox{0.7\hsize}{!}{\includegraphics{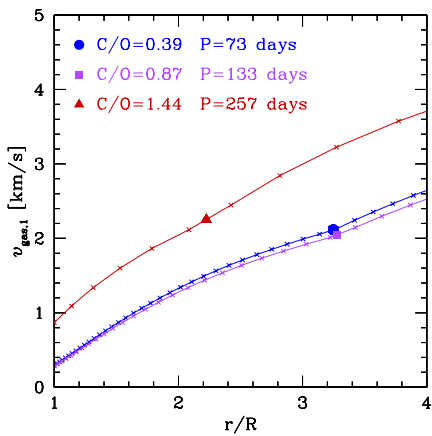}}
\caption{Gas outflow velocity $\upsilon_{\rm gas,1}$ as a function of the radial
distance from the stellar centre. The curves correspond to the pre-flash quiescent
stages of the $2^{\rm nd}$,  $11^{\rm th}$,  and $17^{\rm th}$ thermal pulses, 
experienced by the TP-AGB evolutionary model with initial mass 
$M_{\rm i} = 2.6\,{\rm M}_{\odot}$ and metallicity $Z_{\rm i} = 0.017$. 
For the three cases the filled symbols (circle, square, and triangle) 
mark the corresponding position
of the sonic point, where the gas velocity equals the isothermal 
speed of sound. See the text for more details.}
\label{fig_vgas}
\end{figure}

\subsection{The gas outflow velocity}
\label{ssect_vgas}
Another quantity of interest 
is the average radial increase of the gas outflow  velocity, $\upsilon_{\rm gas,1}$, which is
calculated in the framework of  the Parker theory of stellar winds \citep{Parker_65}, and following
the prescriptions of \citet{Cherchneff_etal92},  which take into account 
the presence of shocks and radiation pressure on dust.
The  condensation radius is assumed to coincide with the Parker radius,  where
the gas velocity equals the isothermal speed of sound given by Eq.~(\ref{eq_cs}).

The Parker solution for the average gas velocity is illustrated  in 
Fig.~\ref{fig_vgas}, which shows the radial profile of   $\upsilon_{\rm gas,1}$ for three
TP-AGB models with different photospheric C/O ratios (0.39, 0.87, 1.44), that may be considered 
representative of the  chemical classes M, S, and C. 
Starting from the stellar photosphere, the outflow is initially subsonic, with  low velocities  
typically $\la 1-2$ km s$^{-1}$.  The isothermal sound velocity 
($\upsilon_{{\rm s, T}} \simeq 2.0-2.2$ km s$^{-1}$) is reached at the Parker radius 
(equal to $r_{\rm cond}$ by construction), which appears to be located in the range
between $\approx 2.0-3.5\, R$. Note that for the C star model, characterized by the
longest period and highest mass-loss rate,  $r_{\rm P}$ sits closer to the star than the 
other two M and S models.  
Beyond the sonic point, the outflow is accelerated due to radiation pressure on dust grains. 

We call attention that a careful consideration of the dust issue is beyond the scope of this
initial work.  The knowledge of $\upsilon_{\rm gas,1}$ has mainly 
a practical relevance in the present context,  
since we use it for defining  the radial grid of mesh points over which the integrations 
of the shock-dynamics and non-equilibrium chemistry are performed, as described 
in Sect.~\ref{ssect_integration}.
The crosses on the curves mark the positions reached after one pulsation period by a gas parcel 
that is moving outward  with a velocity $\upsilon_{\rm gas,1}(r)$.
Over the distance from 
$r=1\, R$ to  $r=5\, R$,  we typically deal with a few tens to several tens of radial mesh-points
for each CSE.

\subsection{Pulsation-induced shocks}  
\label{ssect_pulshock}
The time-dependent structure of the shocked atmosphere and "inner wind" have been modelled
 according to the works of \citet{WillacyCherchneff_98, Duari_etal99, Cherchneff_06,
Cherchneff_11,  Cherchneff_12}. These 
studies apply the semi-analytical formalism of \citet{BertschingerChevalier_85}, 
where the shocked gas initially cools by emitting radiation in a thin layer 
just after the shock front.
Then, 
the gas enters the post-shock deceleration region where
its velocity decreases under the action of the stellar gravity, before being shocked again.
Three main assumptions are adopted.

First, the Lagrangian trajectory of any fixed fluid element is strictly periodic
with zero mass loss. This is reasonable in the inner CSE regions,
where the gas remains gravitationally bound to the star \citep{GustafssonHofner_04}.
Periodicity also means that, at any time, after being accelerated up 
to a maximum height above the shock formation layer, 
the gas element is decelerated downward until it recovers its original unperturbed position.  
In presence of a mass flux moving outward with an unperturbed velocity $\upsilon_{\rm gas,1}$, 
this is not exactly true, and the particle orbits are not exactly closed.
 However, we have verified that, after one pulsation period,  the radial displacement,  
$\Delta r = \upsilon_{\rm gas,1}\times P$,  of the fluid element with respect 
to the original position $r$ remains quite small in most cases, the relative deviation 
$\Delta r/r$ remaining typically below few $10^{-1}$. 

Second, the shock waves are radiative, so that the  gas quickly cools by emission of radiation 
in a thin region behind the shock front.
The post-shock values  $T_2$ and $\rho_2$ are fixed by the  Rankine-Hugoniot conditions \citep{Nieuwenhuijzen_etal93}.

The quantity $M_1$  is the Mach number,  defined as
\begin{equation}
M_{\rm 1} = \frac{\Delta\upsilon}{\upsilon_{\rm s}}\,,
\end{equation} 
where the shock amplitude is defined by Eq.~(\ref{eq_dvr}), and the sound velocity is given by
\begin{equation}
\upsilon_{\rm s} = \sqrt{\gamma_{\rm ad}\frac{k T_1}{\mu_1}}\,,
\end{equation} 
which holds in the case of a classical ideal gas with mean molecular weight $\mu_1$
($k$ is the Boltzmann constant).

Third, during the post-shock deceleration phase  
the effect of the interaction between gas and radiation,
hence of possible radiative losses, is implicitly taken in consideration by introducing an arbitrary 
 ``effective'' adiabatic exponent $\gamma_{\rm ad}^{\rm eff}$.
As a matter of fact, the adiabatic exponent 
can significantly deviate from the classical predictions (e.g.  $\gamma_{\rm ad}=1.4$ for a diatomic
perfect gas), when otherwise neglected 
 non-ideal phenomena, such as ionization, dissociation of molecules, and radiation pressure, 
 are suitably taken into account.
In this respect, a thorough analysis is provided by \citet{Nieuwenhuijzen_etal93}.
Therefore, in view of a more general description, 
in place of the classical adiabatic index  $\gamma_{\rm ad}$ 
we opt to use $\gamma_{\rm ad}^{\rm eff}$ as a 
free parameter to be inferred from observations. 
An ample discussion about $\gamma_{\rm ad}^{\rm eff}$ and its calibration is
given in Sect.~\ref{sssect_gameff}.

Following \citet{BertschingerChevalier_85},  for a given combination of stellar parameters
($M, T_{\rm eff}, R, P, \mu$, $T_1$, $\rho_{1}$), all the properties of the shock that emerges at the radius
$r=r_{\rm s}$ are completely determined once one specifies two parameters, namely:
the  ``effective'' adiabatic exponent $\gamma_{\rm ad}^{\rm eff}$,
and the pre-shock Mach number $M_1$. 
An example application of the shock model is presented and discussed
in Sect.\ref{ssect_integration}.

\begin{figure*}
\begin{minipage}{0.32\textwidth}
\resizebox{0.9\hsize}{!}{\includegraphics{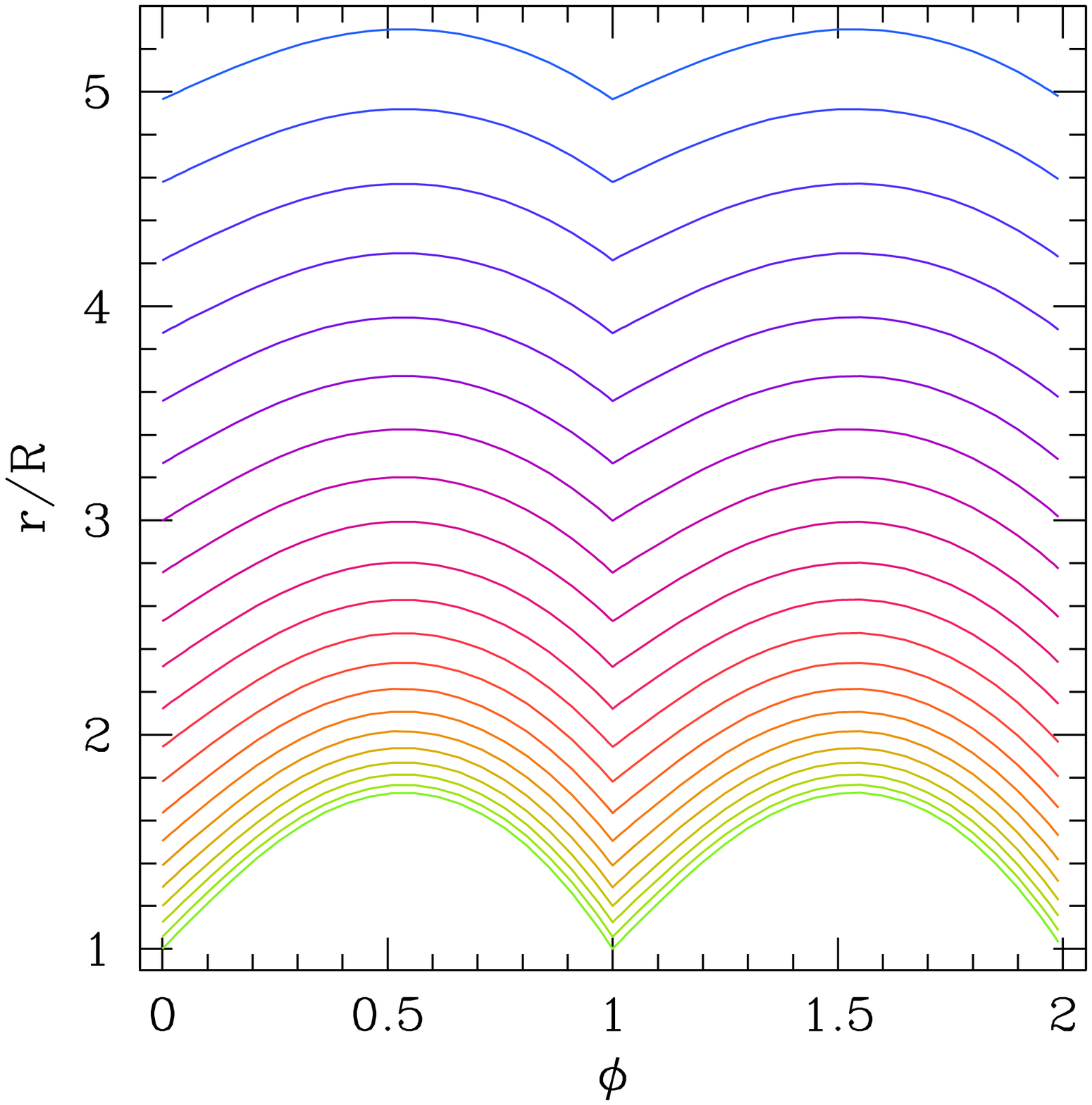}}
\end{minipage}
\hfill
\begin{minipage}{0.32\textwidth}
\resizebox{0.9\hsize}{!}{\includegraphics{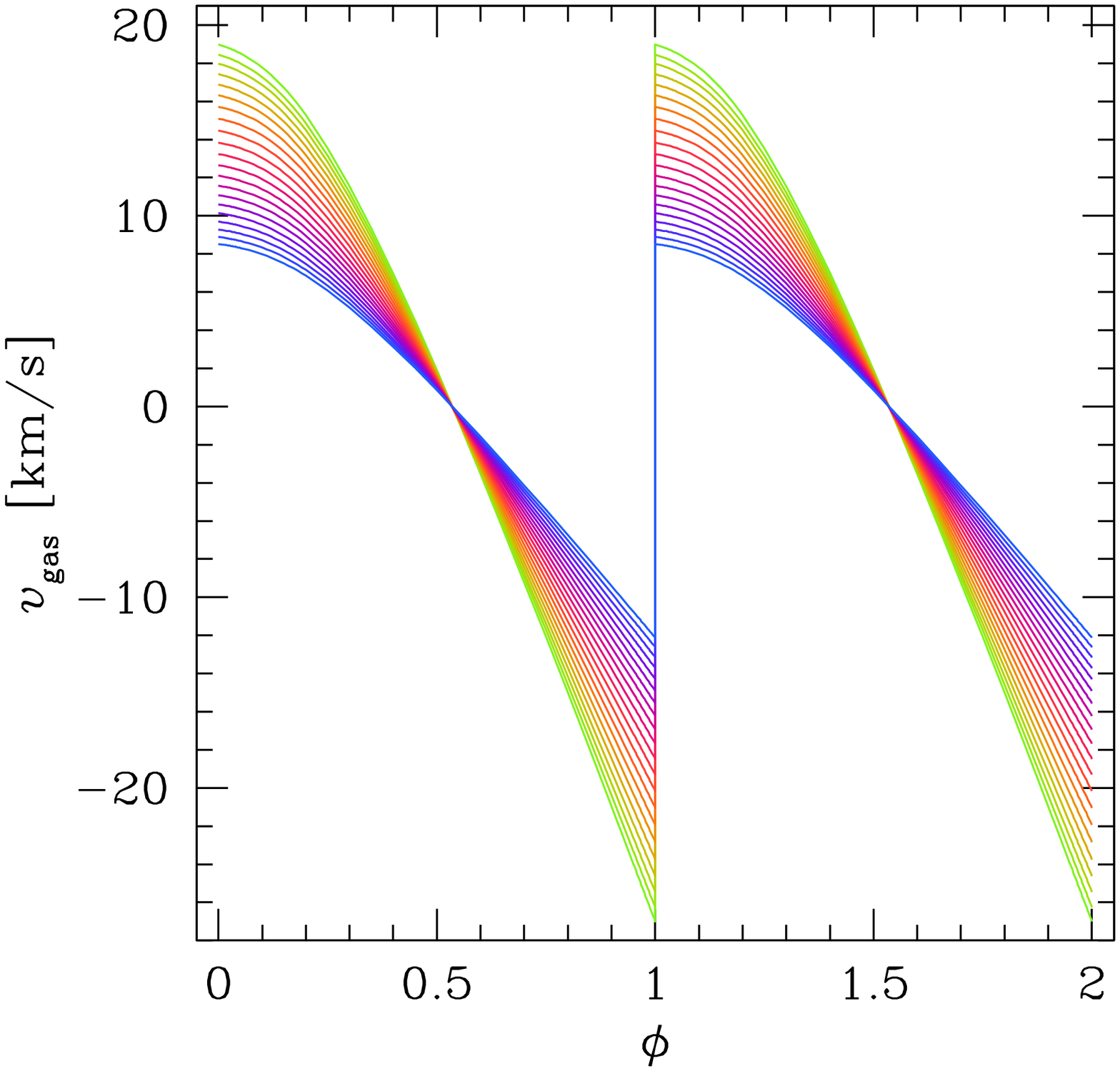}}
\end{minipage}
\hfill
\begin{minipage}{0.32\textwidth}
\resizebox{0.9\hsize}{!}{\includegraphics{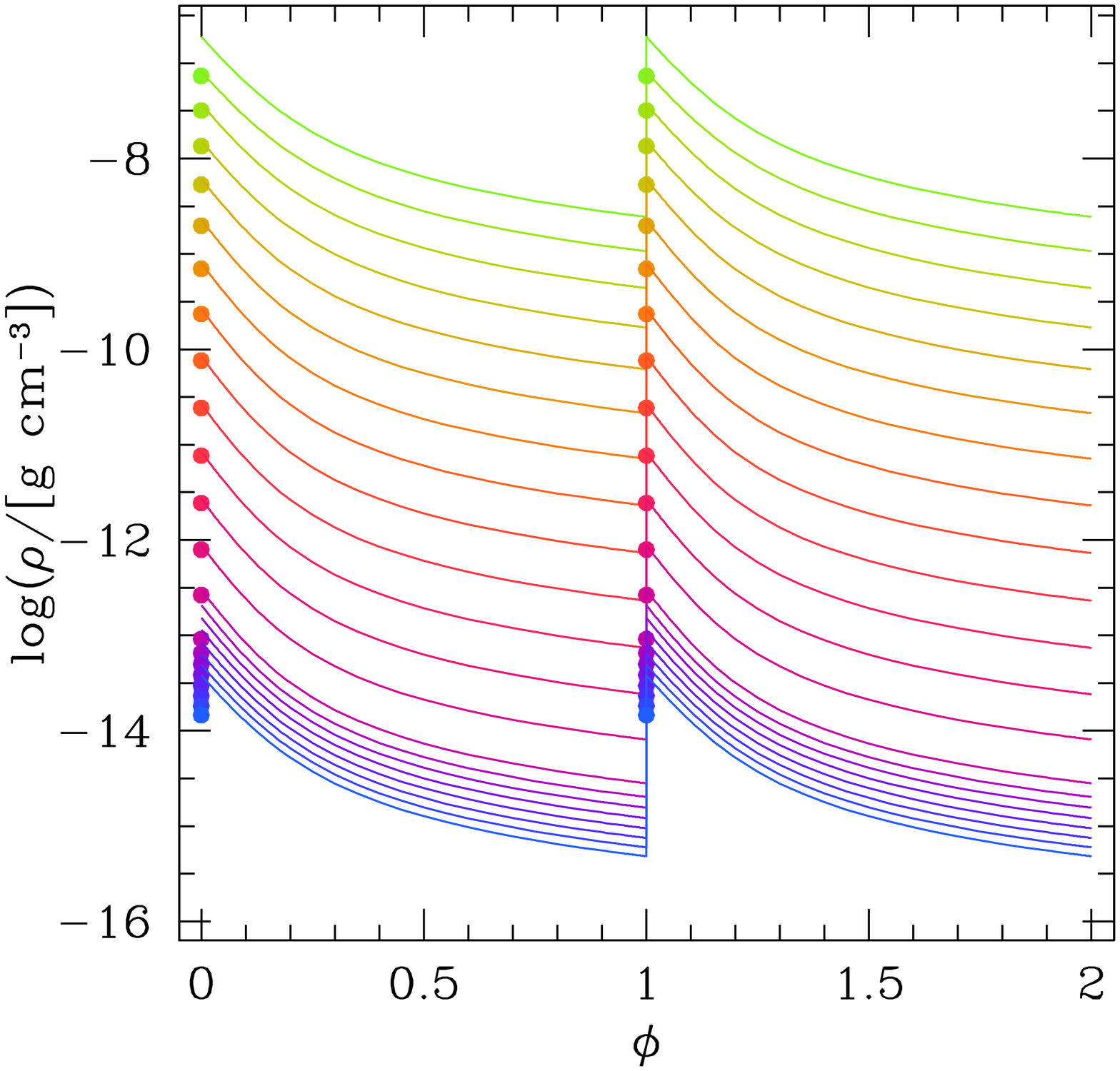}}
\end{minipage}
\hfill
\begin{minipage}{0.32\textwidth}
\resizebox{0.9\hsize}{!}{\includegraphics{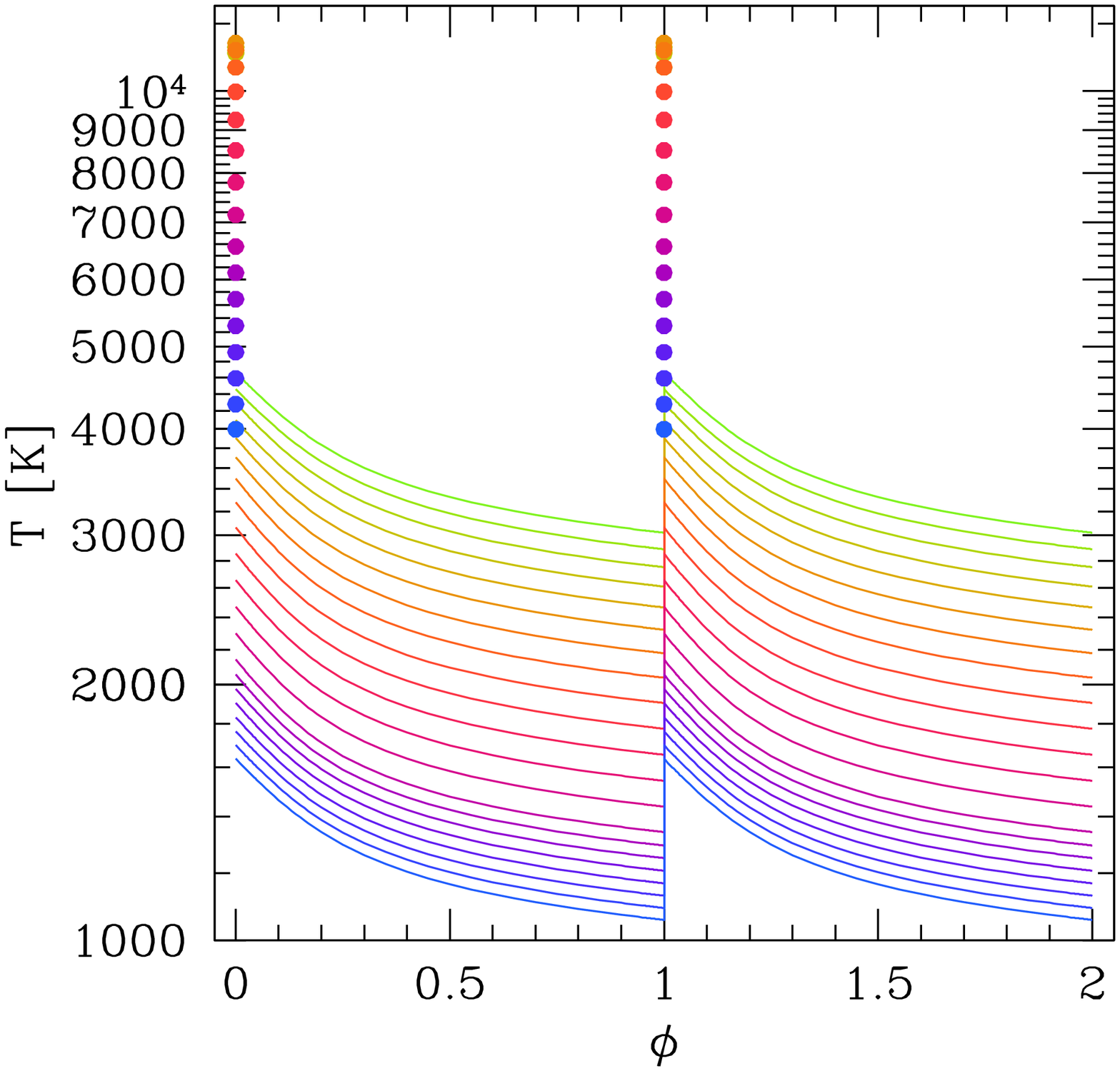}}
\end{minipage}
\hfill
\begin{minipage}{0.32\textwidth}
\resizebox{0.9\hsize}{!}{\includegraphics{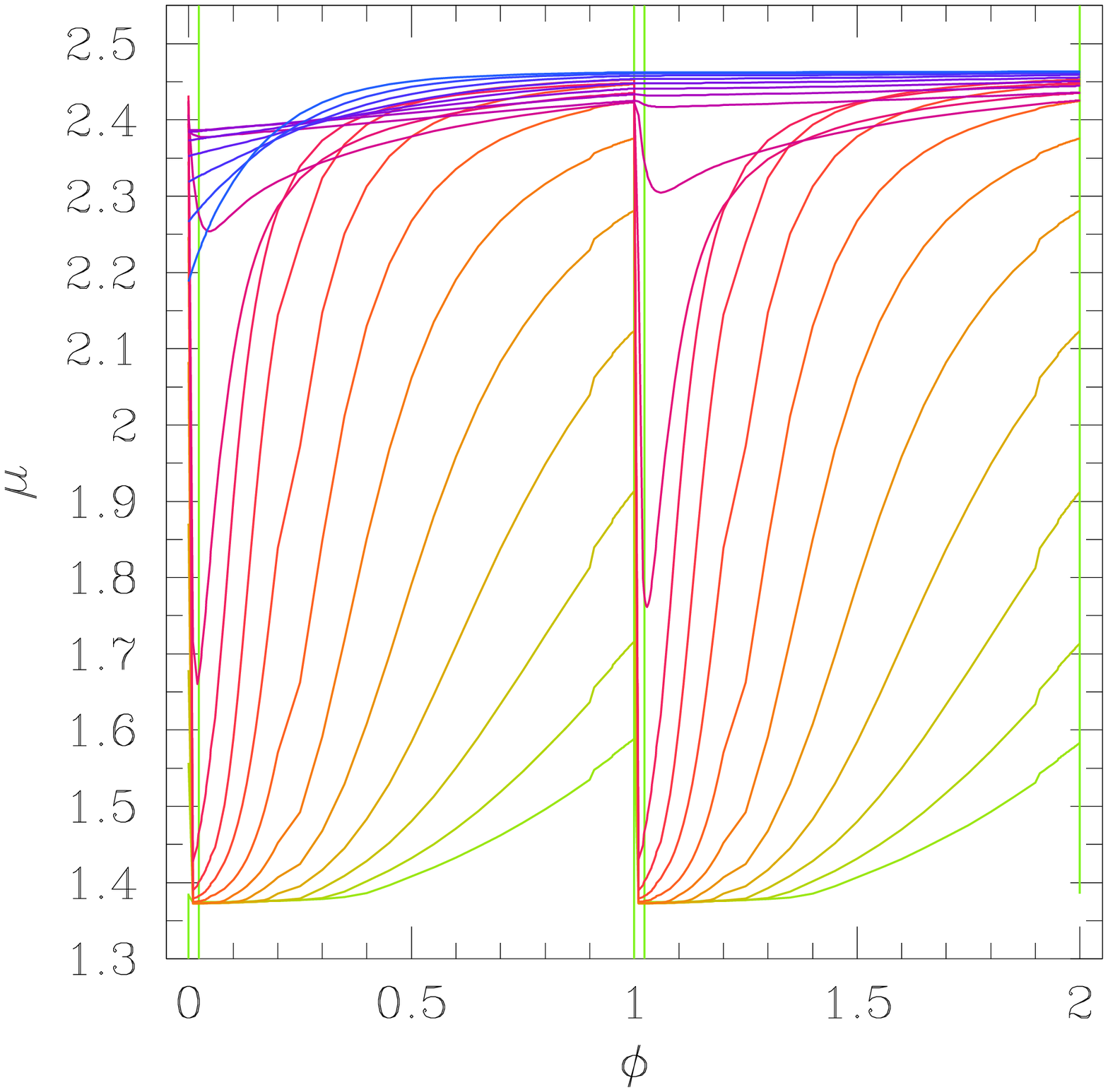}}
\end{minipage}
\hfill
\begin{minipage}{0.32\textwidth}
\resizebox{0.9\hsize}{!}{\includegraphics{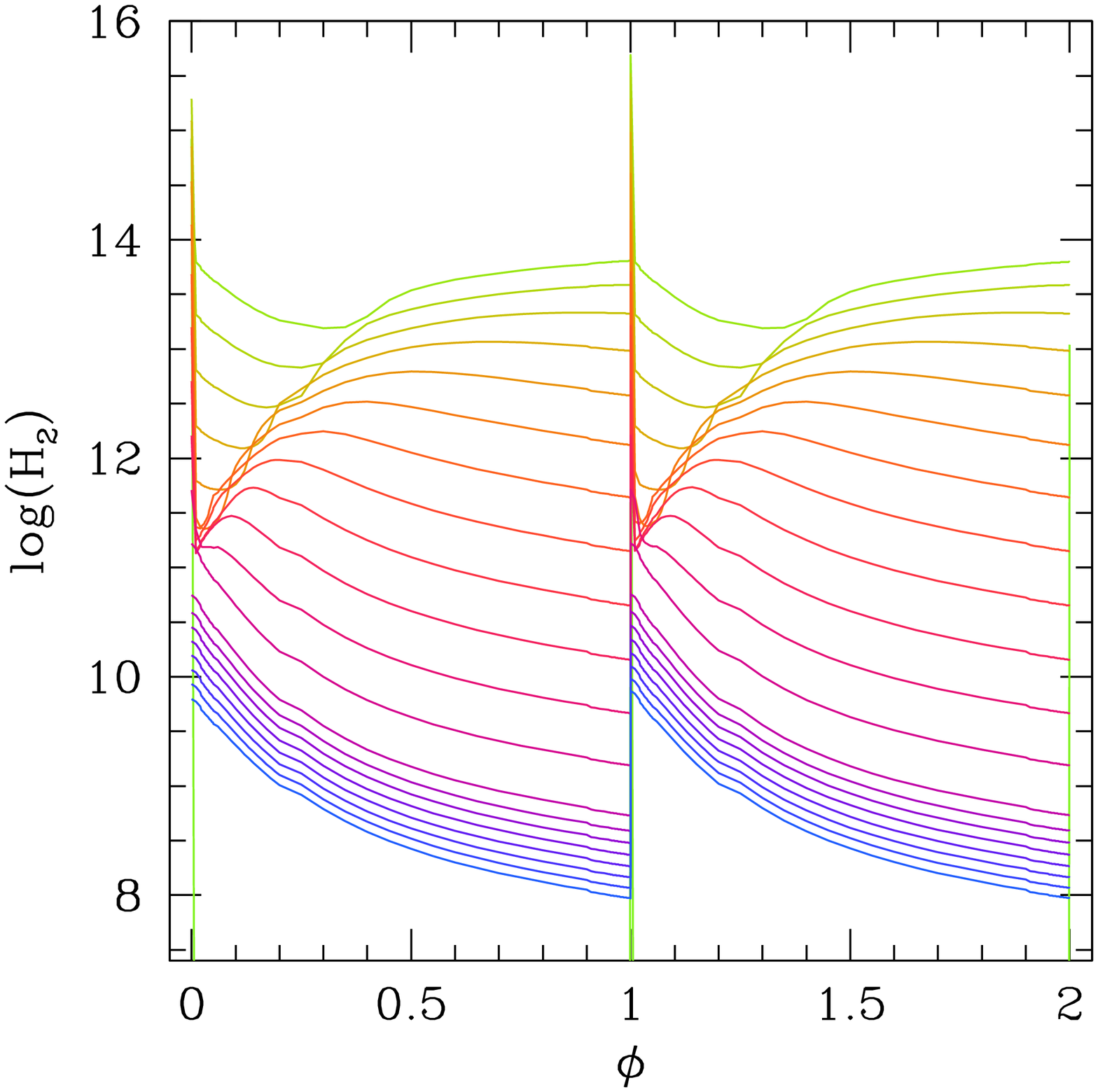}}
\end{minipage}
\caption{ 
Shocked gas properties during the adiabatic expansion, as a function of the phase over two 
pulsation cycles, caused by periodic shock propagation.
In each panel, the bunch of  lines (with colours gradually changing along 
the sequence green-red-blue) 
corresponds to a grid of shocks that develop at  increasing radial distance from 
$r=1\, R$ to $r=5\,R$, in steps of $\Delta r= \upsilon_{\rm gas,1} \times P$.
From top-left to bottom-right the six panels show:
(a) the Lagrangian trajectories of a gas parcel, (b) the gas velocity with respect to
the stellar rest-frame, (c) the mass density and (d) the temperature (the shock-front values
determined by the Rankine-Hugoniot conditions are indicated by filled circles),
(e) the mean molecular weight (in atomic mass units), and (f) the logarithm of 
number density of molecular hydrogen (in cgs units). 
The first shock is assumed to emerge
at  $r_{\rm s,0} = 1\, R$.  The ``effective'' adiabatic exponent 
is taken $\gamma_{\rm ad}^{\rm eff} = 1.1$.
Our dynamic model predicts the  
initial shock amplitude $\Delta\upsilon_0 = 47$ km/s at $r_{\rm s,0}= 1\, R$, the initial 
pre-shock Mach number $M_1 = 10.447$, and 
the  dynamic parameter is $\gamma_{\rm shock} = 0.896$.
The shocked atmosphere corresponds to the quiescent stage just preceding 
the $13^{\rm th}$ thermal pulse of a carbon-rich TP-AGB model, with current mass 
$M=2.311\,,{\rm M}_{\odot}$ (initial mass $M_{\rm i} = 2.6\,,{\rm M}_{\odot}$), 
effective temperature $T_{\rm eff}=3020$ K, luminosity $\log(L/{\rm L}_{\odot})=3.899$, 
pulsation period in the fundamental mode $P_0 = 301$ days, 
photospheric C/O$\,=1.06$, 
mass-loss rate $\dot M = 4.34 \times 10^{-7}\, {\rm M}_{\odot} {\rm yr}^{-1}$, dust condensation
radius $r_{\rm cond} = 2.78\, R$.
}
\label{fig_phi}
\end{figure*}

\subsection{Molecular chemistry}
\label{sect_chem}
The evolution of the gas chemistry through the inner zone of the CSE is followed 
for 78 species (atoms and molecules) using a chemo-kinetic network that
includes $389$ reactions. The network is  essentially the same one as  set up 
by \citet{Cherchneff_12}  in her study of the carbon star  IRC+10216.
It includes termolecular, collisional/thermal fragmentation,
neutral-neutral,  and radiative association reactions and their reverse paths.
The chemical species under consideration are listed in Table~\ref{tab_chem}.

The integration in time of the system of stiff, ordinary, coupled differential
equations  is carried out with the publicly available 
chemistry package KROME\footnote{www.kromepackage.org} \citep{Grassi_etal14}. 
For comparison, 
we will explore also the predictions for the molecular chemistry under the assumption 
of instantaneous chemical equilibrium (hereafter ICE), 
computed with the \texttt{\AE SOPUS} code  \citep{MarigoAringer_09}, which
includes more than 500 molecular  and 300 atomic species.

Unless otherwise specified, in the following we will use the notation 
\begin{equation}
f_{x} = n({x})/[0.5\, n({\rm H}) + n({\rm H}_2)]
\end{equation}
to express the fractional abundance of a given molecule $x$, where
$n(x)$, $n({\rm H})$, and  $n({\rm H}_2)$ are the number densities of
the molecule itself, atomic and molecular hydrogen , respectively.
The denominator $0.5\, n({\rm H}) + n({\rm H}_2)$ gives 
the maximum number density of molecular hydrogen under the hypothesis that 
all available  hydrogen atoms  (not involved in other molecules) were locked in it.
Clearly $[0.5\, n({\rm H}) + n({\rm H}_2)]$ converges to $n({\rm H}_2)$ at lower gas 
temperatures.
The quantity   $f_x$ is built for the practical purpose of  
comparing  model predictions  with observations, 
in which the chemical abundances are usually normalised to that of molecular hydrogen,
under the assumption that all hydrogen is trapped into H$_2$.

\begin{table}
\centering
\caption{Molecular species included in the chemical network
taken from \citet{Cherchneff_12}.
Involved  chemical elements are: H, He, C, N, O, F, Na, Mg, Al, Si, P, S, Cl, K, and Fe.}
\small \begin{tabular}{l} 
\hline 
\hline 
\multicolumn{1}{c}{O-bearing molecules}  \\
\hline 
OH, H$_2$O, CO,  HCO, O$_2$, SiO, SO,  \\
NO, CO$_2$, MgO, FeO, H$_2$OH \\ 
\hline 
\multicolumn{1}{c}{C-bearing molecules} \\
\hline 
CN, HCN, C$_2$H$_2$, C$_2$H$_3$, CH$_2$, C$_2$,\\
CS, SiC, CH, CH$_3$, HCP, CP, C$_3$, \\
C$_2$H,  C$_4$H$_3$,  C$_4$H$_2$, Si$_2$C$_2$, C$_3$H$_2$, \\
C$_3$H$_3$, C$_4$H$_4$, C$_6$H$_5$, C$_6$H$_6$, Si$_2$C$_2$ \\
\hline 
\multicolumn{1}{c}{Other molecules} \\ 
\hline 
H$_2$, NH, N$_2$, NH$_2$, NH$_3$, PH, P$_2$, \\
 PN, SiN, SH, H$_2$S, MgH, MgS,  Mg$_2$, \\ 
 FeH, FeS, Fe$_2$, HF, F$_2$, ClF, AlH, \\
 AlCl, NaH, NaCl, HCl, Cl$_2$, KH, KCl \\
\hline 
\end{tabular} 
\label{tab_chem}
\end{table}

\subsection{Chemo-dynamic integrations}
\label{ssect_integration}
At any stage along the TP-AGB, the stellar parameters at the photosphere $(r=R)$ predicted
by \texttt{COLIBRI}, together with 
the ICE chemical abundances of atoms and molecules provided by 
\texttt{\AE SOPUS},  
are used as initial conditions for the chemo-dynamic integrations of the inner CSE. 
Beyond the photosphere we adopt essentially the same Lagrangian formalism as 
described  by \citet{WillacyCherchneff_98, Duari_etal99, Cau_02,  Cherchneff_06, 
Cherchneff_11, Cherchneff_12}, to whom we refer for all the details. 
A few important modifications are introduced.

At any fixed value of the radial distance $r$,
the chemistry of a gas parcel is solved over two pulsation periods, 
which is enough to check for the periodicity of thermodynamic  and abundance parameters.  
The  molecular concentrations after two pulsation cycles at $r$ are then used as input for the next
two chemical integrations at  $r+\Delta r$, once  re-scaled 
to the new pre-shock gas density.
We take the incremental step $\Delta r =P \times \upsilon_ {\rm gas,1}(r)$, 
that is the distance traveled
by the gas element during one pulsation period with the gas velocity $\upsilon_{\rm gas,1}$,
computed as detailed in Sect.~\ref{ssect_vgas}.

Each of the two integrations includes both 1) the 
``very fast chemistry''  of the chemical cooling 
layer introduced by \citet{WillacyCherchneff_98}, and 2) the prolonged chemistry activity 
during the post-shock hydrodynamic cooling.
In both regimes, the non-equilibrium chemistry is solved with the \texttt{KROME} routines. 

Following \citet{WillacyCherchneff_98}  the length of the thin layer 
at the shock front is determined 
by the dissociation of molecular hydrogen due to collisions 
with hydrogen atoms, and it is calculated with
\begin{equation}
\ell_{\rm diss}=\frac{\upsilon_{\rm shock}}{n_2(\rm{H}) \mathscr{R_{\rm{H, H_2}}}}\,,
\end{equation}
where $\upsilon_{\rm shock}$ is the shock velocity in the stellar rest-frame,
$n_2(\rm{H})$ is the post-shock concentration of hydrogen 
predicted by the Rankine-Hugoniot conditions, and 
$\mathscr{R_{\rm{H, H_2}}}$ $[\rm{cm}^{3}s^{-1}]$ is the rate of the
reaction H+H$_2 \longrightarrow$ 3H.
The dissociation length $\ell_{\rm diss}$ varies significantly with the distance from the
star, typically ranging from $\simeq$ few cm to $10^7$ cm as one moves from the inner 
regions close to the photosphere, 
across the interval $1 \la r/R \la 5$.
The link between the end of the chemical cooling stage and the beginning of
the hydrodynamic relaxation is performed as detailed in \citet[][]{Cau_02}.
  
We note that applying a Lagrangian formalism over the radial extension of the inner CSE,
it is possible to obtain an Eulerian-type profile of the chemical species which can 
eventually  be compared with the measured concentrations.

In this initial study we do not consider any process of dust condensation and growth,
which we postpone to follow-up works, and we focus on the hydrogen cyanide molecule,
for which observational evidence suggests that the formation is not affected  
by dust grain formation \citep{Schoier_etal13}.

For illustrative purposes, Fig.~\ref{fig_phi} shows the solution for the chemo-dynamic structure 
of an AGB pulsating stellar atmosphere traversed by periodic shocks, over two pulsation periods.
In the Lagrangian framework, the variations of the fluid variables 
(radius, velocity, density, temperature, mean molecular weight, and molecular hydrogen abundance) 
are plotted as a function of the pulsation phase along the decelerating particle trajectory

The outward  propagation of the shock is followed from 
its first appearance  at $r_{\rm s, 0}=1 R$  up to
$r=5\,R$, according to a discrete grid of radial meshes. 
The reference system is the stellar rest-frame, in which the radial distance is measured 
from the centre of the star under the assumption of
spherical symmetry. Positive gas velocities correspond
to the initial upward displacement of the gas parcel driven by the pressure forces, 
while negative gas velocities describe the subsequent downward motion due to the gravitational pull, 
back to  the original location.  

During the post-shock relaxation both the mass density and the temperature are expected to
decrease until $\phi=1$, when the fluid element  recovers its initial values, 
experiences the next shock and the cycle repeats itself.

It is also interesting to examine the evolution of the mean molecular weight $\mu$ of the shocked gas.
For shocks that emerge closer to the star ($r < 3\,R$), $\mu$ is seen to increase with the pulsation
phase $\phi$, due to the progressive increment of the molecular concentrations
relative to the atomic abundances, which instead dominate at $\phi \simeq 0$ (where
$\mu \approx 1.35-1.40$).
 This fact is explained by the decrease of the temperature
 during the post-shock gas expansion. At larger distances ($r > 3\,R$), the temperature
at the beginning of the dynamic relaxation is  low and  
the gas is mainly in molecular form already at $\phi\simeq 0$. In these models the 
mean molecular weight remains almost constant during the whole pulsation cycle, 
attaining values ($\mu \approx 2.40-2.45$), 
that are typical of a gas chemistry dominated by molecular hydrogen and carbon monoxide.

\section{CSE abundances of hydrogen cyanide}
\label{sect_hcn}
We follow the chemistry of several molecular species in our model (see Table~\ref{tab_chem}).
However, the discussion of all trends and results is not addressed in this initial paper, and 
we postpone a thorough analysis to follow-up works. 

As a first application we decided to focus on the hydrogen cyanide molecule for two 
main reasons.
First,  we can rely on plentiful data 
of HCN concentrations measured in CSEs of AGB stars, so that a statistical study
is now feasible. 
Second, the HCN abundances have proved an excellent tool for calibrating the 
dynamics of the inner CSEs, without running into the 
uncertainties and complications of the growth of dust grains which, in fact, 
does not affect the chemistry of the HCN molecule.

Indeed, the HCN molecule has been detected in AGB circumstellar envelopes of all chemical types
\citep{Schoier_etal13, Muller_etal08, Marvel_etal05, Bienging_etal00}.
Using a detailed radiative transfer model \citet{Schoier_etal13} have recently derived
the circumstellar HCN abundances  
for a large sample of AGB stars, spanning 
a wide range of properties, mainly in terms of chemical types, mass-loss rates, 
and expansion velocities.

This sample notably enlarges the observed range of the parameter space, and
therefore it  provides a robust statistical test for  chemical models that aim at
investigating the molecular chemistry in the inner CSE regions close to the photospheres
long-period variables.

\begin{figure}
\resizebox{\hsize}{!}{\includegraphics{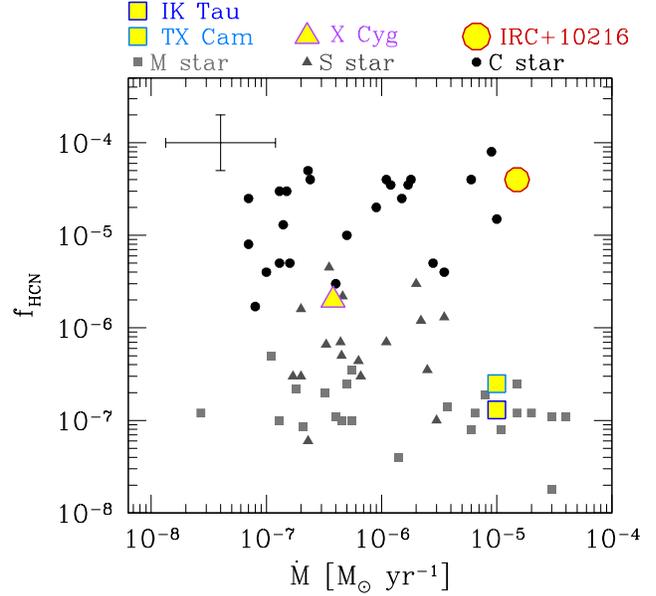}}
\caption{HCN fractional abundances relative to molecular hydrogen, 
$f({\rm HCN}) = n({\rm HCN})/[n({\rm H}_2+0.5 n({\rm H})]$, as a function of the mass-loss rate 
for a sample of 69 AGB stars that include M stars (squares), S stars (triangles), and C stars (circles).
HCN data are obtained from a multi-transition survey of HCN (sub-) millimeter line emission 
carried out by \citet{Schoier_etal13}. 
Typical error bars in the measurements of $f({\rm HCN})$ and $\dot M$ 
are shown. Well-known stars are marked as in the top legenda.}
\label{fig_hcndata}
\end{figure}

The main result of the analysis of \citet{Schoier_etal13} is the strong dependence
of  the HCN abundances on the photospheric C/O ratio, with a clear trend of increasing
HCN/H$_2$ when moving along the spectral sequence of M-S-C.
This property is shown in Fig.~\ref{fig_hcndata}, where the estimated HCN abundances
are plotted as a function of the mass-loss rate for the three chemical types.

The authors also concluded that the observed sensitivity of the HCN abundance on C/O
is in sharp contrast with shock-induced chemistry models, in which the predicted 
HCN abundances are essentially insensitive to the stellar chemical type \citep{Cherchneff_06}.
It should be noted that the results of those shock-induced chemistry models about the 
HCN dependence on the C/O ratio were obtained 
for a single combination of stellar parameters ($M=0.65\,{\rm M}_{\odot}, R=280 R_{\odot},
P=557$ days), and a fixed choice of dynamic parameters ($r_{\rm s,0} = 1 R, \gamma_{\rm shock}=0.89, 
\Delta\upsilon_{0}=25$ km/s).
Furthermore, the results of \citet{Cherchneff_06} for HCN might be obsolete following the 
significant revision of the chemical network carried out by \citet{Cherchneff_11, Cherchneff_12}. 
This is the main reason why we choose to use this later revised network in the present study.

In this work we aim at extending the shock-chemistry analysis 
over a wide range of physical  parameters, most of which are extracted  directly 
from stellar evolutionary calculations for the TP-AGB phase.

In the following we analyse the predictions for HCN abundances in the inner CSEs of TP-AGB stars,
both under the assumption of instantaneous chemical equilibrium, and in the case of non-equilibrium chemistry applied to
periodic shocks.
To this aim, we will test the results for a TP-AGB evolutionary track with initial
mass $M_{\rm i} = 2.6\,{\rm M}_{\odot}$ and metallicity $Z_{\rm i} = 0.017$, 
computed with the \texttt{COLIBRI} code \citep{Marigo_etal13}.
This model is predicted to experience the third dredge-up and to become a carbon star
attaining a final surface C/O$\,=1.55$. 
A summary of the evolutionary properties along the TP-AGB phase 
is given Sect.~\ref{sect_tpagb}, and illustrated in Fig.~\ref{fig_2.6z0.017tpagb}.
Calculations for other TP-AGB  models with different initial masses will be also presented.

In general we choose to compare observations to our $f_{\rm HCN}$ 
abundances in the range $3 < r/R < 5$ because at those distances the HCN chemistry
is frozen out.  At the same time these radii are much smaller than the
typical e-folding radius for HCN inferred from observations, so that
our predictions are directly comparable with the initial abundances
$f_0$, as reported by \citet[][see  their section 3.4]{Schoier_etal13}.

\begin{figure*}
\centering
\begin{minipage}{0.44\textwidth}
\resizebox{\hsize}{!}{\includegraphics{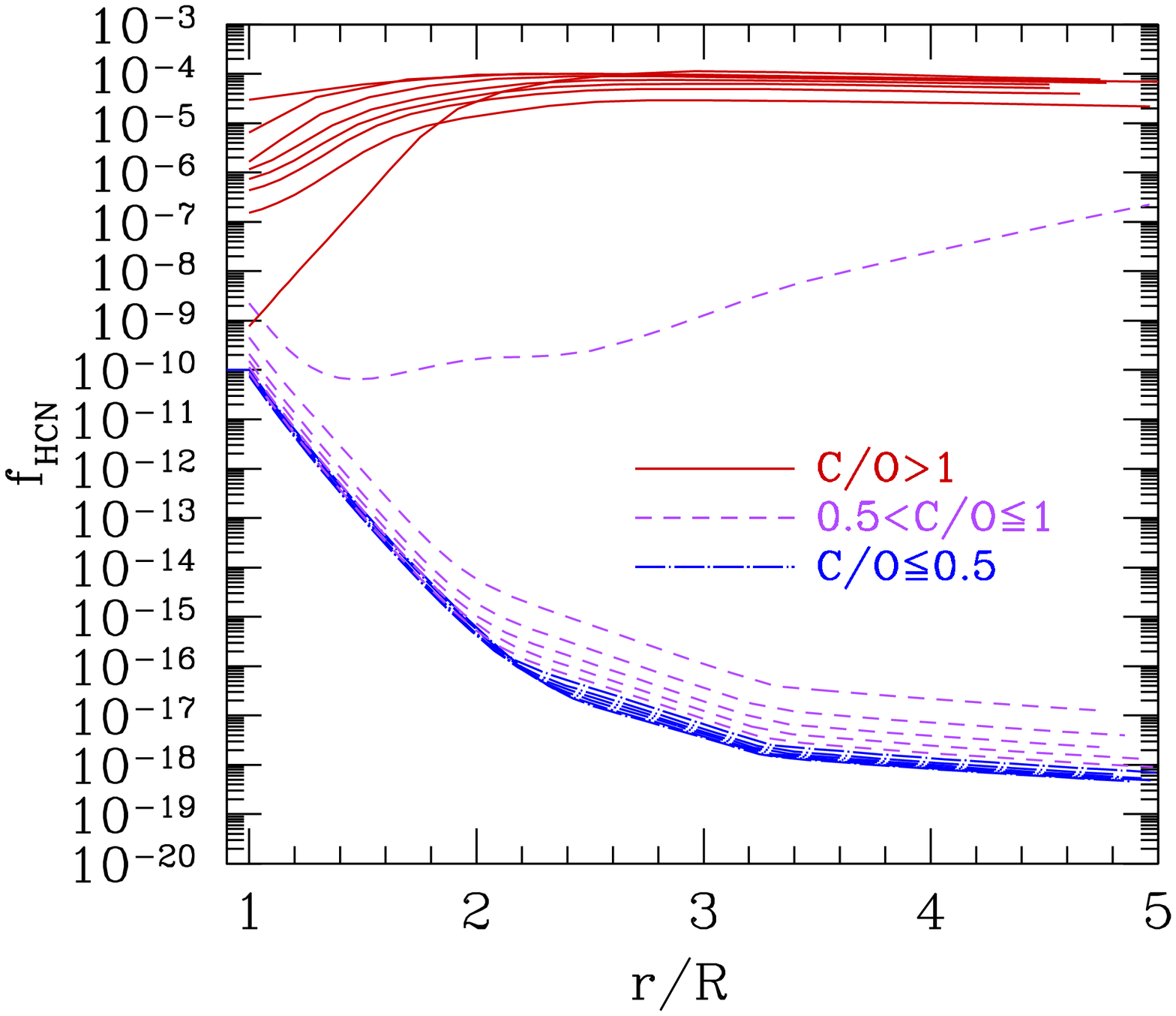}}
\end{minipage}
\begin{minipage}{0.45\textwidth}
\resizebox{\hsize}{!}{\includegraphics{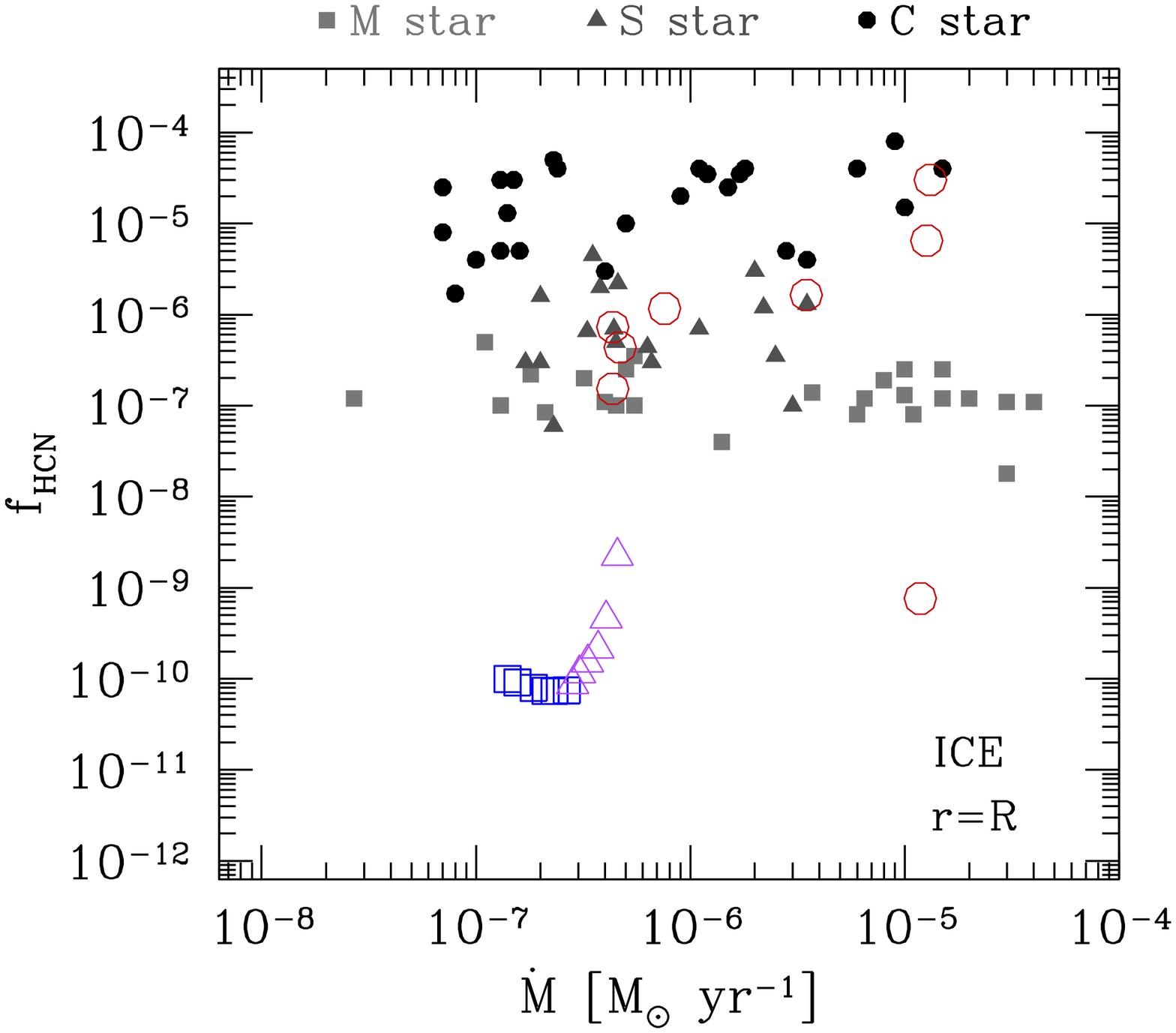}}
\end{minipage}
\caption{Fractional abundance $f_{\rm HCN}$ under the assumption of instantaneous chemical 
equilibrium (ICE) with the \texttt{\AE SOPUS} code \citep{MarigoAringer_09}, corresponding
to our reference model with initial mass $M_{\rm i} = 2.6\,{\rm M}_{\odot}$ and metallicity 
$Z_{\rm i} = 0.017$, which is expected
to experience 20 TPs  evolving along the chemical sequence M-S-C due to the third dredge-up.
At each radius chemistry predictions are shown after the completion 
of two consecutive pulsation cycles.
{\em Left panel}:  Radial profiles of  HCN/H$_2$ concentrations across the $r=1-5\,R$ 
circumstellar envelope. Each line corresponds to the stellar evolutionary stage 
just preceding a thermal pulse.  
{\em Right panel}: Fractional abundances $f_{\rm HCN}$ 
at the photosphere (empty symbols) under the ICE assumption, 
as a function of the mass-loss rate during
the entire TP-AGB evolution. 
Models are denoted with blue squares for C/O $\le 0.5$,
purple triangles for $0.5 < $ C/O $\le 1.0$, and red circles for 
C/O $> 1.0$.
Observed data are the same as in Fig.~\ref{fig_hcndata}.}
\label{fig_ice}
\end{figure*}

\subsection{Equilibrium chemistry}
Several studies have already demonstrated that the chemistry that
controls the inner wind of AGB stars is not at equilibrium owing to
the presence of shocks \citep[e.g.,][]{WillacyCherchneff_98, 
Duari_etal99, Cherchneff_06, Cherchneff_12}, but as a first test, we consider the
case of the instantaneous chemical equilibrium (ICE), which we compute
with the\texttt{\AE SOPUS} code \citep{MarigoAringer_09}.
Figure~\ref{fig_ice} shows the predicted abundances of HCN relative to molecular
hydrogen as a function of the distance from the stellar centre, for the reference 
model with initial mass $M_{\rm i} = 2.6\,{\rm M}_{\odot}$ and metallicity 
$Z_{\rm i} = 0.017$. 

As expected, the equilibrium abundance of HCN is remarkably sensitive to the 
C/O ratio. As long as the star remains oxygen-rich with C/O$\,<0.9$, the HCN concentration
is quite low and decreases significantly while moving away from the stellar
photosphere (left panel of Fig.~\ref{fig_ice}). Correspondingly, at any radial distance the
predicted ICE abundances are lower than the measured ones for M stars 
by several orders of magnitudes.
The comparison with the photospheric abundances for the equilibrium chemistry 
(right panel of Fig.~\ref{fig_ice}) underlines that even in the most favorable conditions 
of high gas density, the predicted $f_{\rm HCN}$ for M and S stars ($\simeq 10^{-10}$) 
is already much lower than the observed data by few orders of magnitude.

As the C/O ratio increases due to the third dredge-up events and enters the critical narrow range 
regime  $0.9\le$ C/O $<1.0$ the equilibrium abundances of 
HCN abruptly rise,  spanning a remarkably wide range. However, even the  highest values 
(HCN/H$_2\simeq 10^{-9}$ for C/O $\simeq 0.967$ at the 12$^{\rm th}$ thermal pulse) 
keep well below the observed median concentration for S stars (HCN/H$_2 \simeq7 \times  10^{-7}$).

Eventually, as soon as the star becomes carbon rich (C/O $>1$)
a remarkable increase of the HCN abundances  (by several order of magnitudes)  takes place  
and the predicted equilibrium abundances 
reach values that appear to be  consistent with the measured data for carbon stars,
starting from $r/R > 2$.

In view of the above, we may conclude that the ICE chemistry is not suitable
for explaining the observations of HCN in the CSEs of AGB stars with different 
chemical types, as already mentioned by \citet{Cherchneff_06}.

\subsection{Non-equilibrium chemistry: dependence on main shock parameters}
In the following we will analyse the HCN abundances predicted by calculations of  
non-equilibrium chemistry regulated by the occurrence of pulsation-induced shocks
which propagate across the CSE.

In particular, we will investigate the dependence of the results on two fundamental parameters, 
namely the shock formation radius $r_{\rm s,0}$, and the effective adiabatic 
index $\gamma_{\rm ad}^{\rm eff}$ (see Sect.~\ref{ssect_pulshock}).
In fact, both quantities control the excursions in density and temperature 
experimented by a gas parcel shocked by pulsation.

\subsubsection{The effective adiabatic index}
\label{sssect_gameff}
In the shock formalism adopted here \citep{BertschingerChevalier_85, WillacyCherchneff_98} 
$\gamma_{\rm ad}^{\rm eff}$ is the key input parameter which, together the
pre-shock Mach number $M_1$ determines the temperature $T_{\rm c}$ 
and the density $\rho_{c}$ at the top of the deceleration region, whence
the hydrodynamic expansion phase starts.
In general, the higher $\gamma_{\rm ad}^{\rm eff}$, the larger $T_{\rm c}$ and 
$\rho_{c}$ are attained. 
They correspond to the pulsation phase $\phi=0$ in panel c) and d) of Fig.~\ref{fig_phi}.
\begin{figure*}
\centering
\begin{minipage}{0.38\textwidth}
\resizebox{\hsize}{!}{\includegraphics{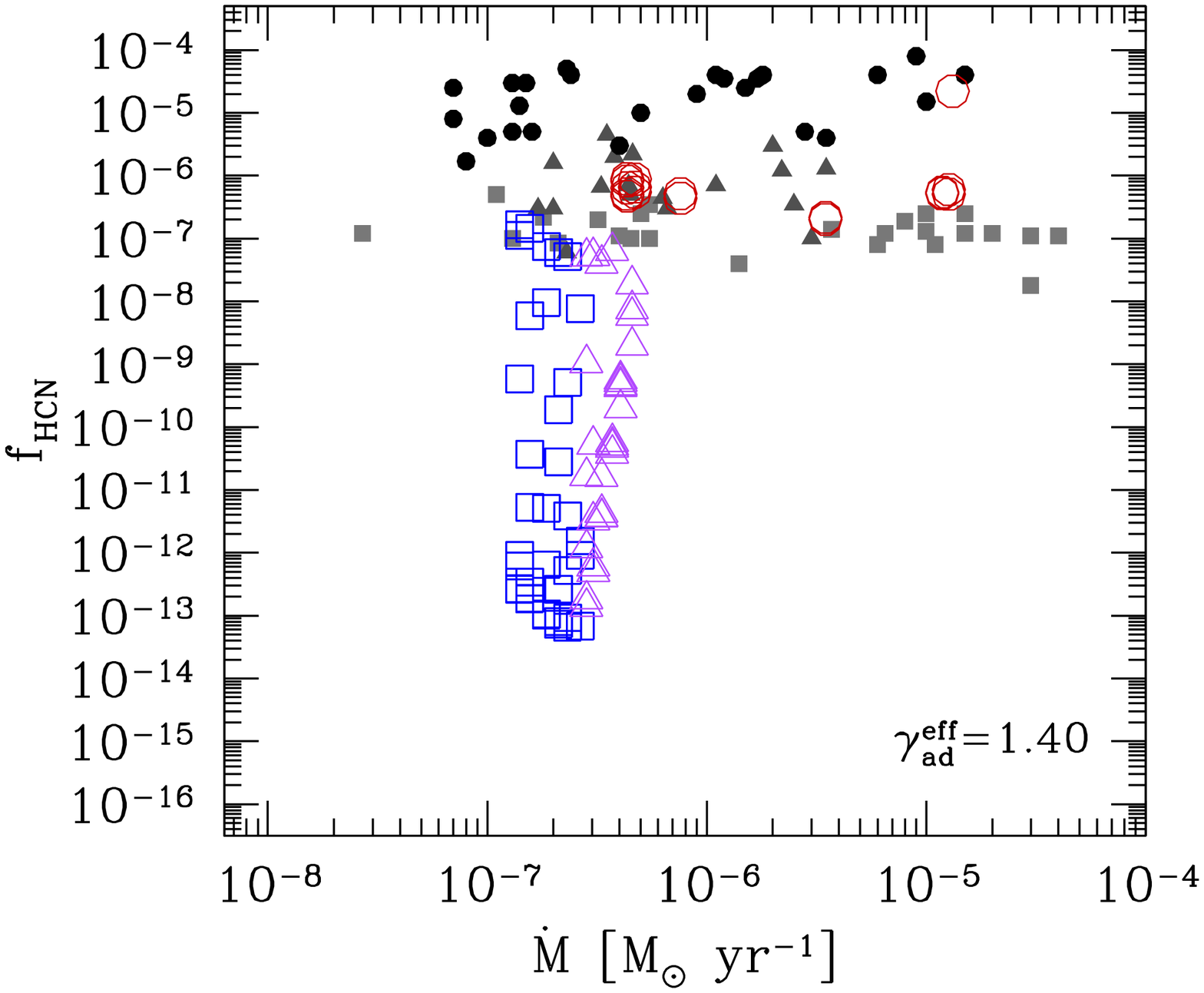}}
\end{minipage}
\begin{minipage}{0.38\textwidth}
\resizebox{\hsize}{!}{\includegraphics{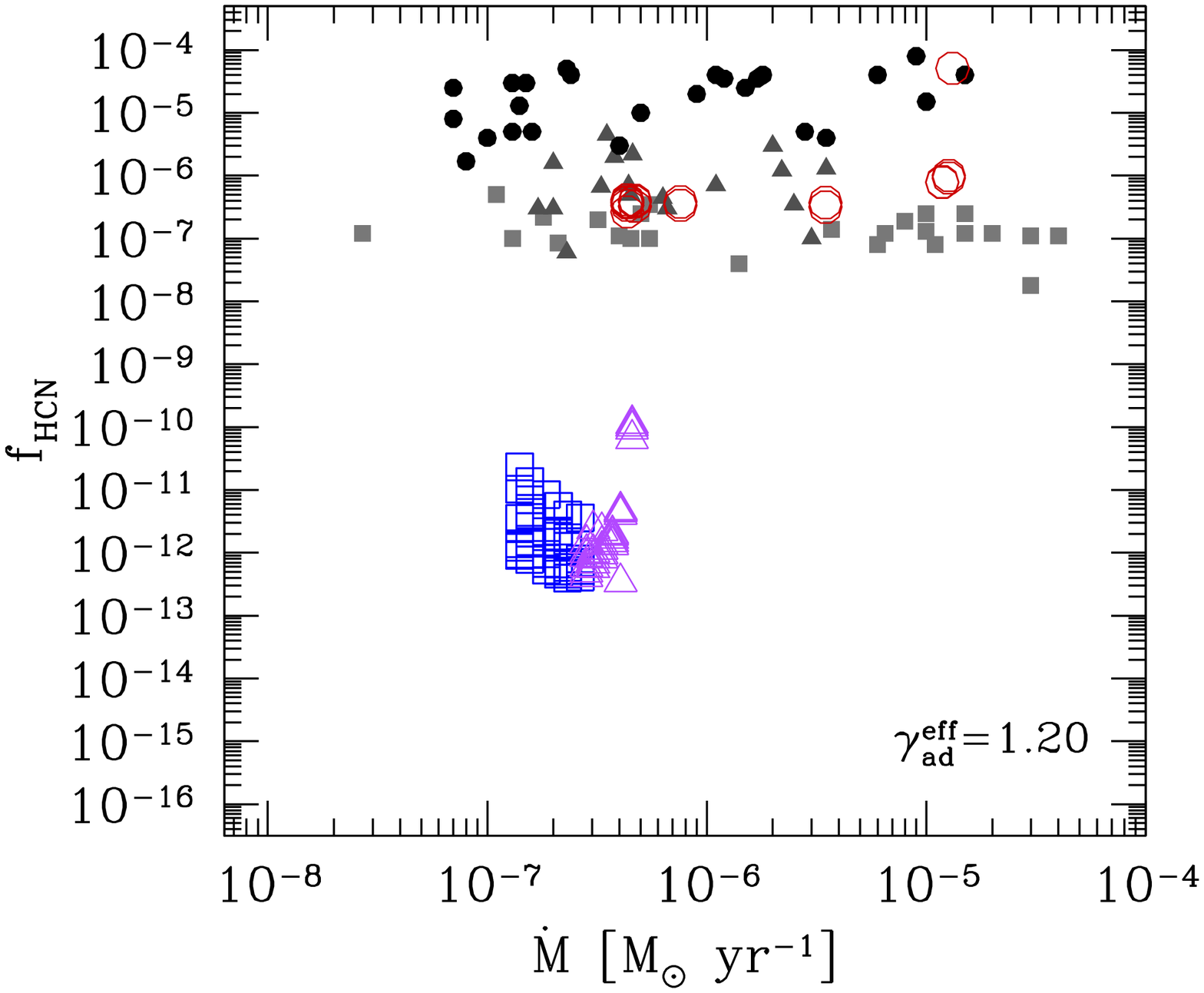}}
\end{minipage}
\begin{minipage}{0.38\textwidth}
\resizebox{\hsize}{!}{\includegraphics{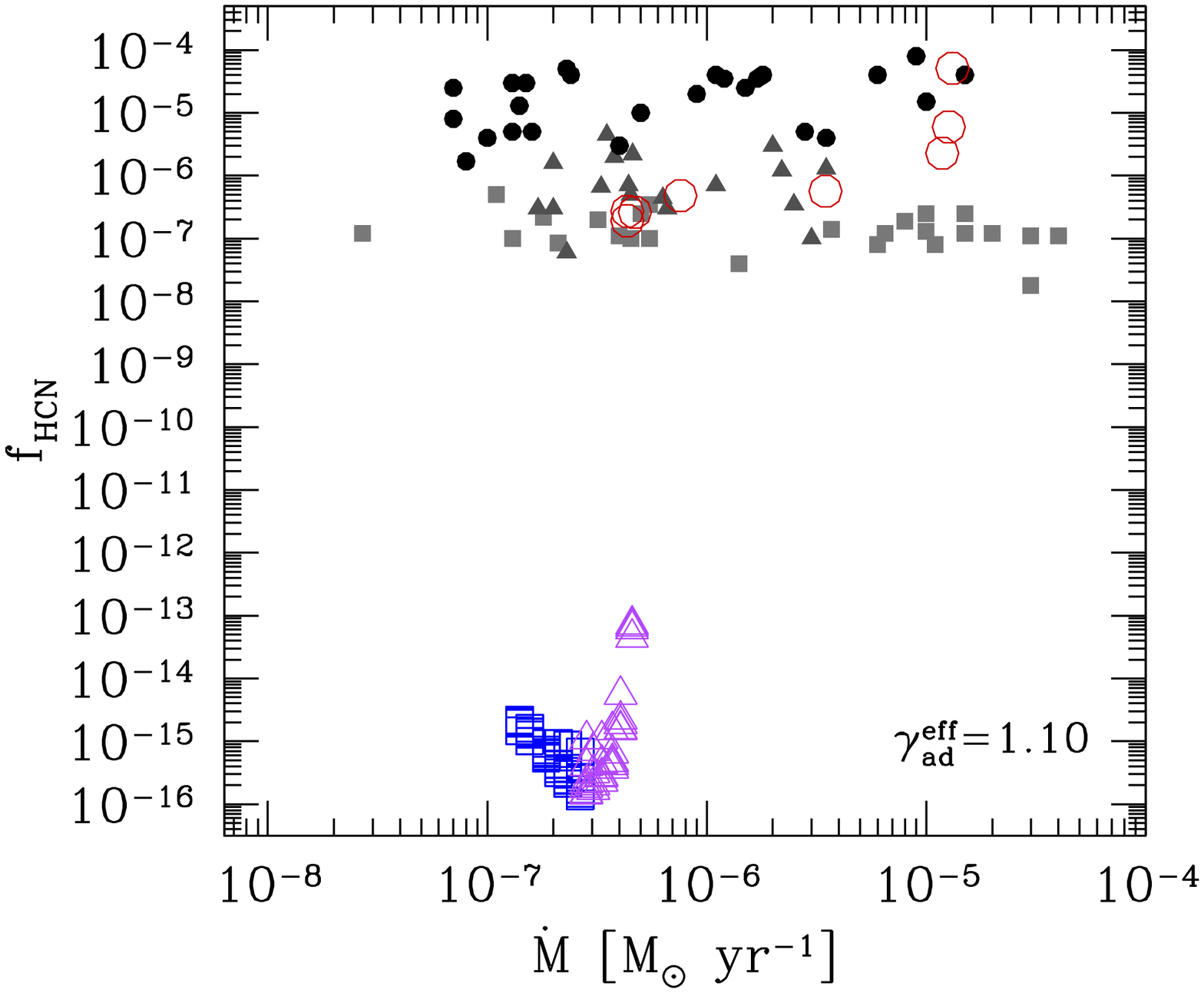}}
\end{minipage}
\begin{minipage}{0.38\textwidth}
\resizebox{\hsize}{!}{\includegraphics{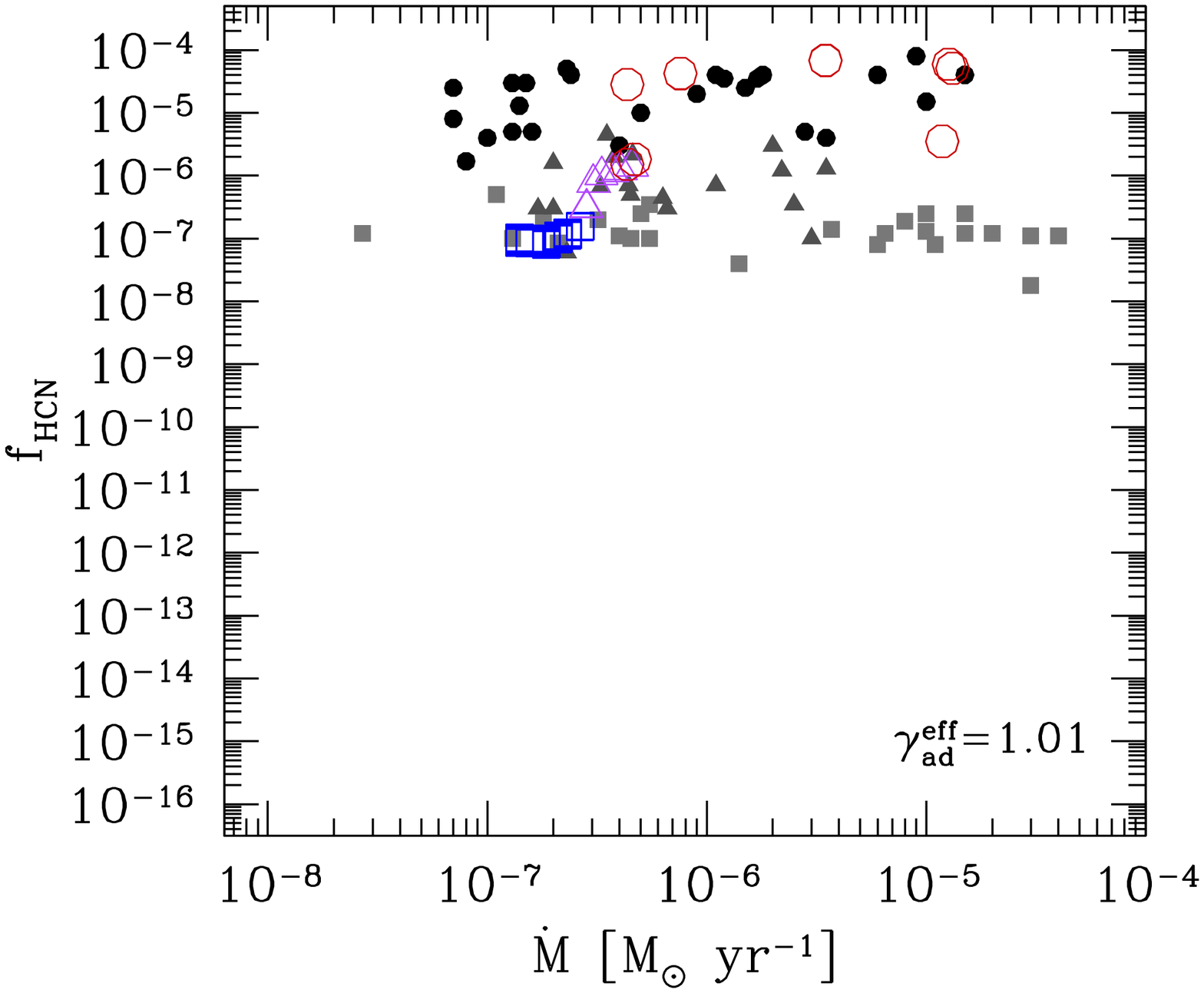}}
\end{minipage}
\caption{Comparison between observed fractional abundances $f_{\rm HCN}$  and predictions
of shock-chemistry models at varying the effective adiabatic 
index $\gamma_{\rm ad}^{\rm eff}$. In all calculations we assume $r_{\rm s,0}/R= 1$.
Open symbols correspond to predicted abundances
over the entire TP-AGB evolution of a model with initial
mass $M_{\rm i} = 2.6\,{\rm M}_{\odot}$ and metallicity $Z_{\rm i} = 0.017$, which is expected
to evolve along the chemical sequence M-S-C due to the third dredge-up.
At given mass-loss rate model abundances are shown across the distance range $r=3-5\, R$.
Note the striking agreement with the data for shock models that assume 
$\gamma_{\rm ad}^{\rm eff}=1.01$, which points to
a prevalent isothermal character of the shocks in the inner CSE. 
}
\label{fig_hcn_gamad}
\end{figure*}

We have explored a few choices of  the effective adiabatic index, namely
 $\gamma_{\rm ad}^{\rm eff}= 1.4, 1.2, 1.1, 1.01$.
The highest value is often adopted in previous studies \citep[e.g.,][and references therein]{Cau_02}. 
The results are illustrated in the four panels of Fig.~\ref{fig_hcn_gamad}, corresponding to 
our reference TP-AGB stellar model.
A few considerations can be drawn.

For all three values $\gamma_{\rm ad}^{\rm eff}> 1$ the HCN concentrations
are always below the measured data at any value of the photospheric C/O ratio.
The theoretical underestimation is more pronounced for the M and S stars, 
amounting to several orders of magnitudes.
In all cases the radial profile of HCN abundance is characterised by large variations,
which explains the sizable scatter of the predictions over a distance range $3\la r/R \la 5$,  
at a given mass-loss rate.
A significant improvement of the model-data comparison 
is obtained when lowering the effective adiabatic index 
to  $\gamma_{\rm ad}^{\rm eff}\simeq 1$. 

We note that lowering $\gamma_{\rm ad}^{\rm eff}$ 
from 1.1 to 1.01 produces a large increase of $f_{\rm HCN}$ 
for M and S models. Such a dramatic change is related to the
evolution of temperature and density during the post-shock relaxation,
which affects the HCN formation through the H$_2$/H ratio (see the
extensive discussion in Sect.~\ref{ssect_hcn}). When  $\gamma_{\rm ad}^{\rm eff}$ 
is relatively high, the
higher temperature at early phases destroys/prevents the formation of
H$_2$, whereas at later phases (when the temperature would be suitable
for H$_2$ formation), the rate of H$_2$ formation is low because of the low
density reached during the post-shock relaxation.  When $\gamma_{\rm ad}^{\rm eff}$
 is close to one (radiative shocks), the cooling is so efficient that the
temperature is suitable for H$_2$ formation at almost all phases.

Adopting $\gamma_{\rm ad}^{\rm eff} \simeq 1$
the predicted HCN abundances over the $3\la r/R \la 5$ radial distance and
during the TP-AGB evolution match the observed data points very well.
In particular, the model recovers the observed stratification of the HCN
concentrations that are seen to increase, on average, when moving along the
chemical sequence of M, S, and C stars.
This result supports the empirical fact that the photospheric C/O  ratio is one of the  major
factors determining the CSE chemistry of HCN, and is at variance with previous 
conclusions \citep{Cherchneff_06}.

Moreover, we note that for $\gamma_{\rm ad}^{\rm eff}\simeq 1$ 
the theoretical dispersion of the HCN abundances becomes quite narrow 
at varying radial distance in the $3\la r/R \la 5$ interval. This finding means that 
the shock-chemistry for HCN tends to freeze out at increasing distance 
(see also the left panel Fig.~\ref{fig_hcn_rad}), and is consistent
with the observational evidence that the HCN concentrations attain typical values
which do not change dramatically, especially within a given chemical class 
of mass-losing AGB stars.

In conclusion, our tests suggest that in order to explain the 
CSE concentrations of HCN in the framework of the non-equilibrium 
chemistry, the pulsation-induced shocks in the inner zone around  AGB stars  
should be described by a low value of the effective adiabatic index, that is  
$\gamma_{\rm ad}^{\rm eff} \simeq 1$.

It is quite interesting to interpret this finding in terms of physical implications. 
If radiative processes are not important, $\gamma_{\rm ad}^{\rm eff}$ is determined by 
thermodynamic equilibrium of the gas. For instance, for a gas mostly consisting of hydrogen, 
$\gamma_{\rm ad}^{\rm eff}=5/3$ for atomic hydrogen,
$\gamma_{\rm ad}^{\rm eff}=1.4$  for molecular hydrogen without vibrational degrees, 
$\gamma_{\rm ad}^{\rm eff}=9/7$  for molecular hydrogen with vibrations excited
\citep{ZeldovichRaizer_66}.

Deviations toward lower values for $\gamma_{\rm ad}^{\rm eff}$ 
may actually occur if radiative processes become significant in the thermal relaxation 
behind the propagating shock waves, so that the assumption of 
thermodynamic equilibrium cannot be safely applied.
\begin{figure}
\centering
\resizebox{\hsize}{!}{\includegraphics{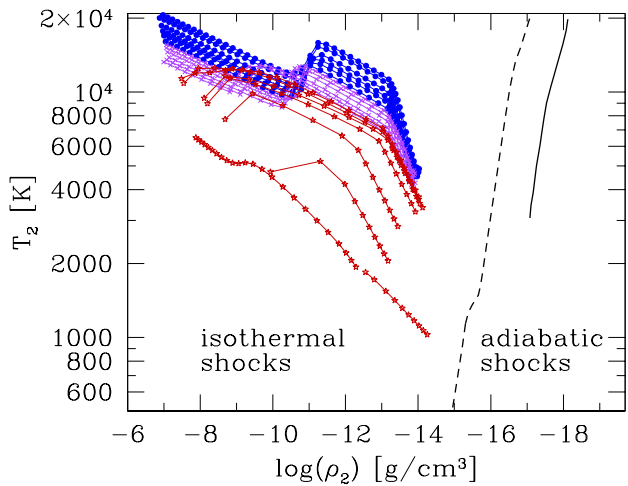}}
\caption{Post-shock density and temperature of the gas, immediately behind the shock front, as determined
by the Rankine-Hugoniot conditions. Each track shows  the evolution of the post-shock conditions as the
shock propagates across the circumstellar envelope from $r=1 R$ to $r=5 R$, and refers to the quiescent condition (pre-flash luminosity
maximum) just prior the occurrence of each thermal pulse experienced 
by the $M_{\rm i}=2.6\, {\rm M}_{\odot}, Z_{\rm i}=0.017$ model.
Symbols are color-coded as a function of the photospeheric C/O ratio,
blue for M stars, purple for S stars, and red for C stars.
The diagonal solid and dashed lines roughly mark the separation 
between shocks of predominantly isothermal
and predominantly adiabatic character, following the detailed analysis of 
\citet[][see text for more details]{Woitke_97}.}
\label{fig_rho2t2}
\end{figure}

An indication that this may indeed be the case for Mira variables was derived by 
\citet{BertschingerChevalier_85} from their model fits to the observed 
CO $\Delta\upsilon=3$ excitation temperature as a function of the phase for four
stars (o Ceti, T Cep, $\chi$ Cyg, R Cas). In all cases they obtained  
$\gamma_{\rm ad}^{\rm eff}\simeq 1.1$.

In this context, \citet{Woitke_97} provided a  thorough discussion about the character 
of the thermal relaxation behind the shock,  which is critically affected by 
the time scale necessary for the  gas to radiate away its excess of internal 
energy dissipated by the shock.
In brief, the thermal behavior of the gas in the CSEs of pulsating stars depends on the
comparison between the radiative cooling time scale $\tau_{\rm cool}$ 
and the stellar pulsation period $P$:
\begin{itemize}
\item If $\tau_{\rm cool} \ga P$ then the gas cannot
radiate away the energy dissipated by one shock within one pulsation period, 
and therefore it cools adiabatically.\\
\item If $\tau_{\rm cool} \ll P$ the gas quickly relaxes to radiative equilibrium
behind the shock before the next shock wave hits it, so that the dominant character
is that of an isothermal shock.
\end{itemize} 

On the temperature-density diagram one can conveniently define a ``critical cooling track'' 
along which the maximum radiative cooling time is $\tau_{\rm cool} = 1$ yr, considering 
one year as a typical period of a Mira variable. Two such critical lines are drawn in
 Fig.~\ref{fig_rho2t2} according to the computations of \citet{Woitke_97}, 
for two choices of the radiation temperature, i.e.  $T_{\rm rad} = 0$ K (dashed line), 
and  $T_{\rm rad}=3000$ K (solid line). 

Gas elements that are shocked to the left of the critical cooling tracks can reestablish radiative
equilibrium before the next shock arrives (dominant isothermal character),
those which are shocked to the right of the critical cooling track will be hit by the
next shock before radiative equilibrium can be achieved (dominant adiabatic character).

We notice that the post-shock values of temperature $T_2$ and density $\rho_2$ predicted 
with our shock models 
typically lie in the region where $\tau_{\rm cool} \ll P$.
As a consequence, we expect that the periodic shocks that hit the gas 
in the high-density inner circumstellar envelope (within few R) present a dominant
isothermal character.

In the framework of the shock formalism adopted in the present study, isothermality 
corresponds to assuming a low value  for $\gamma_{\rm ad}^{\rm eff}$, close to unity,
which is in excellent agreement with the findings  we have derived from the analysis 
of the HCN concentrations just discussed above.
It is important to emphasize that the isothermal character of the shocks which emerge 
in the inner zone of the CSEs, derived from the detailed physical modelling of \citet{Woitke_97},
 nicely provides  a physically independent support to our calibration analysis
which is, instead, based on an purely chemistry-related ground.

\subsubsection{The shock formation radius}
The shock formation radius $r_{\rm s,0}$ is another critical parameter of our
chemo-dynamic model, since it determines
 the boundary between the static region and the pulsation-shocked region in the 
extended atmosphere,
across which the average pre-shock density profile is expected to decrease with  
two quite different scale-heights, as described in Sect.~\ref{ssect_rho}.

Since the static scale-height $H_{\rm 0}$ is shorter than the dynamic scale-height 
$H_{\rm dyn}$ by a factor  $(1-\gamma_{\rm shock}^2)$, it is clear that the farther the location 
$r_{\rm s,0}$, the lower the densities in the pulsation-shocked region.
The density, in turn, affects the integration of the non-equilibrium chemistry network,
hence the concentrations of the different species.

Studies of pulsation-shocked chemistry dealing with the CSE chemistry 
of individual AGB stars, which are based on the \citet{Cherchneff_etal92} 
formalism for the density, locate 
the shock formation radius at typical distances  $r_{\rm s,0}=1.0, 1.2, 1.3\, R$ 
\citep{Cherchneff_12, Agundez_etal12, Cherchneff_11, Cherchneff_06, AgundezCernicharo_06, 
Cau_02, DuariHatchell_00, Duari_etal99, WillacyCherchneff_98, Cherchneff_etal92}.

On the theoretical ground, early numerical models of pulsating atmospheres 
\citep{WillsonBowen_86,  Bowen_88} show that the position of the material when it first 
encounters the shock wave $r_{\rm s,0}$, the maximum shock amplitude $\Delta\upsilon_{0}$, 
and the pulsation period  $P$  (in a given mode) are inter-related quantities.
In general the larger the maximum shock amplitude, 
hence the higher $\gamma_{\rm shock}$, the deeper in the atmosphere 
the first shock is expected to develop. We see that in many cases  with
higher shock amplitudes, $r_{\rm s,0}$
may be located even below the photosphere.

\begin{figure*}
\centering
\begin{minipage}{0.38\textwidth}
\resizebox{\hsize}{!}{\includegraphics{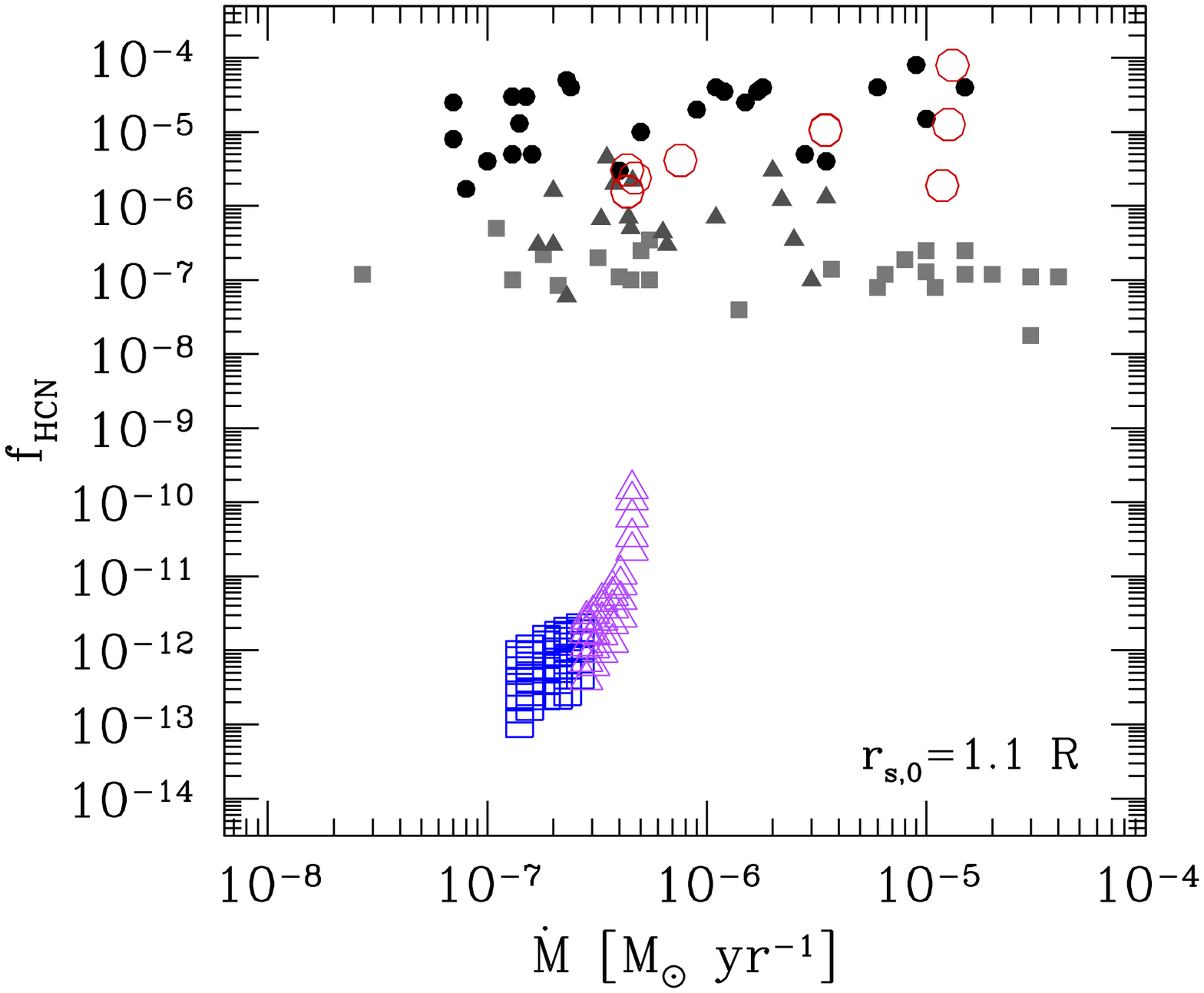}}
\end{minipage}
\begin{minipage}{0.38\textwidth}
\resizebox{\hsize}{!}{\includegraphics{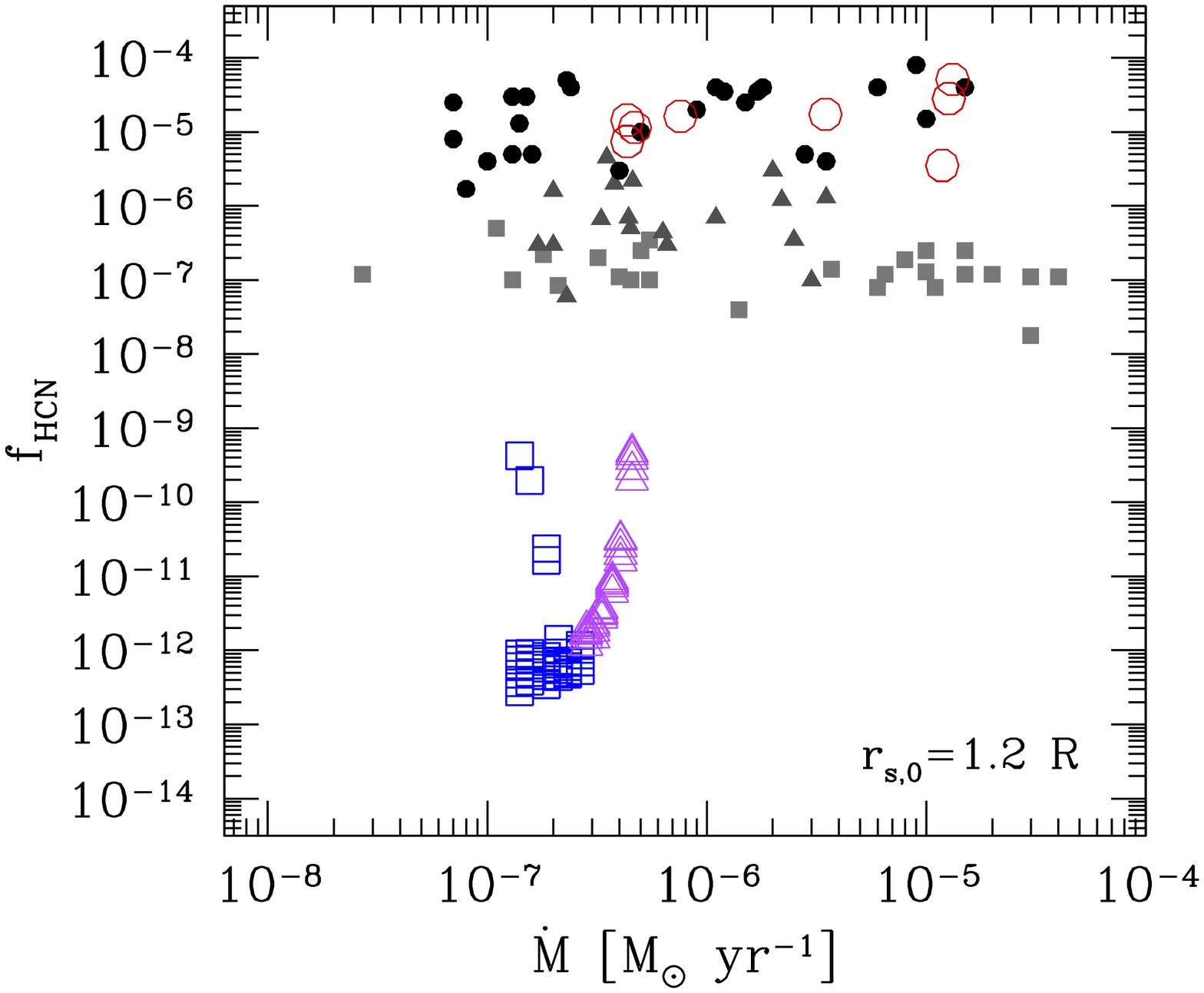}}
\end{minipage}
\caption{Comparison between observed CSE concentrations $f_{\rm HCN}$ and the 
values obtained from shock-chemistry models for two choices of the shock formation radius $r_{\rm s,0}=1.1\, R$ 
(left panel) and $r_{\rm s,0}=1.2\, R$. 
In both computations $\gamma_{\rm ad}^{\rm eff}=1.01$ is used. }
\label{fig_hcn_rs0}
\end{figure*}

We have investigated the effect of the shock formation radius 
on the non-equilibrium HCN abundances, 
testing three values, i.e. $r_{\rm s,0}/R=1.2, 1.1, 1.0$.
For simplicity, we assume that the minimum value for $r_{\rm s,0}$ 
is the photospheric radius.
The predicted concentrations HCN/H$_2$  are displayed 
in Fig.~\ref{fig_hcn_rs0} for $r_{\rm s,0}/R=1.2, 1.1$,
while the results for $r_{\rm s,0}/R=1.0$ are presented in the bottom-right panel
of Fig.~\ref{fig_hcn_gamad}.
In all three cases the results are derived from non-equilibrium chemistry 
computations over the distances $3 \le r/R \le 5 $, 
during  the entire TP-AGB evolution of the  reference
$M_{\rm i}=2.6\, {\rm M}_{\odot}, Z_{\rm i}=0.017$ star.

It is evident that the models with $r_{\rm s,0}/R=1.2, 1.1$ fail to explain 
the observed data for both M and S stars,
predicting HCN abundances that are far too low. Instead, the comparison is satisfactory 
 for C stars.  The main reason for the discrepancy that affects the M and S models
should be ascribed to the gas densities  that are too low across the pulsation-shocked region. 
On the other hand, a very nice agreement for all chemical classes, 
that is at increasing photospheric C/O ratio from below to above unity, 
is achieved by the model with $r_{\rm s,0}/R=1.0$. The same indication is
 obtained when considering  other TP-AGB stellar tracks with different initial masses. 

\begin{figure}
\resizebox{\hsize}{!}{\includegraphics{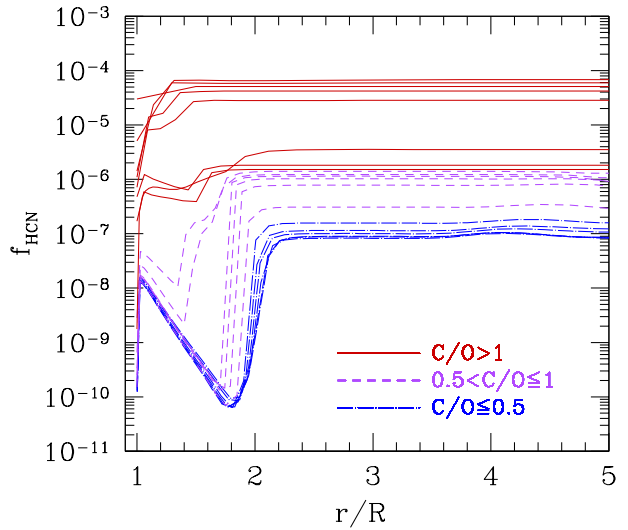}}
\caption{Radial profiles of  $f_{\rm HCN}$ concentration across the $r=1-5\,R$ 
circumstellar envelope, accounting for the non-equilibrium chemistry
affected by the shocks. 
At each radius chemistry predictions are shown after the completion 
of two consecutive pulsation cycles.
The results of the chemo-dynamic integrations are  connected
to detailed TP-AGB evolutionary calculations.
Each line corresponds to the stellar evolutionary stage 
just preceding a thermal pulse, experienced by our  
reference model with initial mass $M_{\rm i} = 2.6\,{\rm M}_{\odot}$ and metallicity  
$Z_{\rm i} = 0.017$. The assumed dynamic parameters are 
$r_ {s,0}=1 \, R$ and $\gamma_{\rm ad}^{\rm eff}=1.01$.}
\label{fig_hcn_rad}
\end{figure}

We notice that while $f_{\rm HCN}$ varies relatively little changing  
$r_{\rm s,0}$ from $1.2\,R$ to $1.1\,R$,  we predict  a dramatic increase of $f_{\rm HCN}$,
by  $\simeq 4$ orders of magnitude, 
if $r_{\rm s,0}$ is made coincide with the photospheric radius $R$.
The reason for the strong jump in the HCN abundance is that eliminating 
the static region (i.e. assuming  $r_{\rm s,0} = 1.0\,R$) shifts up the density profile 
by several orders of magnitude, a shift that makes the HCN formation reactions dramatically faster 
(a similar density shift occurs moving $r_{\rm s,0}$  from $1.2R$ to $1.1R$; 
but this is not enough to reach the threshold density where HCN formation becomes relevant).
This can be better understood from simple considerations.
In the static region (between $R$ and $r_{\rm s,0}$) the density falls exponentially, 
with a scale height $H_0(r) = k T/[\mu (GM/r^2)]$ (see Eq.~\ref{eq_H0}),
whereas in the shock-extended region  this length is longer by a factor 
$1/(1-\gamma_{\rm shock}^2)$ (see Eq.~\ref{eq_Hdyn}), 
which  corresponds to a factor of ~$3-10$. 
Therefore, eliminating the static region (e.g. moving $r_{\rm s, 0}$ from $1.1\,R$ to $1.0\,  R$)
replaces a zone where the density drops quite fast with one where the drop is much milder.
This is important because in the neighbourhood of $r=R$ we have that $H_0(r) \ll r$,
as one can appreciate from the relation $H_0(R)/R \simeq  0.008 (T/3000 {\rm K}) 
(R/200 R_{\odot}) (\mu/1.7 m_{\rm u})^{-1}  (M/ 2M_{\odot})^{-1}$
(where $m_{\rm u}$ denotes the atomic mass unit).
It turns out that moving across a static region from $r=1.0\,R$ to $r=1.1\,R$
(i.e., if $r_{\rm s, 0}=1.1R$), the density drops by a factor 
$\exp(0.1/0.008)\simeq 2.7\times 10^5$.
Instead, if the static region does not exist ($r_{\rm s, 0}=1.0\, R$), the density drops only by a
factor $\exp(0.1/0.037)\simeq 15$ over the same distance (assuming
$\gamma_{\rm shock}=0.89$).  The difference depends on the exact stellar
parameters, but is always quite large, and should persist at larger distances.

In summary, our tests indicate that the non-equilibrium shock chemistry in the inner CSEs
can account for the measured  HCN abundances, provided that the
first shock emerges quite close to the star, already from the photospheric layers.
It is encouraging to realize that this stringent finding, derived from 
a chemistry-based calibration, is in line with other independent analyses carried out both on
theoretical and observational grounds.
Indeed, current detailed models of pulsating atmospheres for Mira variables
indicate that pulsation-driven shocks form in the deep atmosphere,
even below the photosphere, whence they propagate outward
\citep{Nowotny_etal11, Nowotny_etal05, Hoefner_etal03, Hoefner_etal98}.  
Observational
support to an inner atmospheric origin of the shocks comes from the velocity and
temperature structures derived from the vibration-rotation CO
$\Delta\upsilon=3$ emission lines,  which are consistent with the
development of shock waves originating deeper than the line forming
layers \citep{Hinkle_etal84}. Likewise, studies based on envelope tomography of
long period variables suggest that the double absorption lines in
their optical spectra are caused by the propagation of a shock wave
through the photosphere \citep{Alvarez_etal01}.

\subsection{Non-equilibrium chemistry: major routes to HCN formation}
\label{ssect_hcn}
We will discuss here the major chemistry channels involved in the formation of the 
HCN molecule across the shocked inner CSE. The chemo-dynamic integrations are carried
out with the best-fitting parameters, $r_{\rm s,0}/R=1$ and 
$\gamma_{\rm ad}^{\rm eff}=1$.

\begin{figure*}
\centering
\begin{minipage}{0.48\textwidth}
{\resizebox{0.7\hsize}{!}{\includegraphics{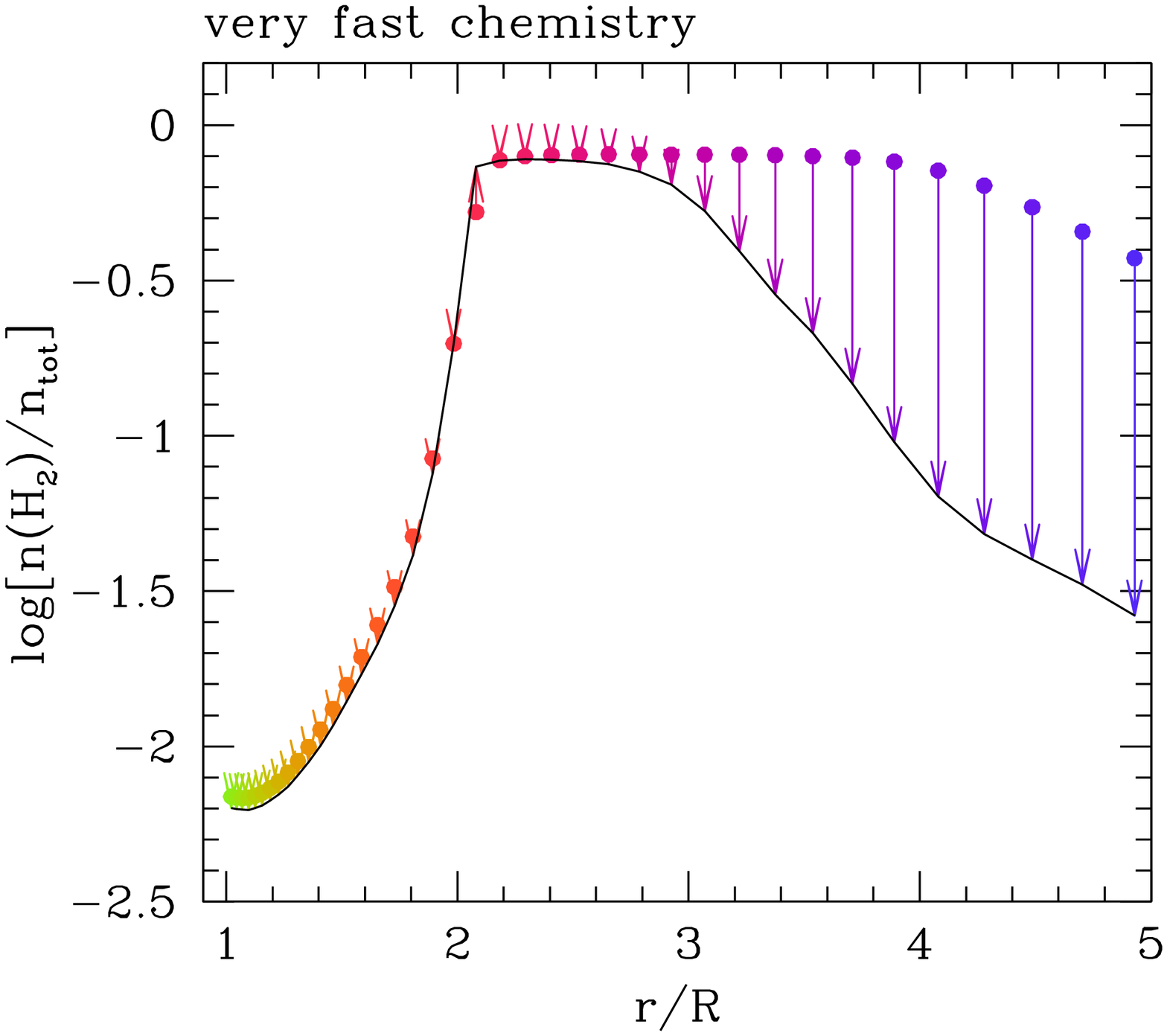}}}
\end{minipage}
\hfill 
\begin{minipage}{0.48\textwidth}
{\resizebox{0.7\hsize}{!}{\includegraphics{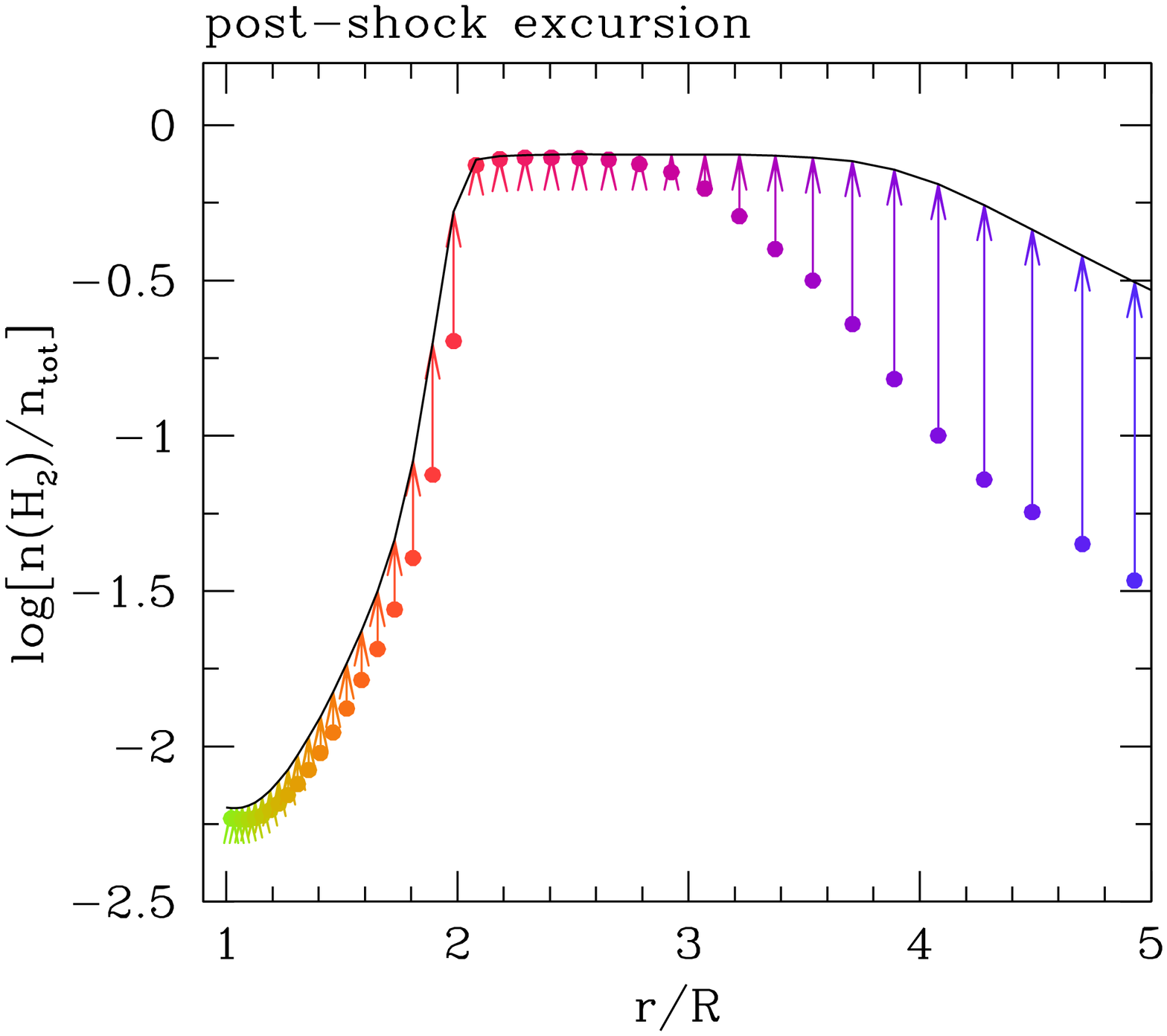}}}
\end{minipage}
\begin{minipage}{0.48\textwidth}
{\resizebox{0.7\hsize}{!}{\includegraphics{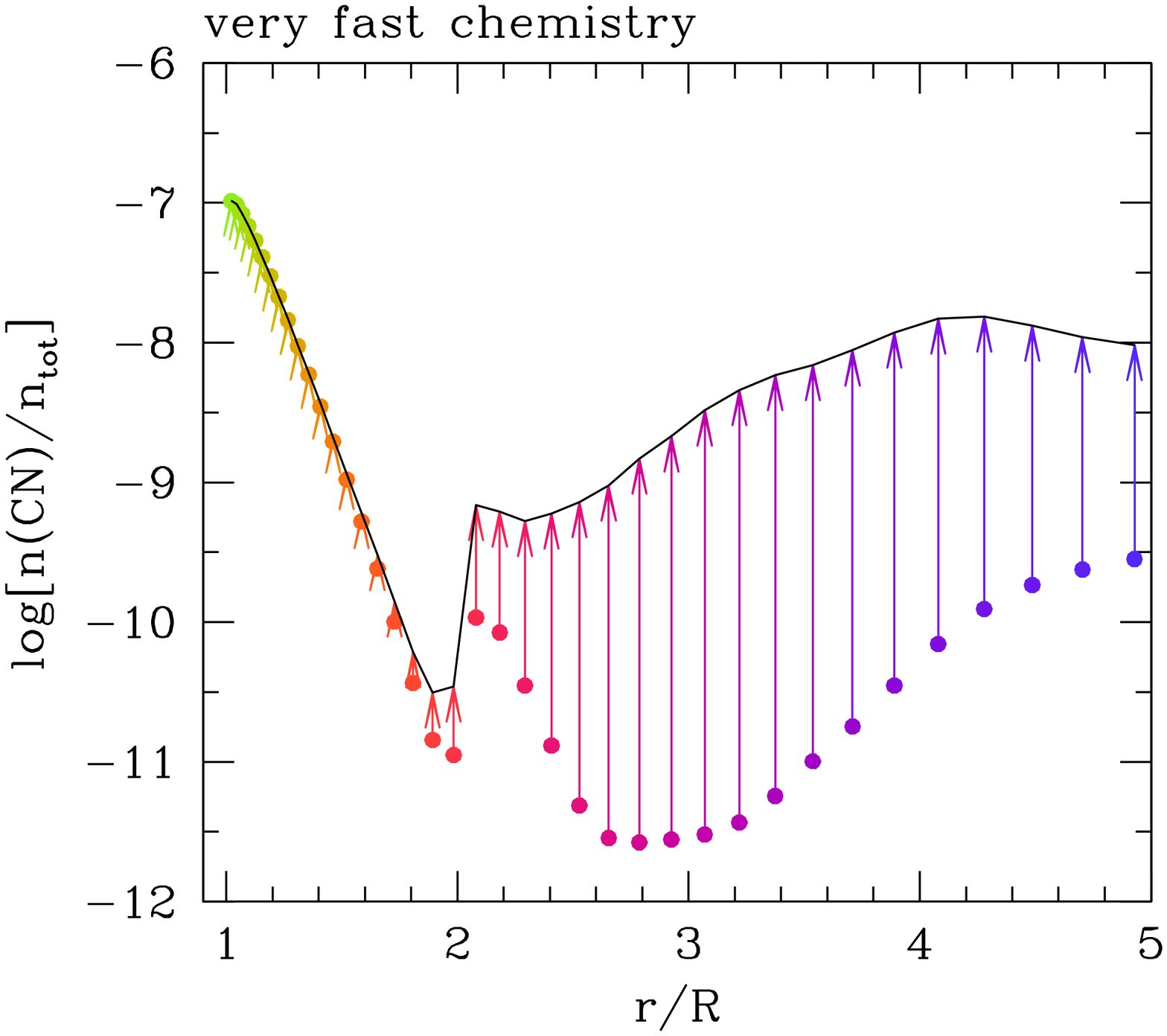}}}
\end{minipage}
\hfill 
\begin{minipage}{0.48\textwidth}
{\resizebox{0.7\hsize}{!}{\includegraphics{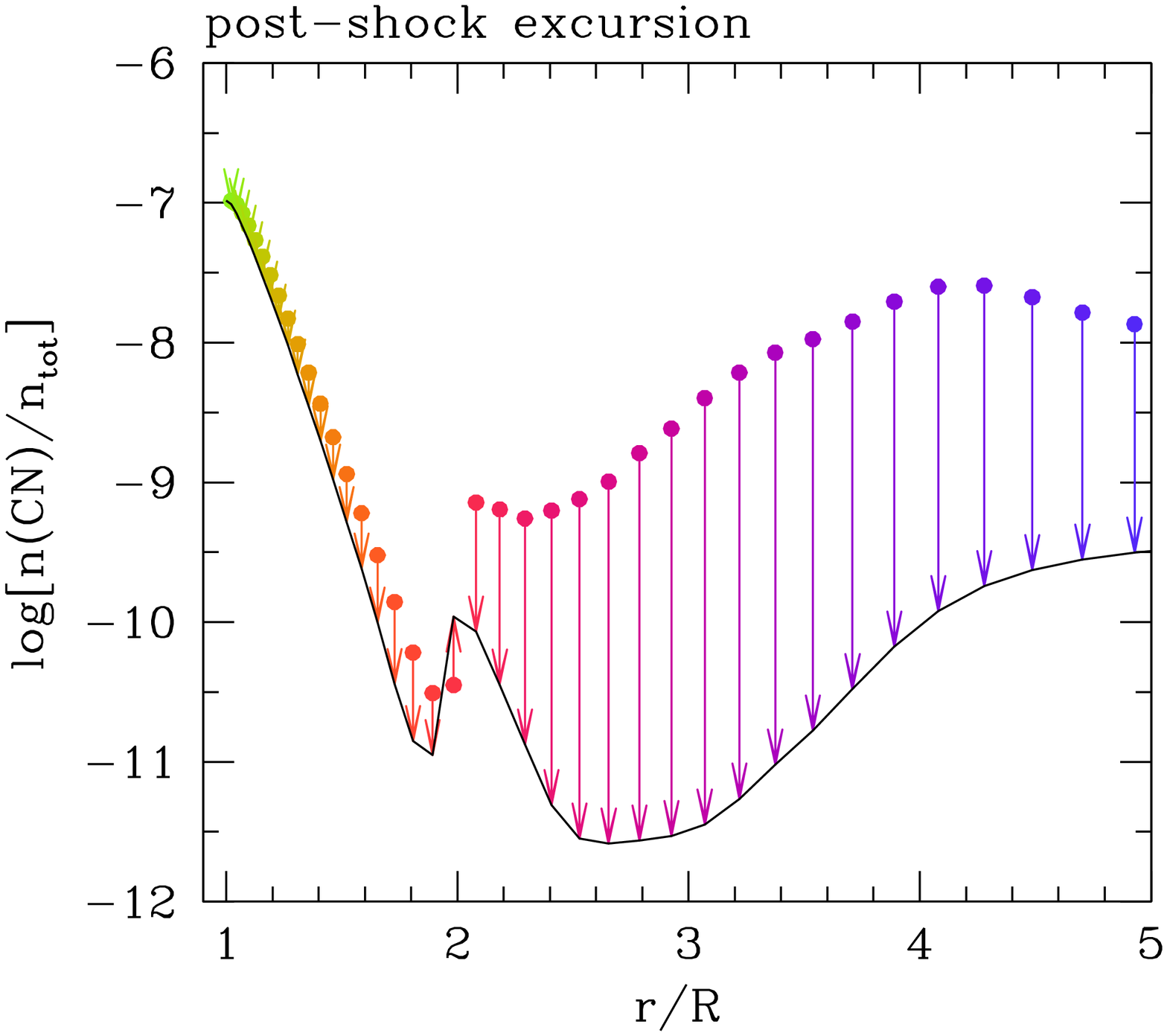}}}
\end{minipage}
\begin{minipage}{0.48\textwidth}
{\resizebox{0.7\hsize}{!}{\includegraphics{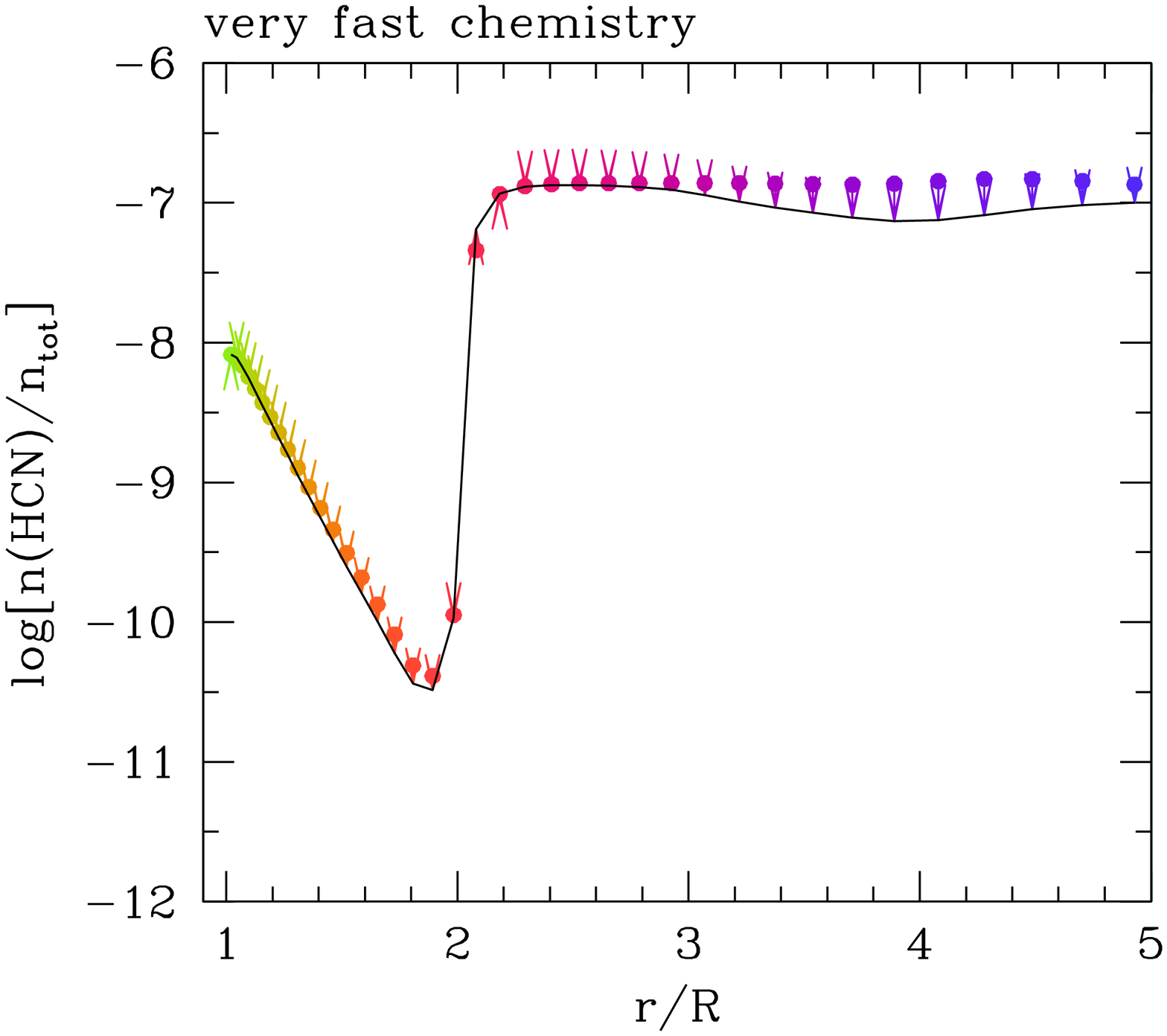}}}
\end{minipage}
\hfill 
\begin{minipage}{0.48\textwidth}
{\resizebox{0.7\hsize}{!}{\includegraphics{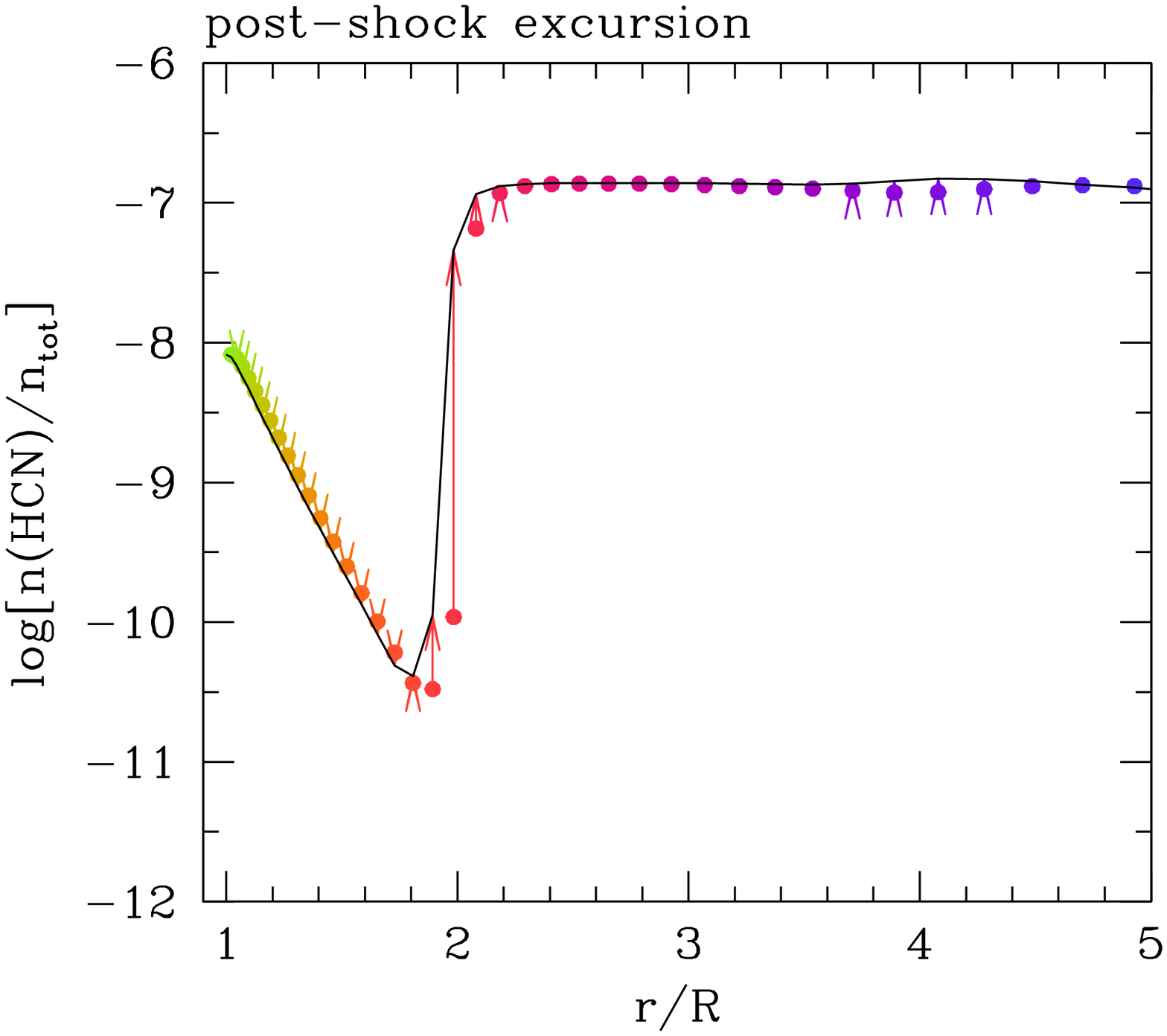}}}
\end{minipage}
\caption{ 
Results of the non-equilbrium chemistry for H$_2$, 
CN, and HCN, that occurs a) in  the  ``very fast chemistry''
zone close   to the shock front (left panels), and b) during the subsequent 
post-shock relaxation (right panels). 
For each molecule the arrows drawn show the total variation in the molecule's fractional abundance, 
relative to the total number density, in the two regimes at any given distance across the CSE.
In the left panels the dots mark the pre-shock abundances while the solid line
connects the abundances reached just before the onset of the relaxation.
These latter are taken as initial values of the post-shock expansion and are drawn with 
dots in the right panels, while the solid line 
connects the abundances at the completion of one pulsation cycle.
Note the steep increase of the H$_2$ abundance for $r/R \la 2$ during the 
post-shock excursion phases, which is crucial for the formation of HCN (see Eq.~\ref{eq_hcneq}).   
The stellar parameters correspond to the
quiescent stage just prior the $4^{\rm th}$ thermal pulse experienced by a  star with initial 
mass $M_{\rm i}=2.6\, M_{\odot}$, metallicity $Z=0.017$, and current photospheric 
C/O$\,=0.41$. }
\label{fig_veryfast}
\end{figure*}

\begin{figure*}
\centering
\begin{minipage}{0.32\textwidth}
{\resizebox{1\hsize}{!}{\includegraphics{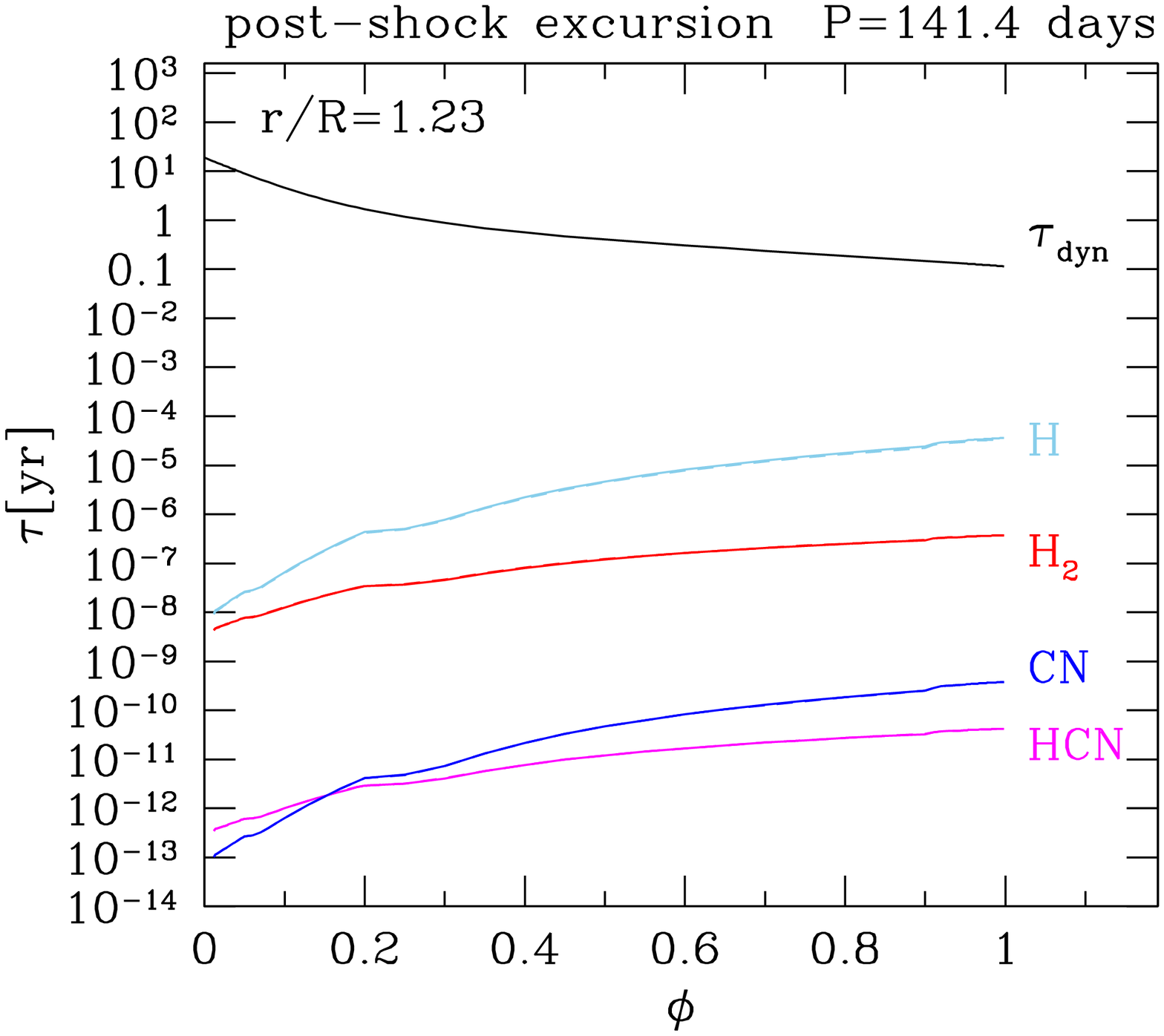}}}
\end{minipage}
\hfill 
\begin{minipage}{0.32\textwidth}
{\resizebox{1\hsize}{!}{\includegraphics{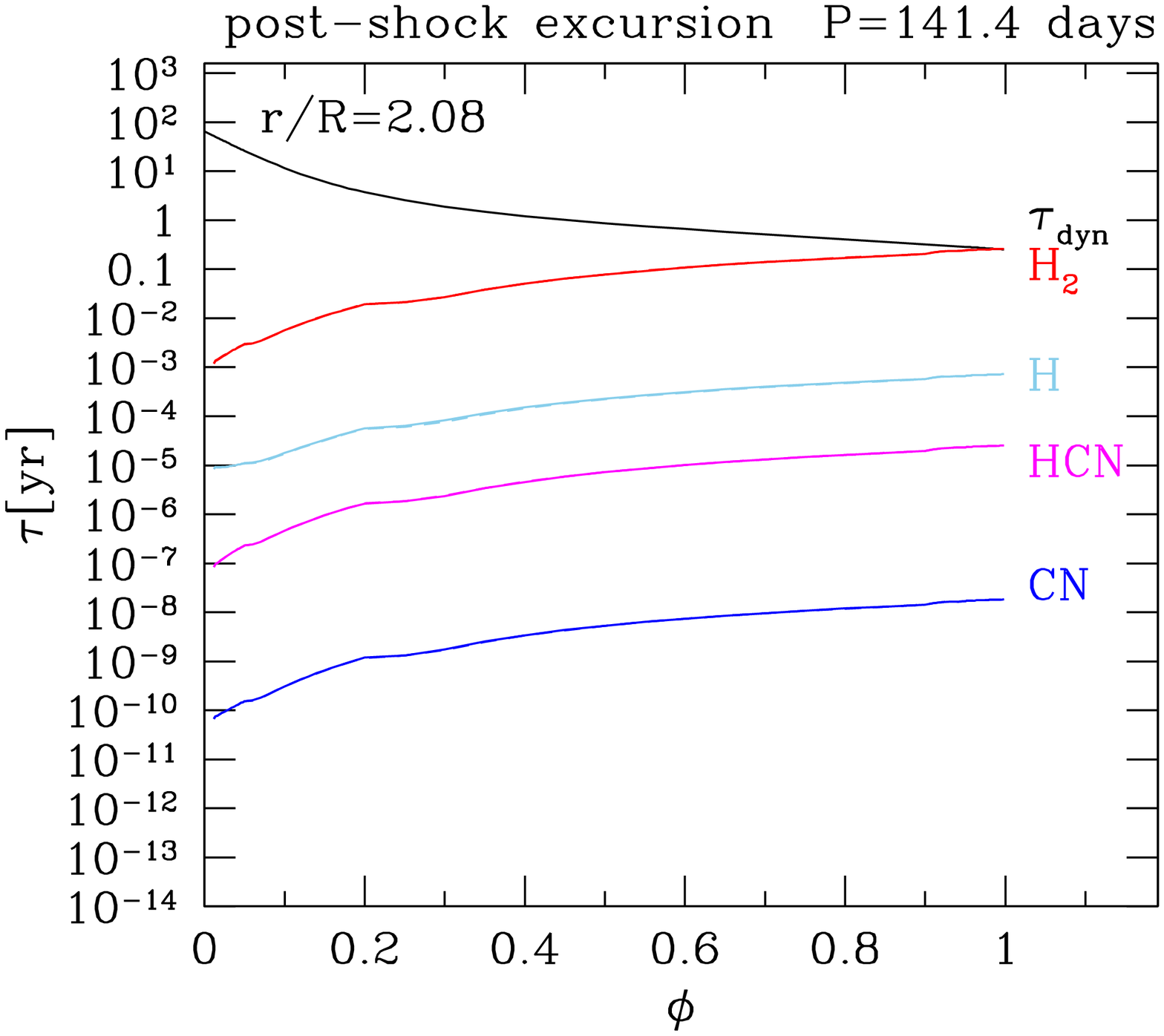}}}
\end{minipage}
\hfill
\begin{minipage}{0.32\textwidth}
{\resizebox{1\hsize}{!}{\includegraphics{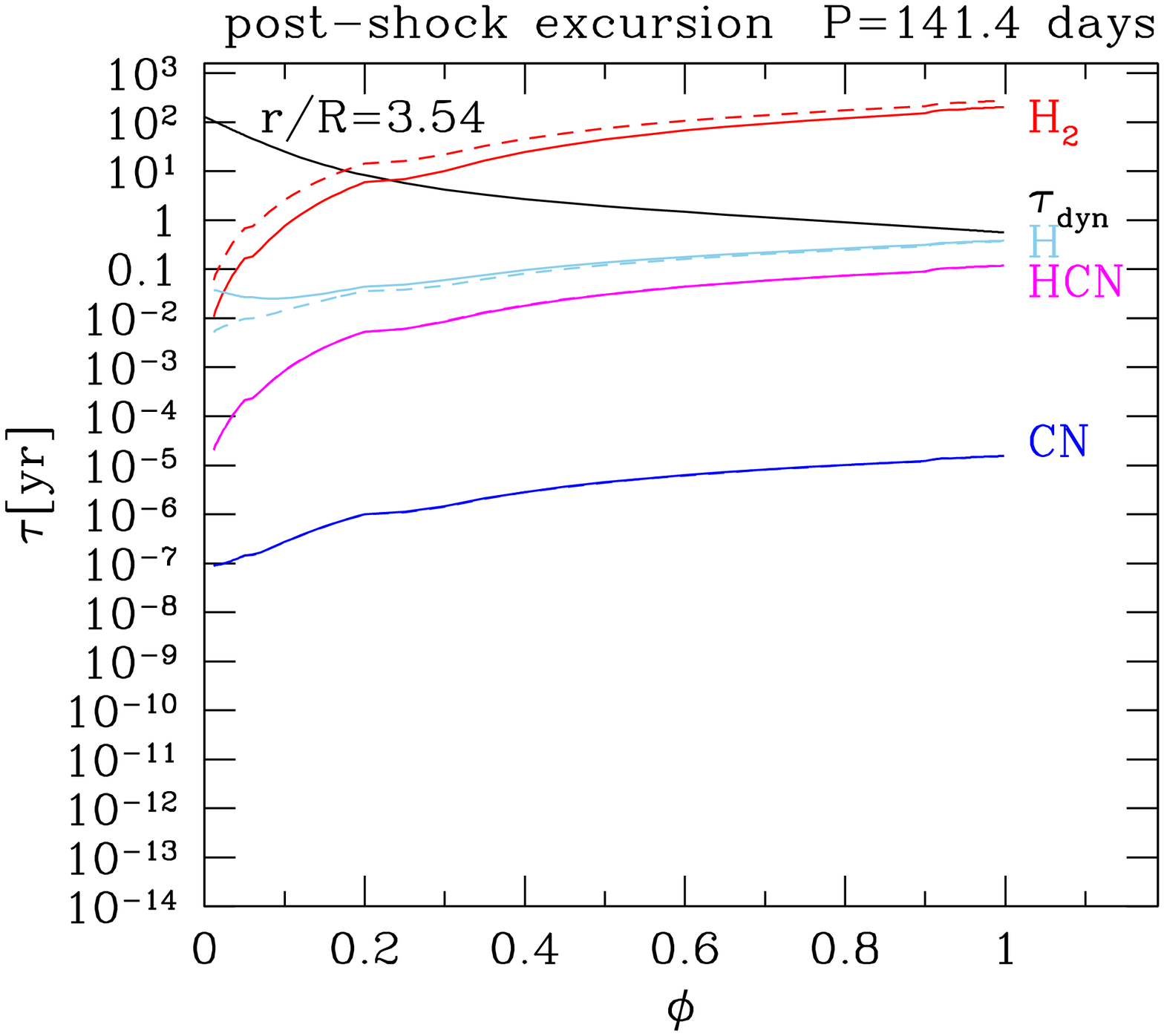}}}
\end{minipage}
\caption{
Relevant time-scales during the post-shock excursion of the gas as a function of the phase 
at three different radial distances.
The stellar parameters correspond to the
quiescent stage just prior the $4^{\rm th}$ thermal pulse experienced by a  star with initial 
mass $M_{\rm i}=2.6\, M_{\odot}$, metallicity $Z=0.017$, and current photospheric 
C/O$\,=0.41$.}
\label{fig_rtau}
\end{figure*}

\begin{figure*}
\centering
\begin{minipage}{0.48\textwidth}
{\resizebox{0.80\hsize}{!}{\includegraphics{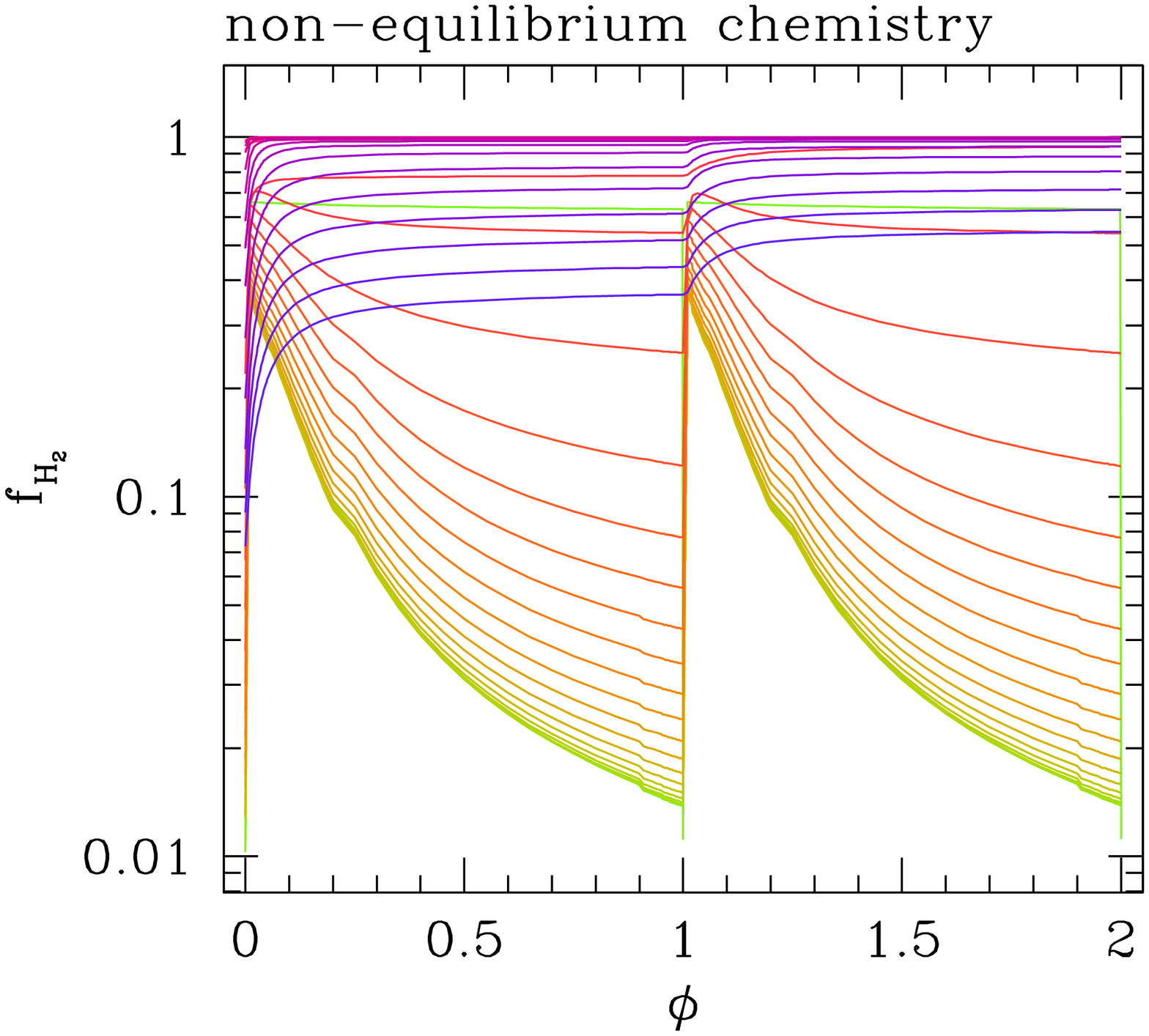}}}
\end{minipage}
\hfill
\begin{minipage}{0.48\textwidth}
{\resizebox{0.80\hsize}{!}{\includegraphics{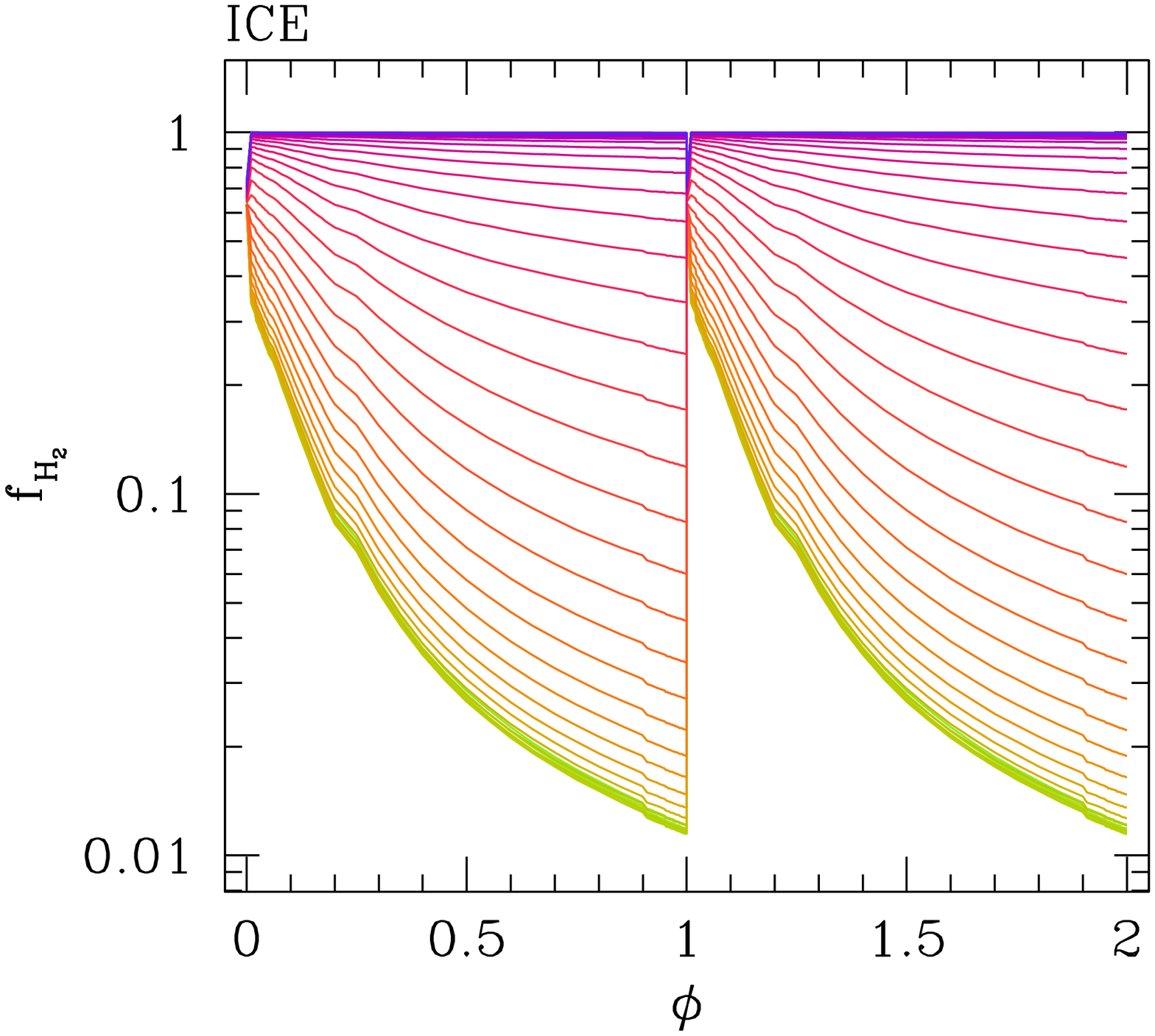}}}
\end{minipage}
\begin{minipage}{0.48\textwidth}
{\resizebox{0.80\hsize}{!}{\includegraphics{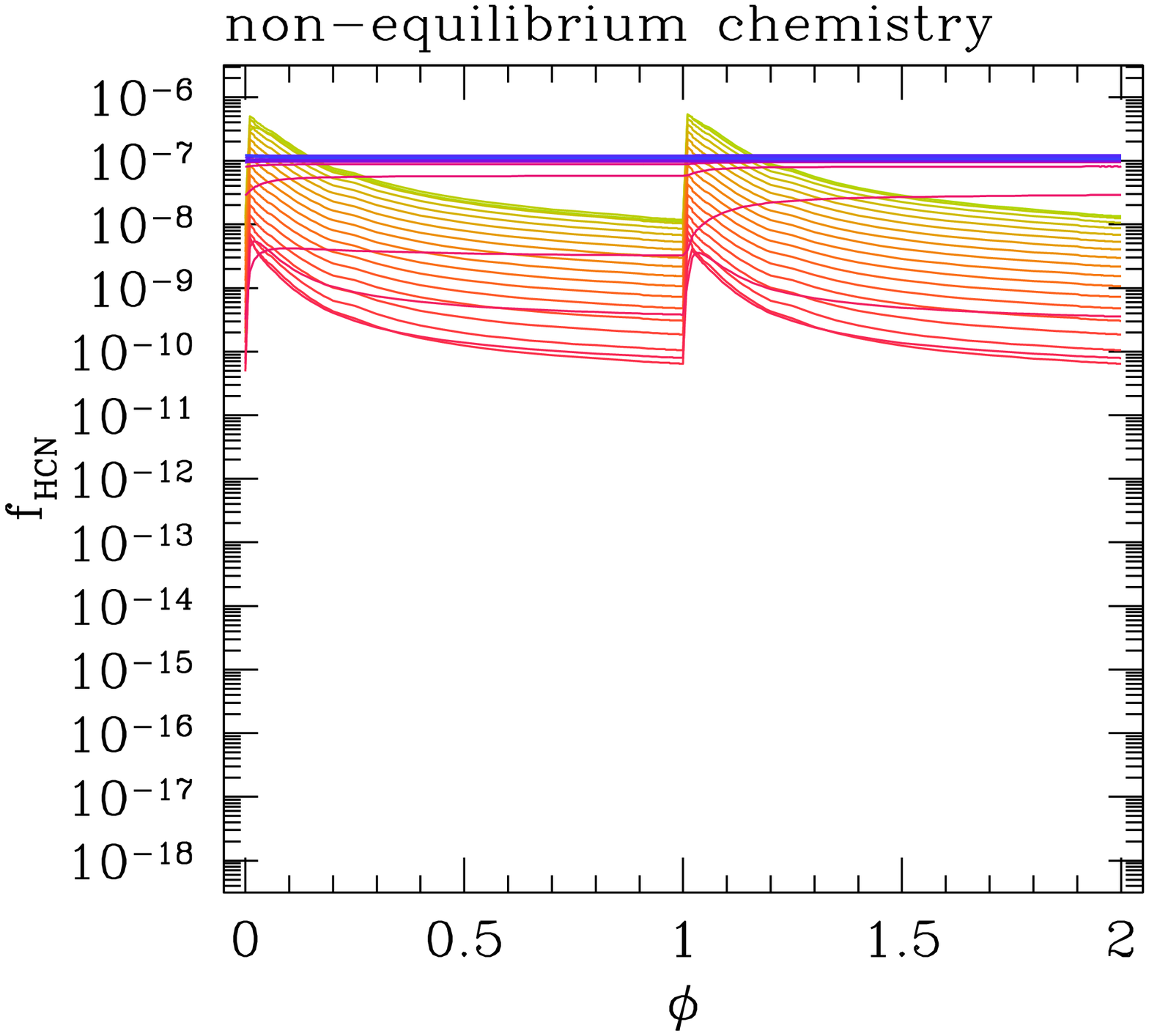}}}
\end{minipage}
\hfill
\begin{minipage}{0.48\textwidth}
{\resizebox{0.80\hsize}{!}{\includegraphics{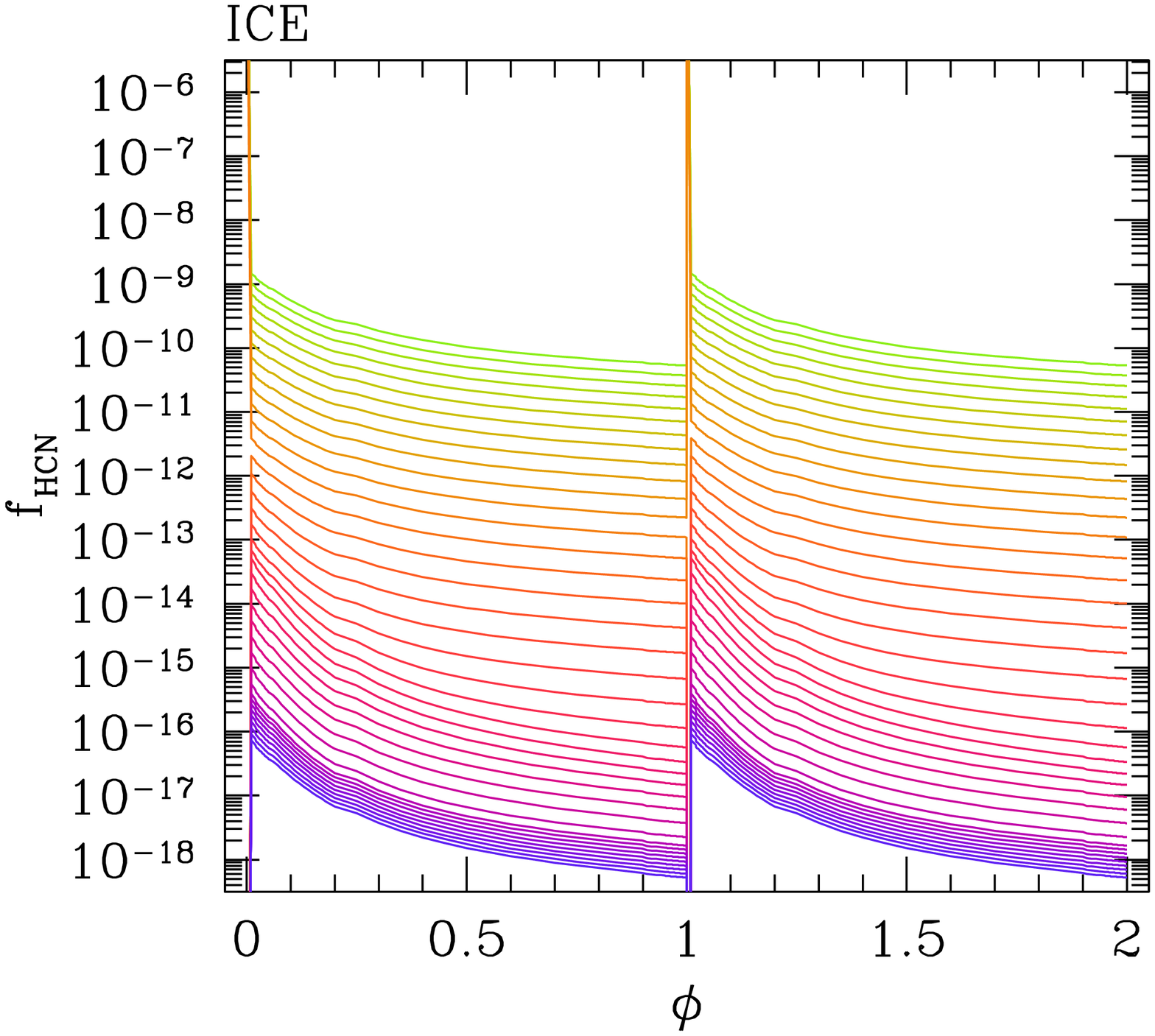}}}
\end{minipage}
\caption{Evolution of the fractional concentrations of H$_2$ and HCN
as a function of the phase  over two consecutive 
pulsation periods during the post-shock relaxation.
Within the adopted Lagrangian framework,
 each line shows the abundance evolution experienced by a parcel of gas  
when it is hit by the outward propagating shock at a specific distance from the star. Each distance of
the spatial grid is defined by a color gradually changing along  the sequence green-red-blue moving 
from $r/R=1$ to $r/R=5$.
The results in the left panels correspond to non-equilibrium chemistry calculations, while
those in the right panels are obtained under the assumption of ICE for all chemical species. 
Calculations refer to the same TP-AGB model as in Fig.~\ref{fig_veryfast}.}
\label{fig_h2_hcn_phi}
\end{figure*}

Figure~\ref{fig_hcn_rad} (left panel) neatly shows that the HCN abundances 
do depend on their photospheric C/O ratio.
For any given C/O, after an initial decrease down to a minimum (much more pronounced for M-stars)
in the region very close to the star,  the HCN concentration starts increasing from
$r/R \approx 2$  until it reaches a well-defined  value which remains almost
constant at larger distances. The level of the plateau
is seen to systematically increase moving through  
the domains of M stars (C/O$\,<0.5$), S stars ($0.5 \la$C/O$\,\la 1.0$), and
C stars  (C/O$\,> 1.0$). 
 
\begin{figure*}
\centering
\begin{minipage}{0.48\textwidth}
{\resizebox{0.85\hsize}{!}{\includegraphics{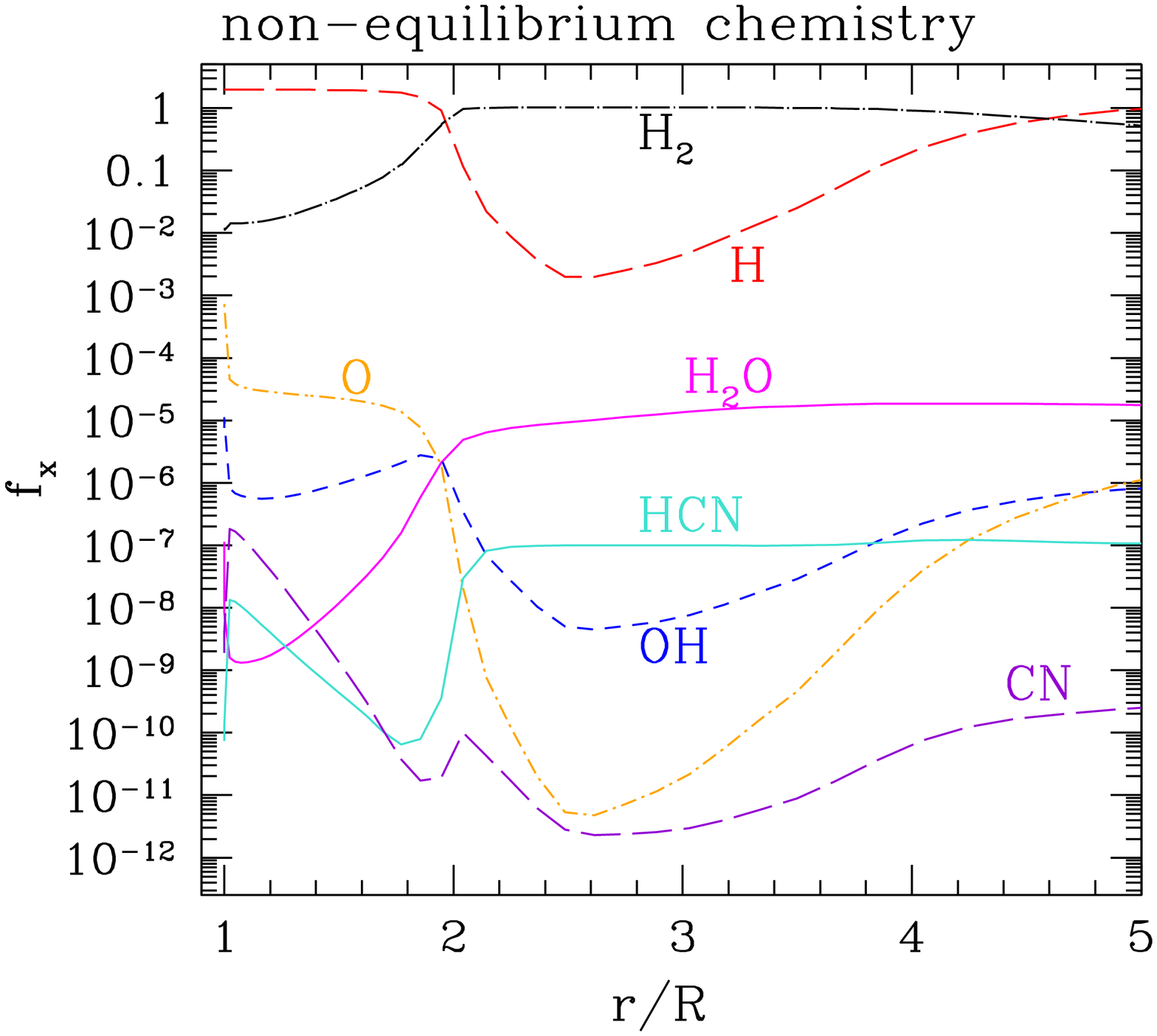}}}
\end{minipage}
\hfill
\begin{minipage}{0.48\textwidth}
{\resizebox{0.85\hsize}{!}{\includegraphics{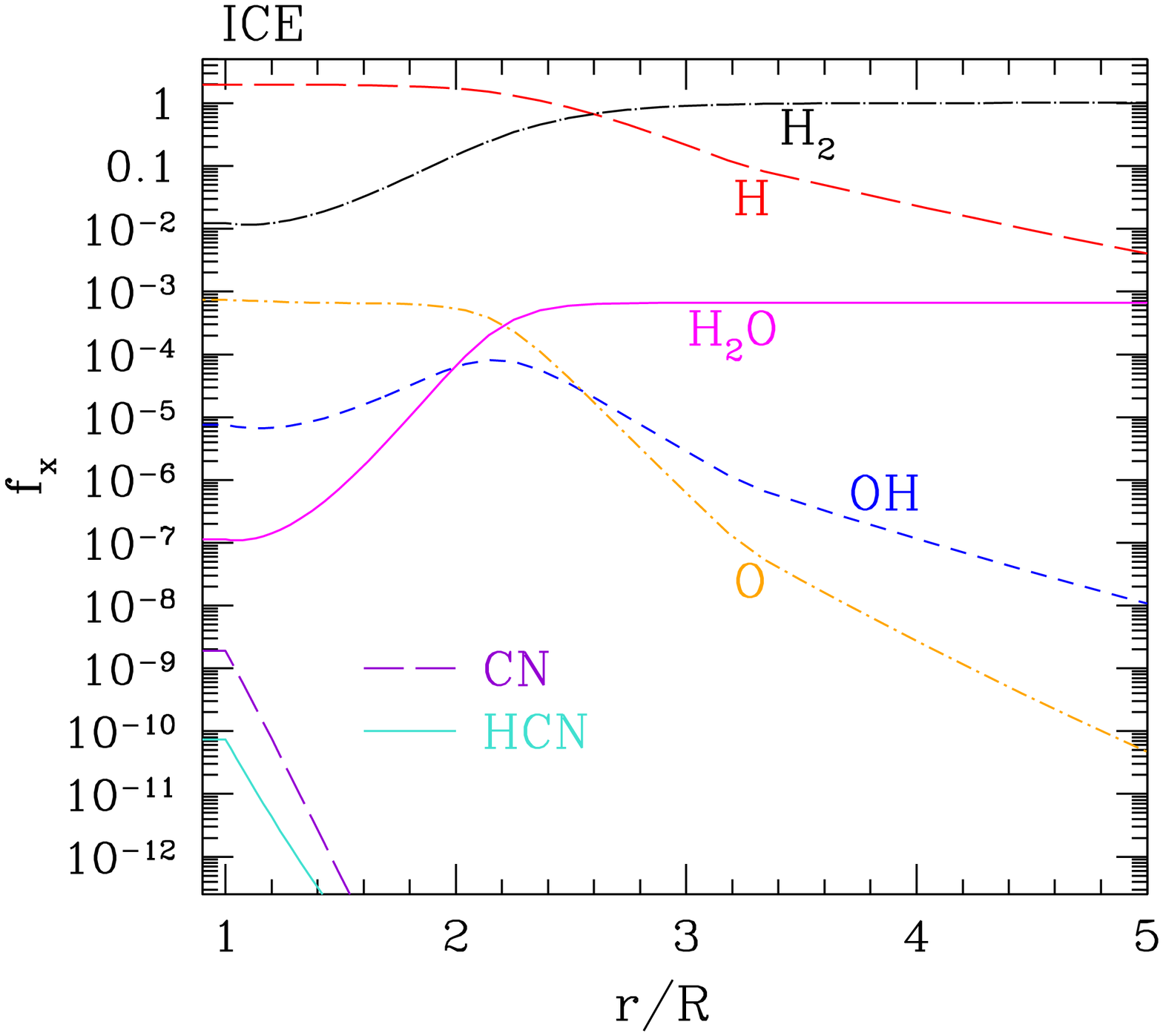}}}
\end{minipage}
\caption{Radial profiles of the fractional concentrations of H, O, OH, H$_2$, H$_2$O, CN, and 
HCN  the distance from $r/R=1$ to $r/R=5$. 
They are obtained connecting  the abundances reached at each mesh point of the CSE
(which defines a shocked fluid element in the Lagrangian scheme)
 soon after  the completion of two pulsation periods ($\phi=2$) after the passage of the shock.
Calculations refer to the same TP-AGB model as in Fig.~\ref{fig_veryfast}.
{\em Left panel}: Results of the integration of the non-equilibrium chemistry network. 
Note the abrupt increase of $f_{\rm{HCN}}$ up to $\simeq 10^{-7}$ 
which takes place  
at  $r/R \simeq 2$, just where $f_{\rm{H}_2}$ rises to one and 
$f_{\rm{H}_2O}$ exceeds $f_{\rm{OH}}$. 
 {\em Right panel}: Results obtained under the assumption of ICE for all 
chemical species. Note the severe  
drop of $f_{\rm{HCN}}$ and $f_{\rm{CN}}$ that starts already close to the star.}
\label{fig_nr}
\end{figure*}

This finding is in contrast with the analysis of \citet{Cherchneff_06}, who predicted
the presence of HCN in large quantities for any C/O ratios.
That circumstance was ascribed to the ``very fast chemistry'' occurring in a very narrow region
just after the shock front itself, as described by \citet{WillacyCherchneff_98}. 

In particular, it was suggested that the formation of CN, the radical responsible
for the bulk of HCN, 
should mainly take place in the fast chemistry region at $r\simeq R$ via the 
reaction:
\begin{equation}
\mathrm{N+CH}\longrightarrow \mathrm{CN+H}\,,
\label{r_n_ch}
\end{equation} 
and the radical CH should be simultaneously produced close to the shock front
by
\begin{equation}
\mathrm{H+C}_2\longrightarrow \mathrm{CH+C}.
\label{r_h_c2}
\end{equation}
The latter reaction has a very high energy activation barrier, which prevents efficient
formation of CH far from the shock front.
This would imply that the production of HCN is essentially controlled by the  
physical conditions of the gas in proximity of the shock front, 
hence being almost insensitive to the photospheric C/O ratio.

In any case, our calculations point at a different conclusion compared to 
 the study of \citet{Cherchneff_06}.
As an illustrative case, we will analyse the results obtained for the shock
chemistry of an oxygen-rich CSE characterized by a photospheric C/O$\, =0.41$
(Fig.~\ref{fig_veryfast}). The stellar parameters refer to  the 
quiescent stage just prior the $4^{\rm th}$ thermal pulse experienced by a  star with initial 
mass $M_{\rm i}=2.6\, M_{\odot}$, metallicity $Z=0.017$.
In this respect it is worth examining 
the evolution of the H$_2$, CN and HCN concentrations that results from  both the 
``very fast chemistry''  at the shock front, and the chemistry 
during the subsequent post-shock relaxation phase.

For all three molecules our models show (left panels of Fig.~\ref{fig_veryfast}) 
that the ``very fast chemistry''  is not efficient 
at small radial distances, $r/R < 2$, where the thickness $\ell_{\rm diss}$ 
of the cooling layer is very small (see Sect.~\ref{ssect_integration}).
At larger radii,  $r/R \ga 2$, the trends are different.
The concentration of H$_2$ is expected to decrease for  $r/R \ga 3$,
while a significant increase of CN takes place already starting from
$r/R \simeq 2$ outward. 
The HCN concentration is little changed compared to the pre-shock value, at each 
distance.
We find that in the ``very fast chemistry'' zone the
two reactions that are mainly responsible for the  production/destruction 
of the CN molecule are
\begin{equation}
\mathrm{HCN+H} \longrightarrow \mathrm{CN+H}_2\,,\,\,\,\,{\rm and}
\label{r_hcn_h}
\end{equation} 
and 
\begin{equation}
\mathrm{CN+H}_2 \longrightarrow \mathrm{HCN+H}\,
\label{r_cn_h2}
\end{equation}
respectively.
Both reactions are also the major processes involved in the chemistry
of HCN at the shock front.
At this stage we cannot provide a precise reason for the differences 
between our results and those of \citet{Cherchneff_06}, as they are 
the combined result of different input stellar and shock parameters.
We just note that the adoption of a low effective adiabatic exponent 
$\gamma_{\rm ad}^{\rm eff}\simeq 1$ leads to lower post-shock 
temperatures, so that reactions with 
high energy activation barrier, e.g., Reaction~\ref{r_h_c2}, are not efficient.

Let us now move to examine the results of the chemistry during the post-shock 
excursion (right panels of Fig.~\ref{fig_veryfast}).
Here the trends appear opposite compared to the ``very fast chemistry''.
In particular, we note that the concentration of H$_2$ starts to increase 
steeply at inner radii, $r/R < 1.5$, quickly reaching a plateau that corresponds to the condition 
in which almost all hydrogen atoms are locked into H$_2$.

The chemistry of H$_2$ has a dramatic impact on the abundance of HCN.   
After an initial drop in the inner regions, the HCN concentration increases suddenly
by a notable amount, just where H$_2$ attains its maximum concentration, 
at $r/R \approx 2$.  Then, HCN levels off at $n({\rm HCN})/n_{\rm tot} \approx 10^{-7}$,
which  is maintained across larger distances.

To understand the trends just described, it is useful to compare the
main times-scales involved.
We denote with $\tau_{\rm dyn}= r/\upsilon_{\rm gas}$ the flow dynamical time-scale, 
where $r$ is the position in the flow, and  $\upsilon_{\rm gas}$ 
is the local flow velocity (see Fig.~\ref{fig_rtau}).
For any given molecule $x$, we also define the chemical time-scale for
destruction 
\begin{equation}
\label{eq_des}
\tau_{\rm des}(x) =\frac{n(x)}{\mathscr{F}_{\rm des}} = \frac{n(x)}
{\sum_{j} n(x)\,n(j) \mathscr{R}_{xj}}\,\,,{\rm and}
\end{equation}
the chemical time-scale for production  
\begin{equation}
\label{eq_prod}
\tau_{\rm prod}(x) =\frac{n(x)}{\mathscr{F}_{\rm prod}} = \frac{n(x)}{\sum_{i,j} n(i)\,n(j) \mathscr{R}_{ij}}\,,
\end{equation}
where, for simplicity, the chemical fluxes for destruction $\mathscr{F}_{\rm des}$, and production 
$\mathscr{F}_{\rm prod}$ [cm$^{-3}$ s$^{-1}$] are shown in the case 
of two-particle collisions with generic rate 
$\mathscr{R}_{ij}$ [cm$^{3}$ s$^{-1}$]. Clearly all the reactions 
of the chemistry network,
that are involved in the production/destruction of a given molecule, must be considered 
in the actual computation of the chemical fluxes.  

As for HCN we derive two main points.
First, in the interval $1\le r/R \le 5$,  at any phase during the post-shock gas expansion, 
$\tau_{\rm des} \simeq  \tau_{\rm prod}$ and  
both ($\tau_{\rm des}$, $\tau_{\rm prod})  \ll \tau_{\rm dyn}$.
This means that the HCN abundance is in chemical equilibrium during 
the post-shock relaxation, which is defined by the exact balance of the chemical
fluxes, $\mathscr{F}_{\rm des}=\mathscr{F}_{\rm prod}$.
Second, the main chemical channels that keep HCN in equilibrium 
are Reactions~(\ref{r_hcn_h}) and (\ref{r_cn_h2}), so that 
 the equilibrium concentration of HCN 
can be well approximated with:
\begin{equation}
n(\rm{HCN})_{\rm eq} \simeq n(\rm{CN})
\frac{n(\rm{H}_2)}{n(\rm{H}) }\frac{\mathscr{R}_{\rm prod,\ref{r_cn_h2}}}
{\mathscr{R}_{\rm des,\ref{r_hcn_h}}}\,,
\label{eq_hcneq}
\end{equation}
where $n(\rm{CN})$, 
$n(\rm{H}_2)$, $n(\rm{H})$ are the number densities [cm$^{-3}$]
of the involved molecules. 

Similarly, the CN molecule attains its equilibrium concentration, corresponding to the 
balance of the chemical fluxes, that is
\begin{equation}
n(\rm{CN})_{\rm eq} \simeq n(\rm{HCN})
\frac{n(\rm{H})}{n(\rm{H}_2) }\frac{\mathscr{R}_{\rm des,\ref{r_hcn_h}}}
{\mathscr{R}_{\rm prod,\ref{r_cn_h2}}}\,.
\label{eq_cneq}
\end{equation}

It is  clear from Eqs.~(\ref{eq_hcneq}) - (\ref{eq_cneq}) that the equilibrium abundances of HCN and CN 
are critically dependent, in opposite way, on the ratio  of molecular  to atomic 
hydrogen,   ${n(\rm{H}_2)}/{n(\rm{H})}$.
It is therefore important to analyse the behaviour of molecular hydrogen during
the post-shock relaxation, at increasing distance form the star.

Figure~\ref{fig_h2_hcn_phi} (top panels) displays the evolution of the ratio  
$f_{{\rm H}_2} \equiv n(\rm{H}_2)/[0.5 n(\rm{H}) + n(\rm{H}_2)]$ as a function of the pulsation phase 
during the post-shock expansion of a fluid element that is hit by a shock 
at increasing distance from the star. 
Comparing the results obtained with the non-equilibrium chemistry network
(left panel) and with the assumption of ICE of all species (right panel), we see 
that in both cases
at given distance (along each line) 
a rapid increase of  $f_{{\rm H}_2}$ takes place in the very initial phases,
followed by a steady decrease during the following post-shock expansion.
This  trend is reversed  at larger distances, where almost hydrogen atoms become
 trapped into H$_2$ and   $f_{{\rm H}_2} \simeq 1$ during the
entire post-shock relaxation, at any phase.
In this respect the major difference  is that while in the non-equilibrium chemistry
case the quantity $f_{{\rm H}_2}$ reaches unity already quite close to the star
($r/R \ga 2$) when the gas density is still relatively high, under the ICE condition 
molecular hydrogen is efficiently formed farther from the star ($r/R \ga 3.5$), so that    
$f_{{\rm H}_2} \simeq 1$  in regions where the density has already dropped considerably.

The evolution of the HCN abundance reflects such a difference in 
the evolution of H$_2$ (bottom panels of Fig.~\ref{fig_h2_hcn_phi}).
In the case of the non-equilibrium chemistry 
 as soon as $f_{{\rm H}_2} \simeq 1$  is attained, the concentration 
of HCN steeply increases leveling out at  $f_{{\rm HCN}}\approx 10^{-7}$,  
which is kept constant at larger radii, independently of the phase.
Conversely, in the ICE case the condition $f_{{\rm H}_2}\simeq 1$ is reached only 
at longer distances,  in low-density regions where the CN abundance
is quite low. Hence the concentration of HCN keeps on
decreasing as the shock propagates outward.

At this point, it is necessary to consider the shock chemistry of H-H$_2$ in more detail.
Examining the time-scales (Fig. ~\ref{fig_rtau}) and the chemical fluxes involved 
we find that in the inner part,  $1\le r/R \la 3$, the H$_2$ abundance reaches a
condition of  chemical 
equilibrium,  but the  main reactions that contribute to $\mathscr{F}_{\rm des}$ and 
$\mathscr{F}_{\rm prod}$ (Eqs.~\ref{eq_des}-\ref{eq_prod})
change with increasing distance, according to the sequence: 
\begin{equation}
\label{r_1}
\mathrm{H+H+H}  \longleftrightarrow  \mathrm{H_2+H}\,\,\,\,\,\,\,\,\,\,\,\,\, 1.0 \le r/R \la 1.2
\end{equation}
\begin{equation}
\label{r_2}
\mathrm{H+OH} \longleftrightarrow \mathrm{H_2+O} \,\,\,\,\,\,\,\,\,\, 1.2\la r/R \la 2.0
\end{equation}
\begin{equation}
\label{r_3}
\mathrm{H+H_2O}  \longleftrightarrow  \mathrm{H_2+OH}\,\,\,\, \,\,\,\,\,\,\,\,\,\, 2.0\la r/R \la 2.4
\end{equation}
\begin{equation}
\label{r_4}
\mathrm{H+AlH}  \longleftrightarrow  \mathrm{H_2+Al}  \,\,\,\,\,\,\,\,\,\, 2.4\la r/R \la 3.2
\end{equation}

Figure~\ref{fig_nr} shows the 
radial abundance profiles  of the involved species (H, O, H$_2$, OH, H$_2$O, CN, and HCN) 
obtained connecting the results soon after  the completion  of two pulsation periods 
 (at phase $\phi=2$) at each mesh point of the spatial grid. 
Similar plots can be built at other  phases.  In general, 
one should be aware that radial abundance  profiles obtained
connecting  loci of equal phases may differ depending on the selected pulsation phase.
This is evident, for instance, looking at Fig.~\ref{fig_h2_hcn_phi}.
However, observations are compared with the predicted HCN abundances that are
reached in the freeze-out  region  ($3 \la r/R \la 5$)  where the phase dependence disappears.

It is evident that the predictions from the non-equilibrium chemistry (left panel) are quite different 
from the ICE results. The passage of shocks favours the efficient  
formation of H$_2$ in regions closer to the star, compared to the ICE case (right panel).
At $r/R\simeq 2$ already $99\%$ of  H atoms are 
locked in molecular hydrogen according to the non-equilibrium chemistry integrations, 
while in condition of instantaneous equilibrium chemistry the fraction of molecular hydrogen is 
much less,   $f_{{\rm H}_2} \simeq 30\%$, and increases to $f_{{\rm H}_2} \simeq  100\%$
at larger distances, from $r/R \ga 3.5$.

In the non-equilibrium case the condition $f_{{\rm H}_2}\simeq 1$ 
is reached at $r/R\simeq 2$ just where the abundance of water  $f_{\rm{H}_2O}$ 
also rises  and quickly levels out to a maximum value. At this point  
Reaction~(\ref{r_3}) overtakes Reaction~(\ref{r_2}) as  main chemical channel
for the production/destruction of H$_2$, and the ratio of molecular-to-atomic hydrogen is well approximated with  
\begin{equation}
\frac{n(\rm{H}_2)}{n(\rm{H})} \approx \frac{n(\rm{H}_2O)}{n(\rm{OH})}
\frac{\mathscr{R}_{\ref{r_3},\rightarrow}}{\mathscr{R}_{\ref{r_3},\leftarrow}} 
\end{equation}
All this concurs to produce  a steep increase of the HCN equilibrium 
concentration at $r/R\simeq 2$, which appears clearly in Figs.~\ref{fig_hcn_rs0}, 
\ref{fig_veryfast}, \ref{fig_h2_hcn_phi}, and \ref{fig_nr}.
For $r/R > 2$,  the fractional concentration $f_{\rm HCN}$ remains
roughly invariant while $f_{{\rm H}_2} \simeq 1$, and eventually the condition 
of chemistry freeze-out is reached  when the gas density and temperature drop and
the dynamical time-scale becomes shorter than chemical time-scale for molecular
hydrogen (see right panel of Fig.~\ref{fig_rtau}).

Completely different results are obtained under the ICE assumption 
(right plot of Fig.~\ref{fig_nr}): the fractional abundances of HCN and CN
exhibit a steadily decreasing trend across the CSE quickly sinking to extremely low
values. 

In conclusion our analysis shows that a proper treatment of the H-H$_2$ chemistry 
in the inner shocked CSE during the post-shock relaxation stages 
is critical for the abundances of HCN and CN. 
In particular, the ICE assumption  is 
inadequate and leads to a systematic under-estimation of the HCN concentration, 
at least for  the chemical mixtures characterized by C/O$<1$.
     
\subsection{Additional effects: HCN abundance evolution 
through the thermal-pulse cycle}
\label{ssect_tpcycle}
The results discussed in the previous sections apply to the quiescent TP-AGB evolution,
so that each envelope integration corresponds to the stellar parameters at the stage just prior
the development of a thermal pulse.

\begin{figure*}
\centering
\begin{minipage}{0.32\textwidth}
\resizebox{\hsize}{!}{\includegraphics{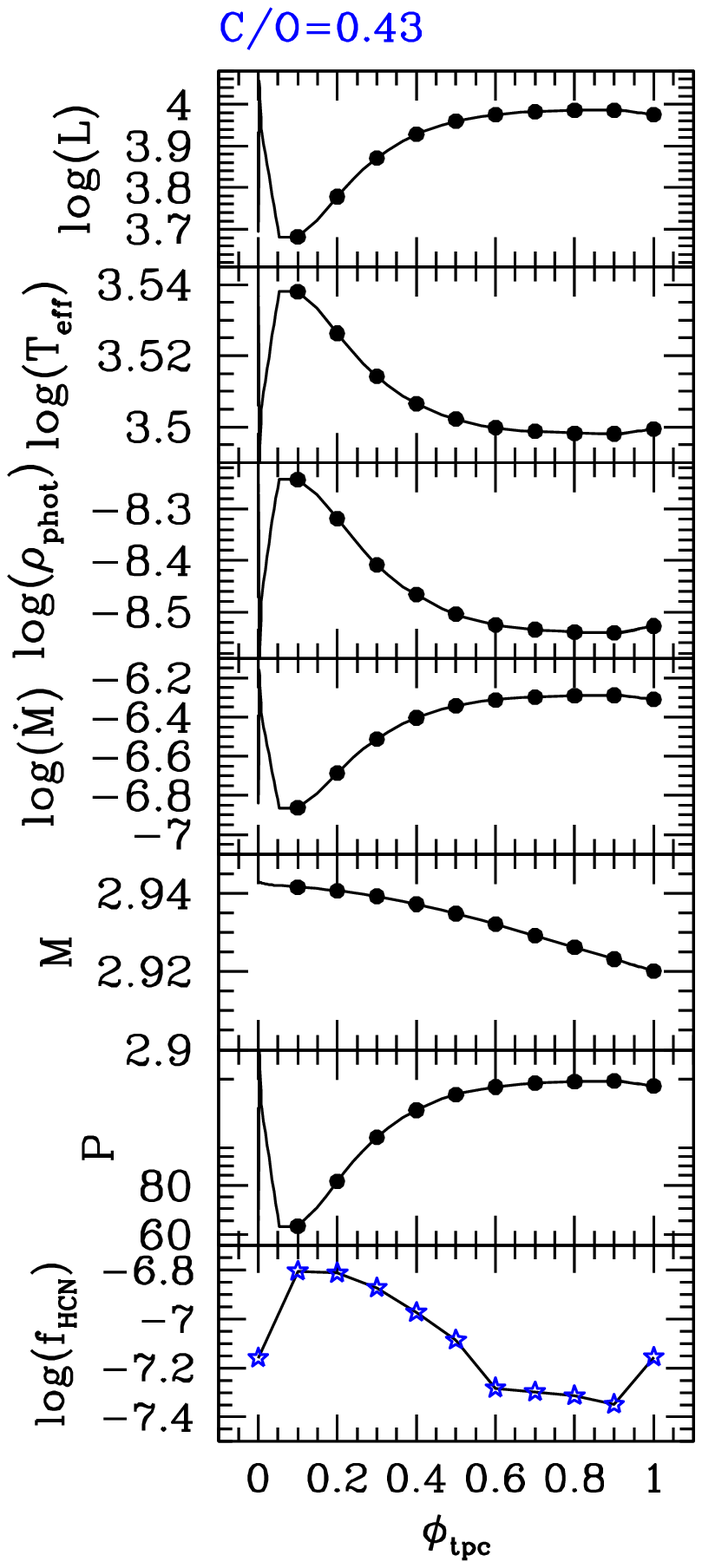}}
\end{minipage}
\hfill
\begin{minipage}{0.32\textwidth}
\resizebox{\hsize}{!}{\includegraphics{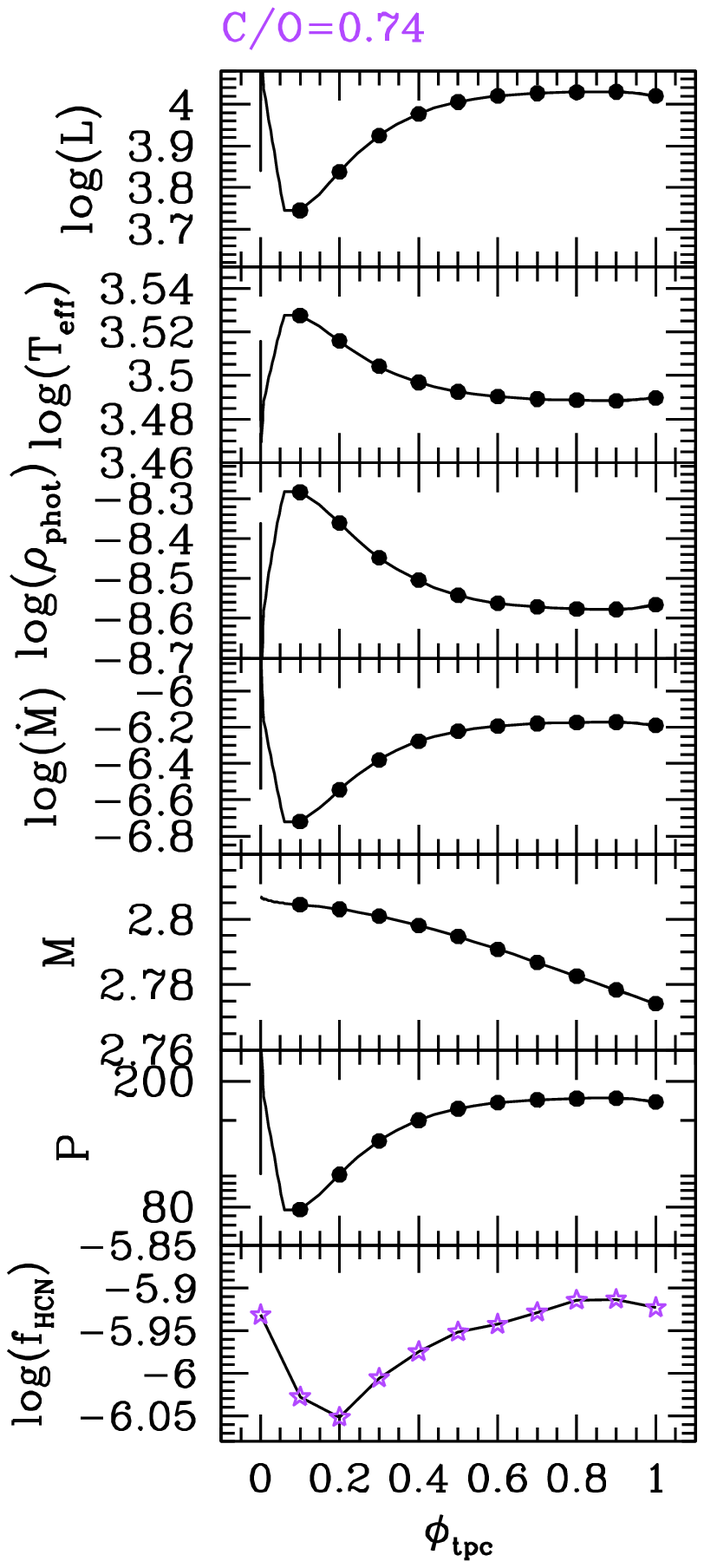}}
\end{minipage}
\hfill
\begin{minipage}{0.32\textwidth}
\resizebox{\hsize}{!}{\includegraphics{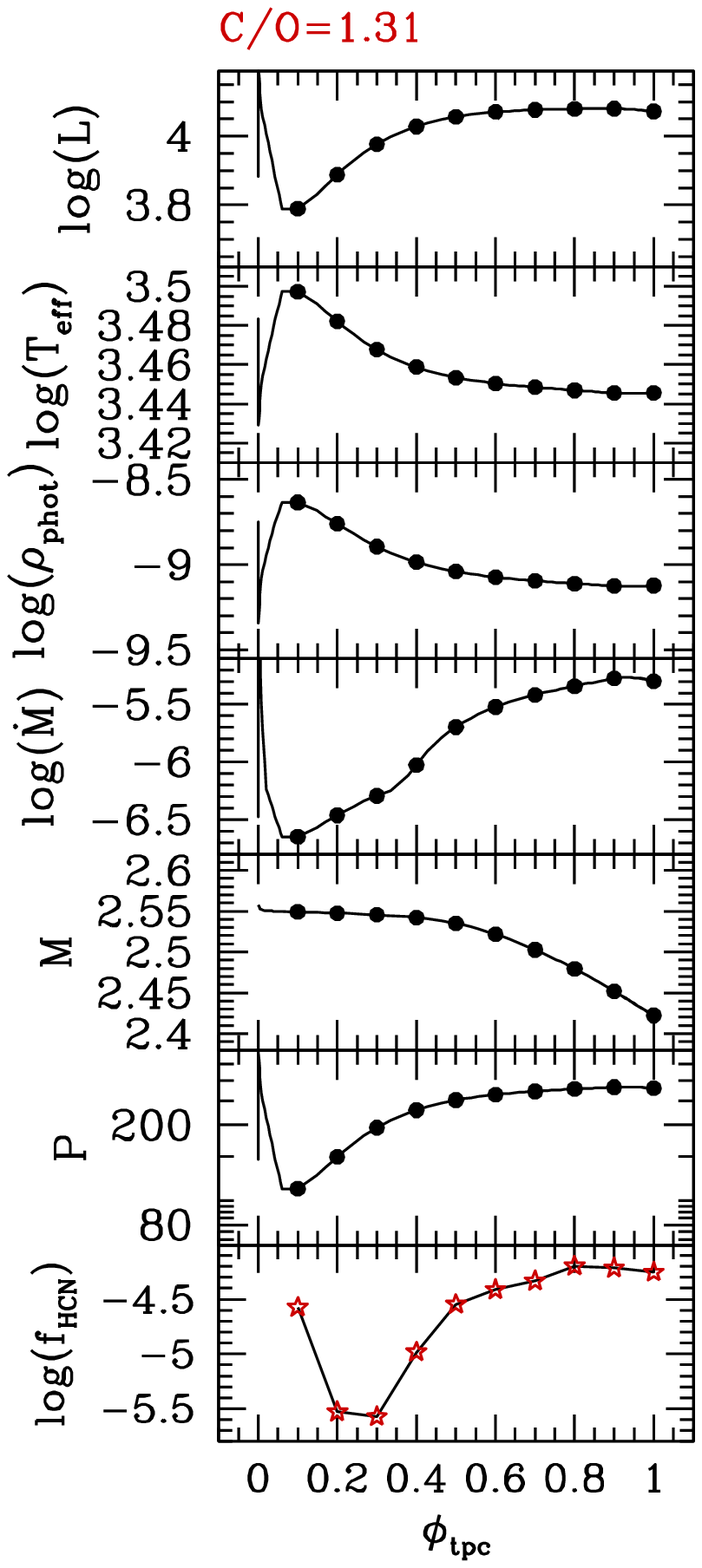}}
\end{minipage}
\caption{Evolution of stellar luminosity $[{\rm L}_{\odot}]$, 
effective temperature $[{\rm K}]$, photospheric mass density 
$[{\rm g\, cm}^{-3}]$, mass-loss rate  $[{\rm M}_{\odot}\,{\rm yr}^{-1}]$,
total mass  $[{\rm M}_{\odot}]$,  pulsation period $[{\rm days}]$ in the fundamental mode,
and $f_{\rm HCN}$ concentration derived from shock-chemistry integrations 
at the radial distance $r/R=3$, 
 during the $5^{\rm th}$, $10^{\rm th}$, $16^{\rm th}$ pulse cycle 
of a TP-AGB model with initial mass $M_{\rm i}=3.0\,{\rm M}_{\odot}$ and metallicity 
$Z_{\rm i} =0.017$. The results are shown as a function of the phase, starting from 
$\phi_{\rm tpc}=0$,  
at the onset of the He-shell flash, up to $\phi_{\rm tpc}=1$,
 at the end of the inter-pulse period, just prior
the development of the next thermal pulse. The rapid post-flash maximum stages 
take place for $\phi_{\rm tpc} \la 0.1$. Black circles and starred symbols 
mark the selected phases  ($0.1 \le \phi_{\rm tpc} \le 1.0$)
at which chemo-dynamic integrations of the inner CSE are carried out 
(see also Fig.~\ref{fig_chempulse}). Note the different morphologies and ranges covered by 
the HCN abundance tracks (bottom panels), depending on the photospheric C/O ratio
and the other stellar parameters. 
}
\label{fig_pulse}
\end{figure*}

However, stellar models show that the TP-AGB phase cannot be represented only by quiescent
stages, since the quasi-periodic occurrence of  thermal instabilities of the He-burning shell 
cause significant variations of the stellar 
structure, modulating the evolution of the luminosity according to a typical profile
that is exemplified in Fig.~\ref{fig_pulse}. Denoting with $\tau_{\rm ip}$ the inter-pulse
period between two consecutive He-shell flashes, we define 
the phase over a thermal pulse-cycle  
as $\phi_{\rm tpc}=t/\tau_{\rm ip}$, where $t$ is the time counted from the occurrence
of a  flash ($\phi_{\rm tpc}=0$) up to the stage just prior the next one 
($\phi_{\rm tpc}=1$).

Looking at Fig.~\ref{fig_pulse}, we see that  during the first stages of a
thermal pulse-cycle, 
a TP-AGB star is expected to have luminosities that may be much fainter
than the quiescent values at $\phi_{\rm tpc}\approx 1$, 
 due to the low nuclear activity, or even extinction, 
of the H-burning shell. 

The duration of these dimmer stages becomes longer for decreasing stellar mass, 
and it can cover up to 30-40$\%$ of the inter-pulse periods in TP-AGB
stars with masses $\la 2\, {\rm M}_{\odot}
$\citep{BoothroydSackmann_88, WagenhuberGroenewegen_98}. It follows that the 
statistical significance of the post-flash sub-luminous phase -- known with the global term of 
``low-luminosity dip'',  cannot be neglected in terms of detection probability.

During the low-luminosity dip the star is characterized by significant variations 
of all other physical quantities, namely higher effective temperatures 
$T_{\rm eff}$, 
shorter pulsation periods $P_{0}$, and lower mass-loss rates $\dot M$. 
Figure~\ref{fig_pulse}  shows the evolution of these quantities 
over the  $4^{\rm th}$, $10^{\rm th}$, and $16^{\rm th}$ 
thermal pulse-cycle experienced by a TP-AGB model with initial mass 
$M_{\rm i}=3.0\, {\rm M}_{\odot}$ and metallicity $Z_{\rm i}=0.017$.

Since $L$, $T_{\rm eff}$, $M$, $P$, and $\dot M$ are the key  input parameters of 
the present chemo-dynamic model for the inner CSE, 
it is worth exploring the  effect of the pulse-driven variations of the stellar 
structure on the pulsation-shocked chemistry of HCN.
To this aim we applied  the chemo-dynamic integrations
for ten values of the pulse-cycle phase, from 
$\phi_{\rm ptc}=0.1$ to $\phi_{\rm ptc}=1.0$.

Figure ~\ref{fig_pulse} provides an example of how the stellar parameters evolve,
together with the resulting HCN non-equilibrium abundances in the inner CSE, 
during three complete thermal pulse-cycles. The pulses were selected to form a sequence
of increasing photospheric C/O ratio (0.43, 0.74, and 1.31) so as to represent 
the three chemical classes of M, S, and C stars.
We note that, indeed, the HCN concentrations are affected by the occurrence
of thermal pulses. In the case with ${\rm C/O} =0.43$, the HCN abundance 
 is  seen to increase during the initial pulse-cycle 
phases up to maximum value reached at $\phi_{\rm tpc} \simeq 0.1$,
 and then it decreases at later phases up to the completion of the inter-pulse period.
For the models with ${\rm C/O} =$ 0.74, 1.31 the trends are reversed, as the HCN
concentrations reach a minimum at $\phi_{\rm tpc} \simeq 0.3$ 
and then increase for remaining part the thermal-pulse cycle.

\begin{figure*}
\begin{minipage}{0.32\textwidth}
\resizebox{\hsize}{!}{\includegraphics{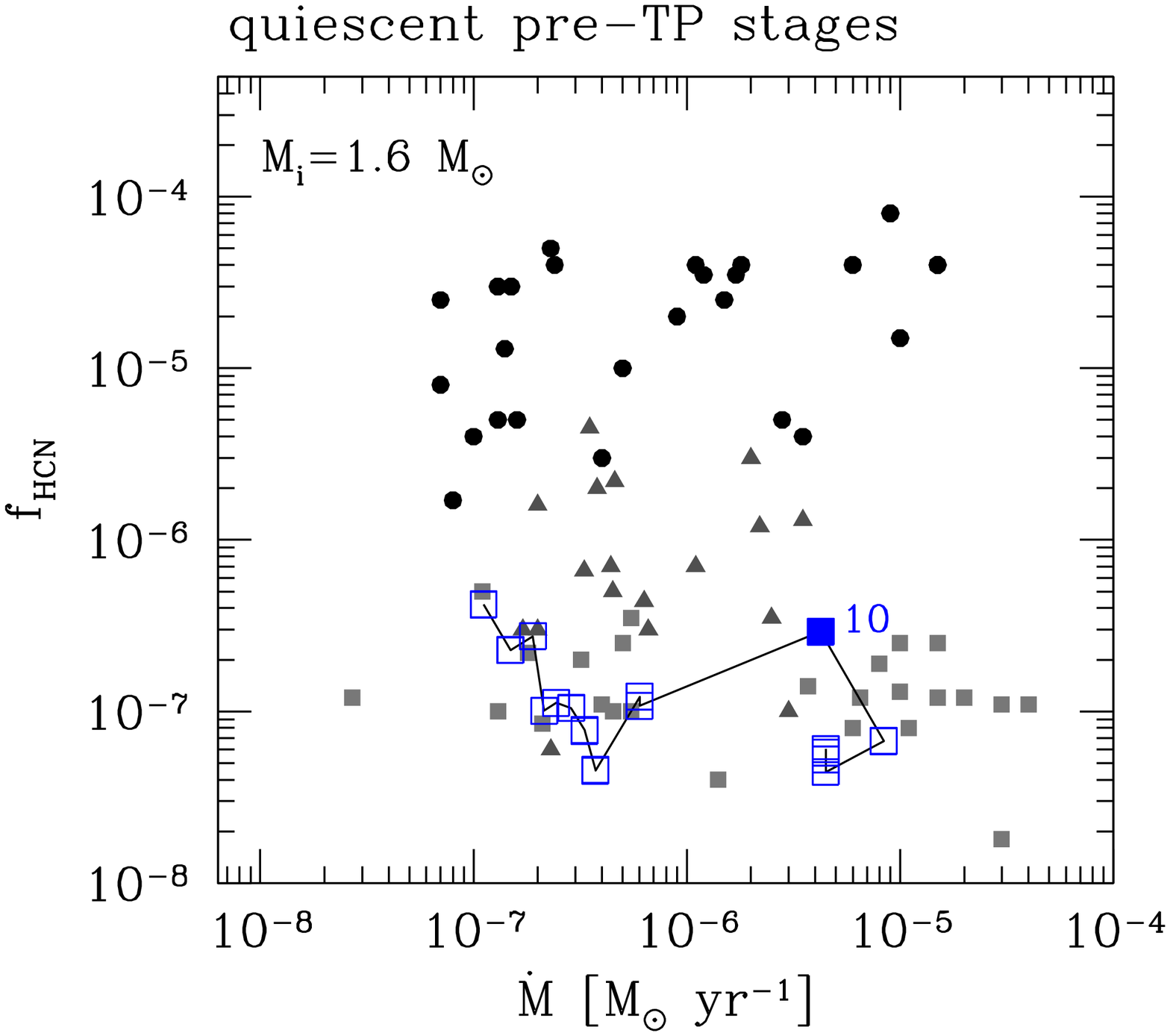}}
\end{minipage}
\hfill
\begin{minipage}{0.32\textwidth}
\resizebox{\hsize}{!}{\includegraphics{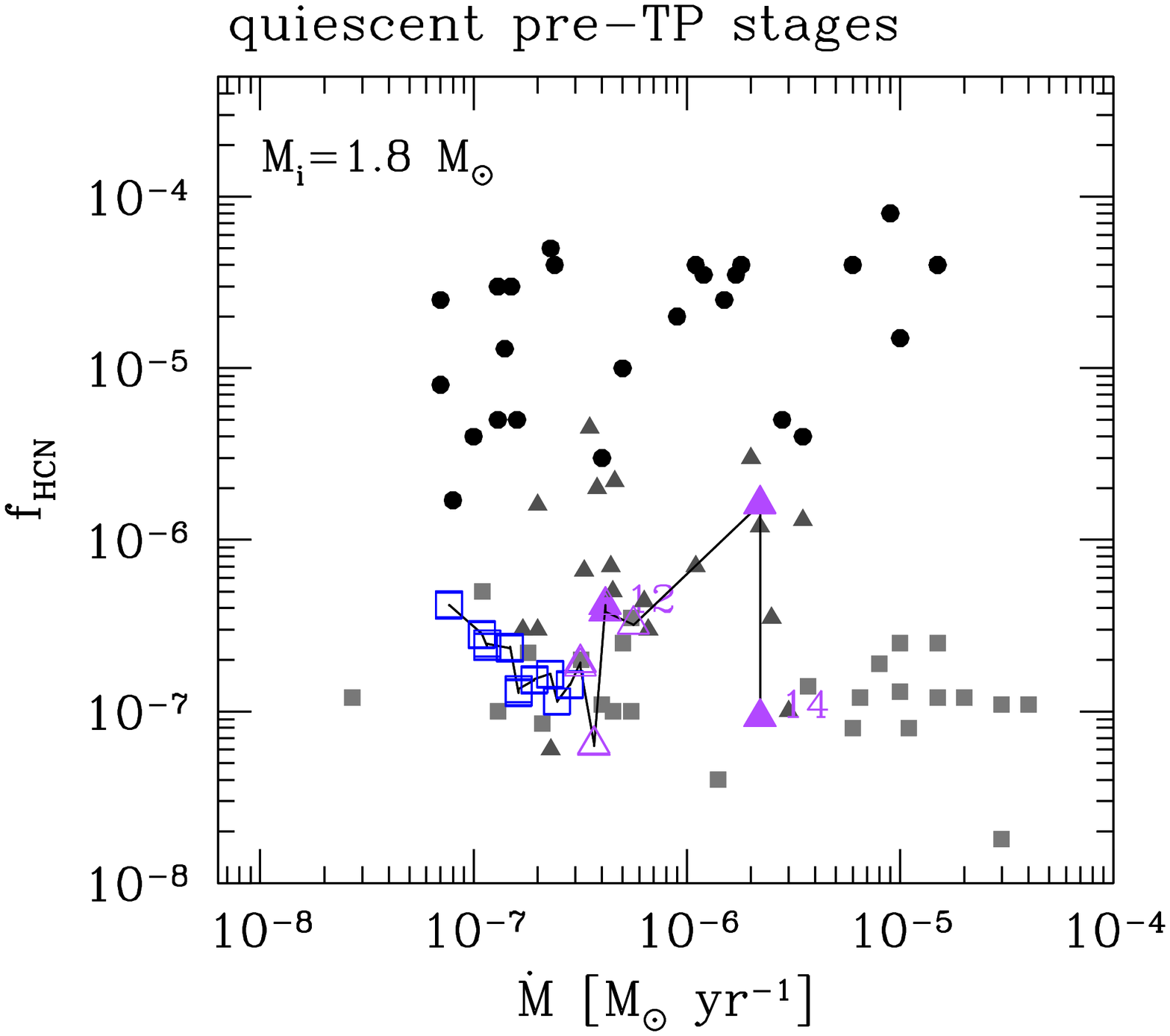}}
\end{minipage}
\hfill
\begin{minipage}{0.32\textwidth}
\resizebox{\hsize}{!}{\includegraphics{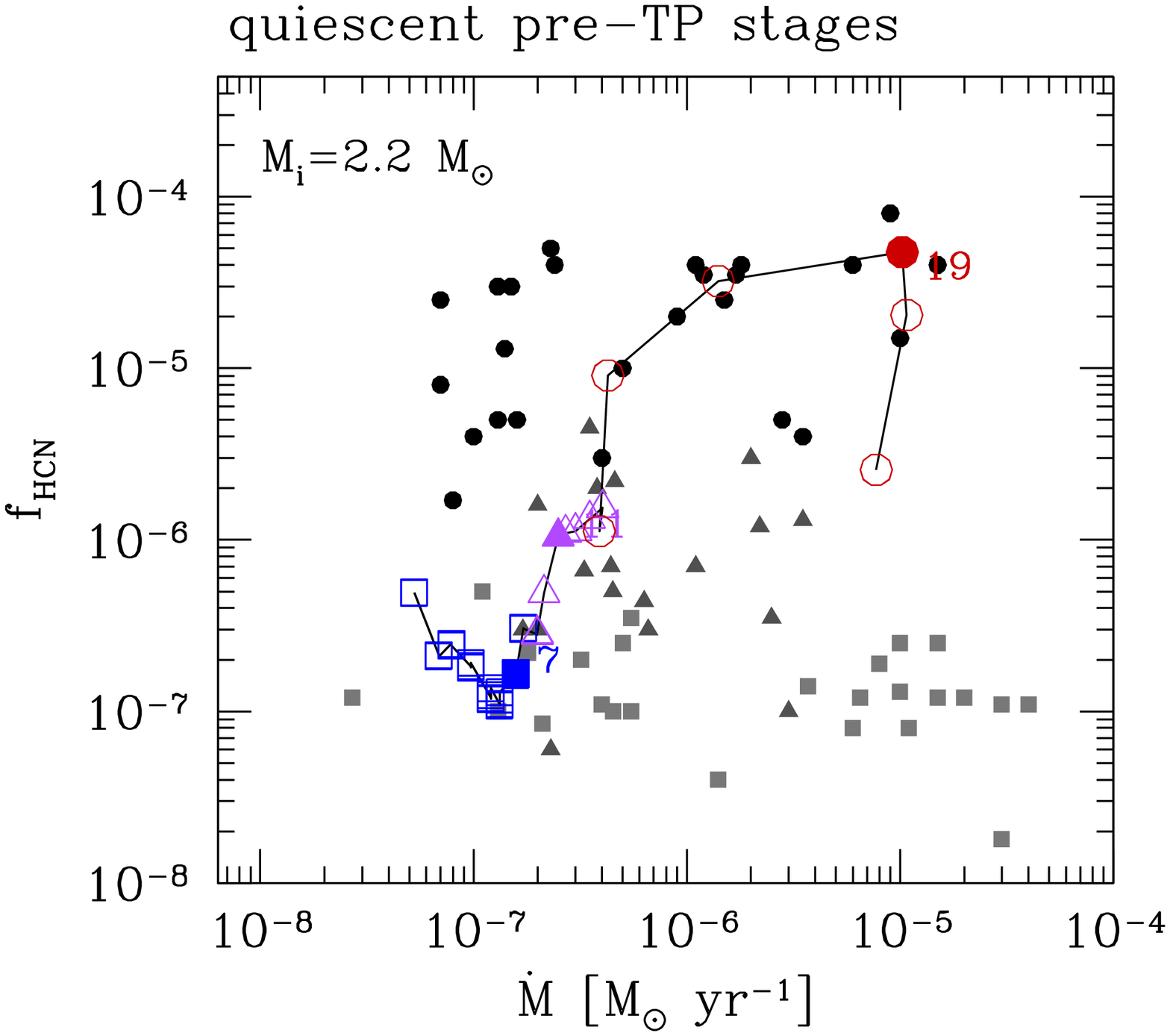}}
\end{minipage}
\begin{minipage}{0.32\textwidth}
\resizebox{\hsize}{!}{\includegraphics{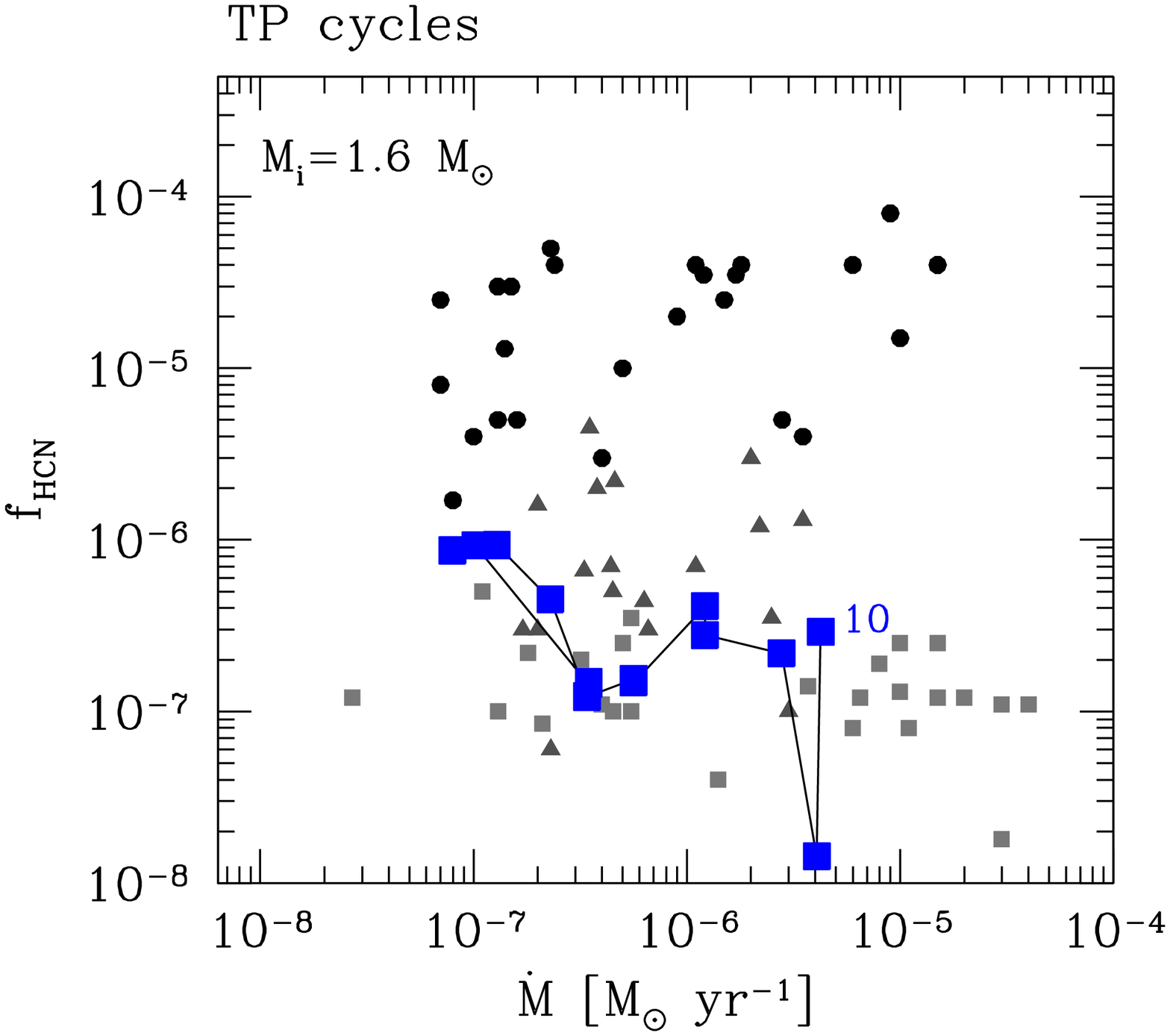}}
\end{minipage}
\hfill
\begin{minipage}{0.32\textwidth}
\resizebox{\hsize}{!}{\includegraphics{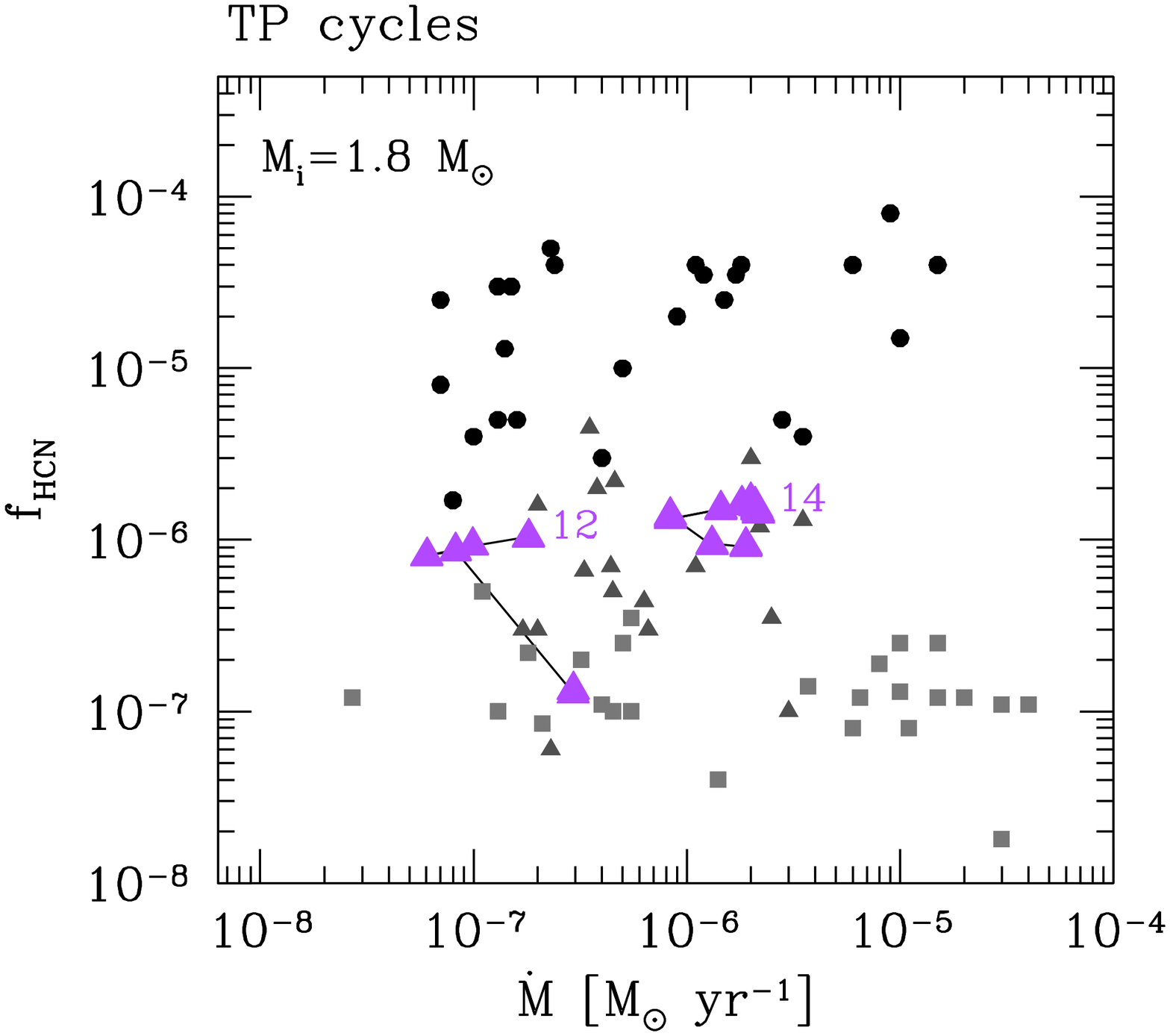}}
\end{minipage}
\hfill
\begin{minipage}{0.32\textwidth}
\resizebox{\hsize}{!}{\includegraphics{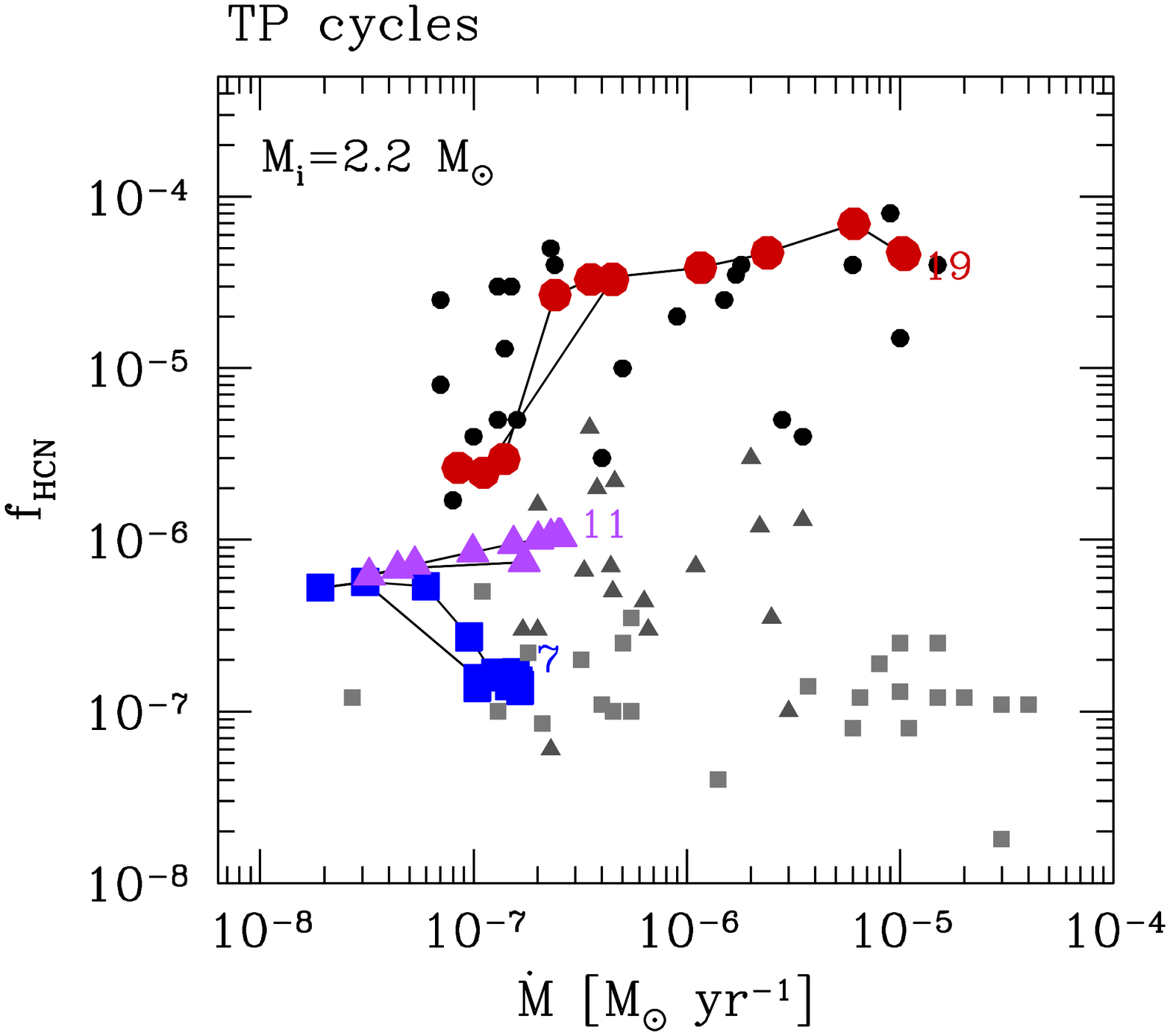}}
\end{minipage}

\begin{minipage}{0.32\textwidth}
\resizebox{\hsize}{!}{\includegraphics{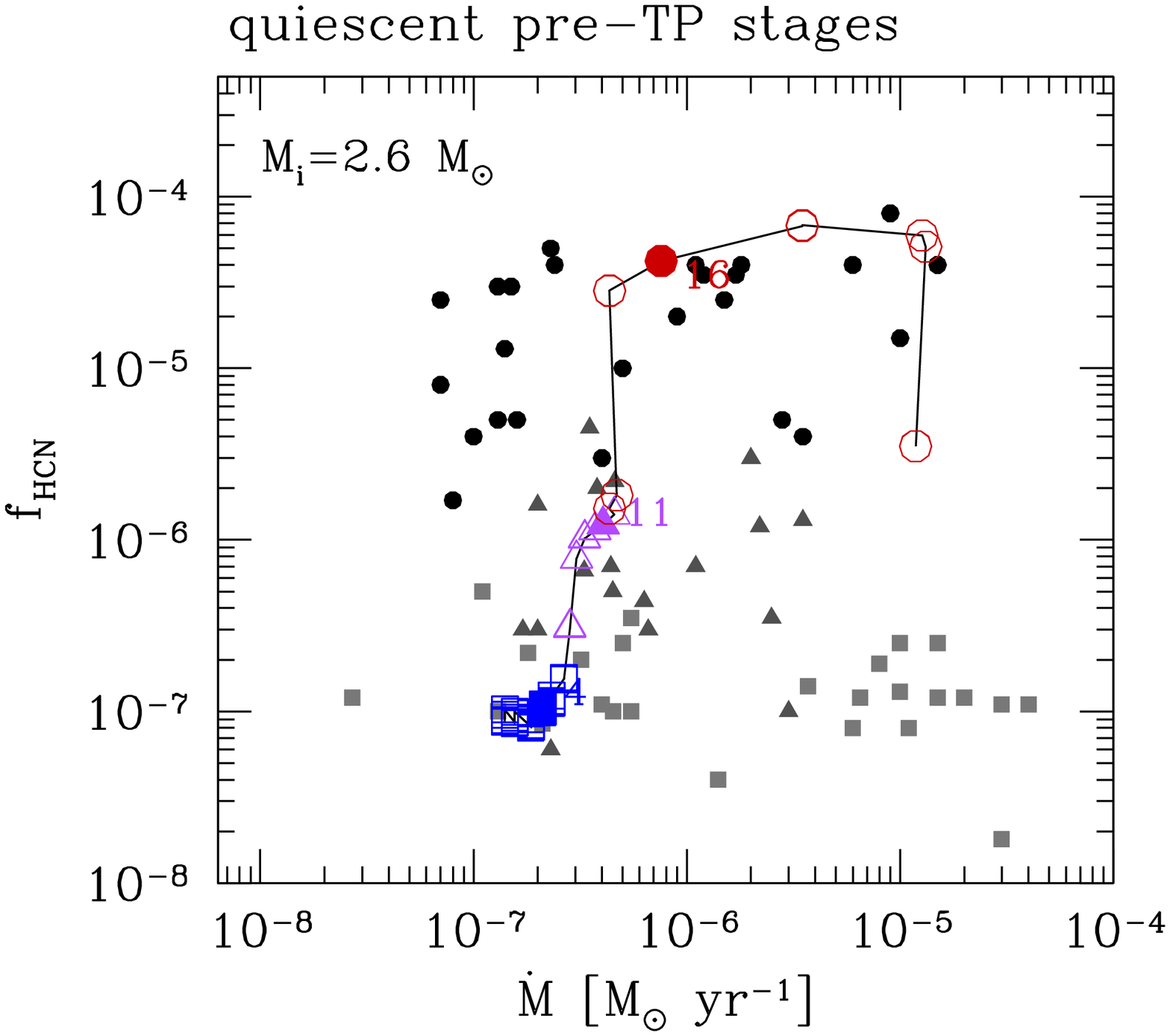}}
\end{minipage}
\hfill
\begin{minipage}{0.32\textwidth}
\resizebox{\hsize}{!}{\includegraphics{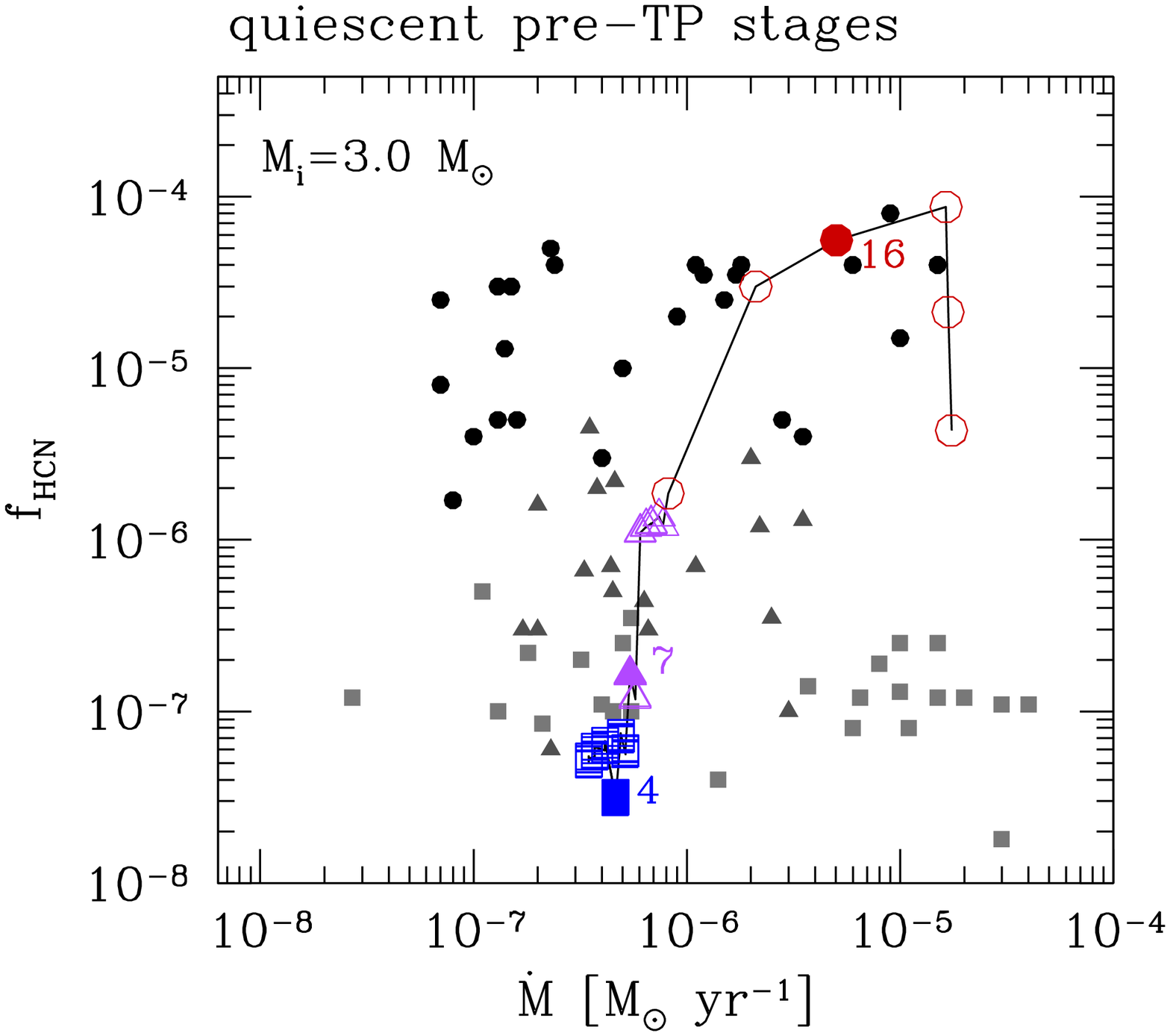}}
\end{minipage}
\hfill
\begin{minipage}{0.32\textwidth}
\resizebox{\hsize}{!}{\includegraphics{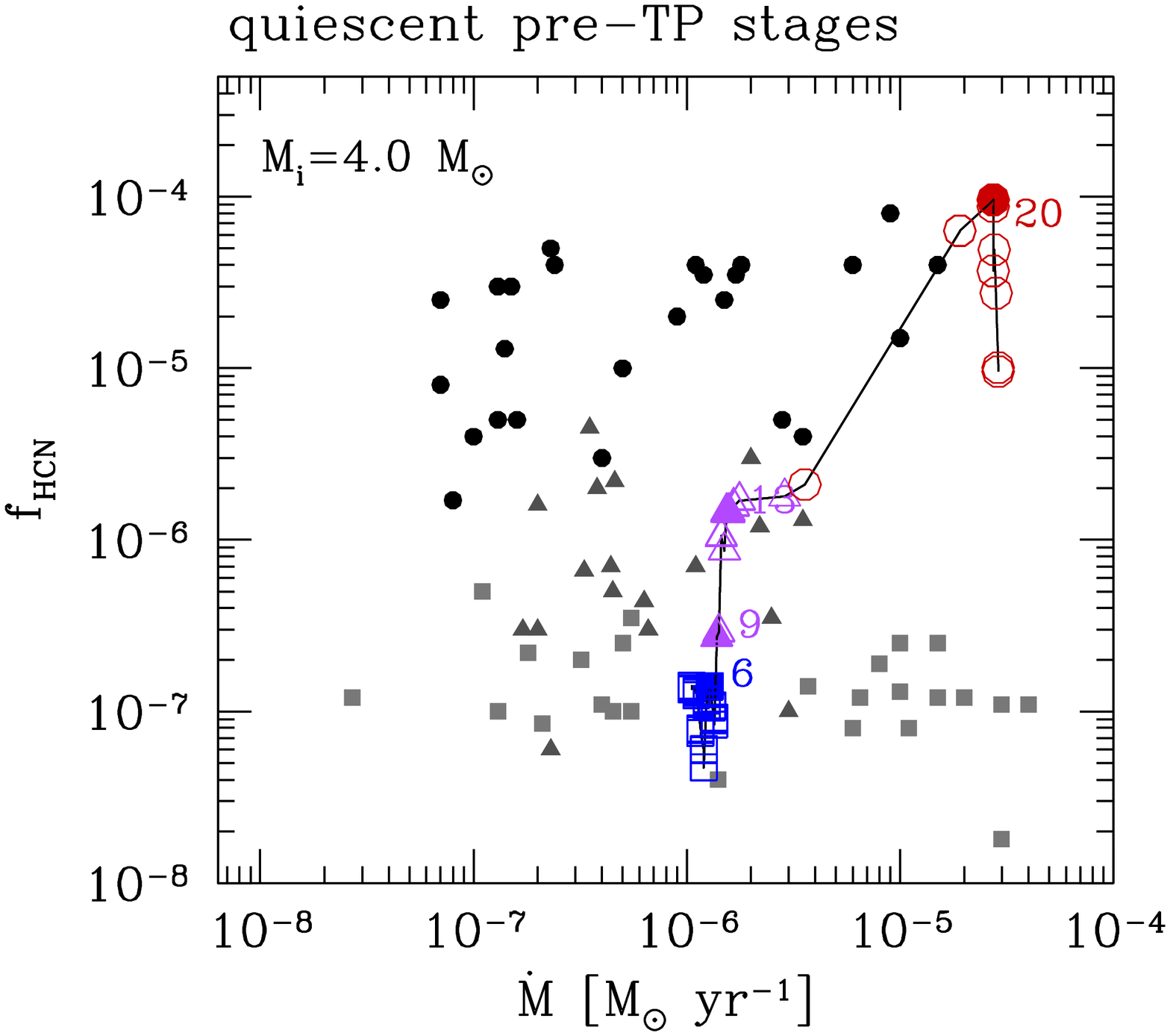}}
\end{minipage}
\begin{minipage}{0.32\textwidth}
\resizebox{\hsize}{!}{\includegraphics{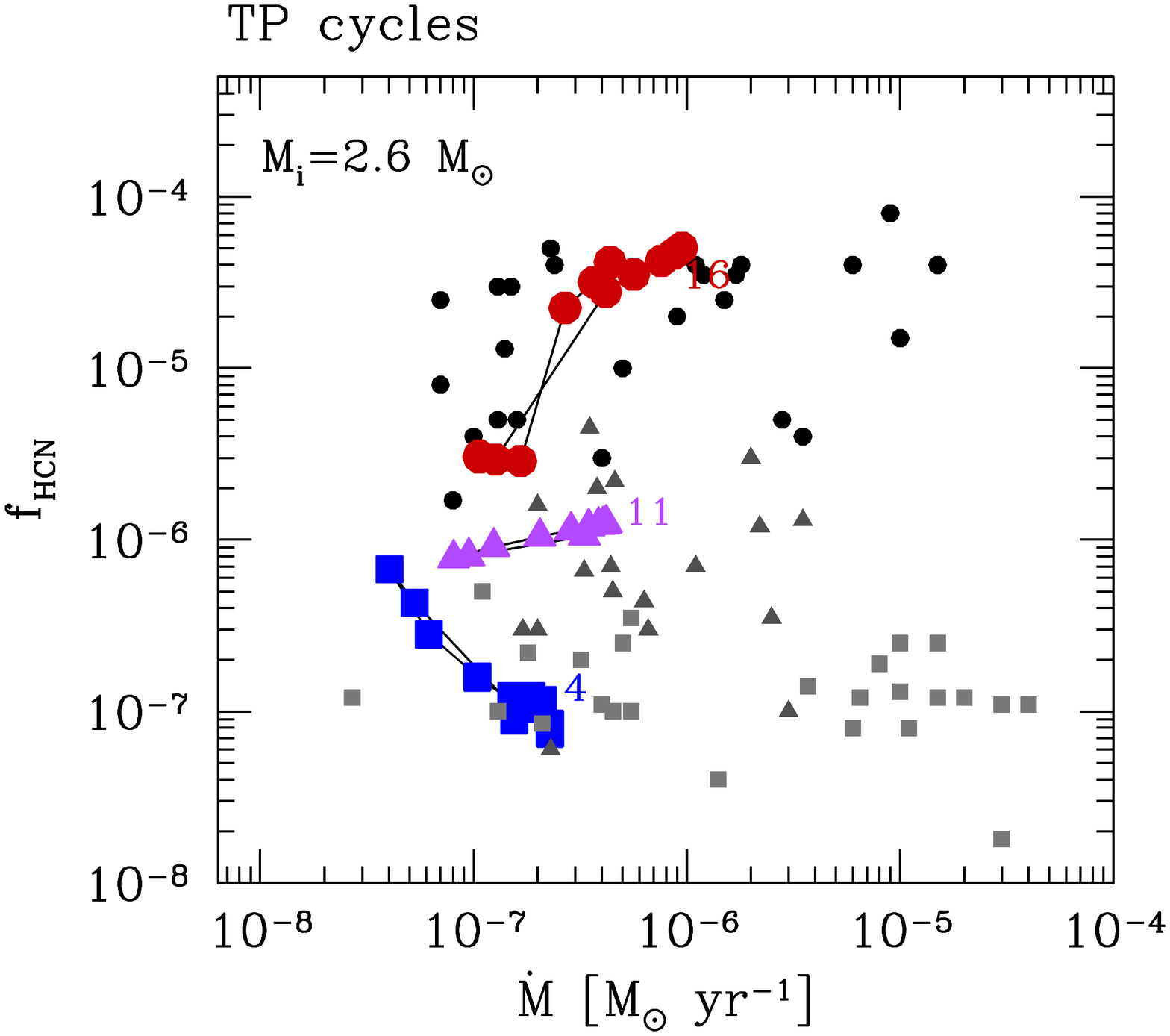}}
\end{minipage}
\hfill
\begin{minipage}{0.32\textwidth}
\resizebox{\hsize}{!}{\includegraphics{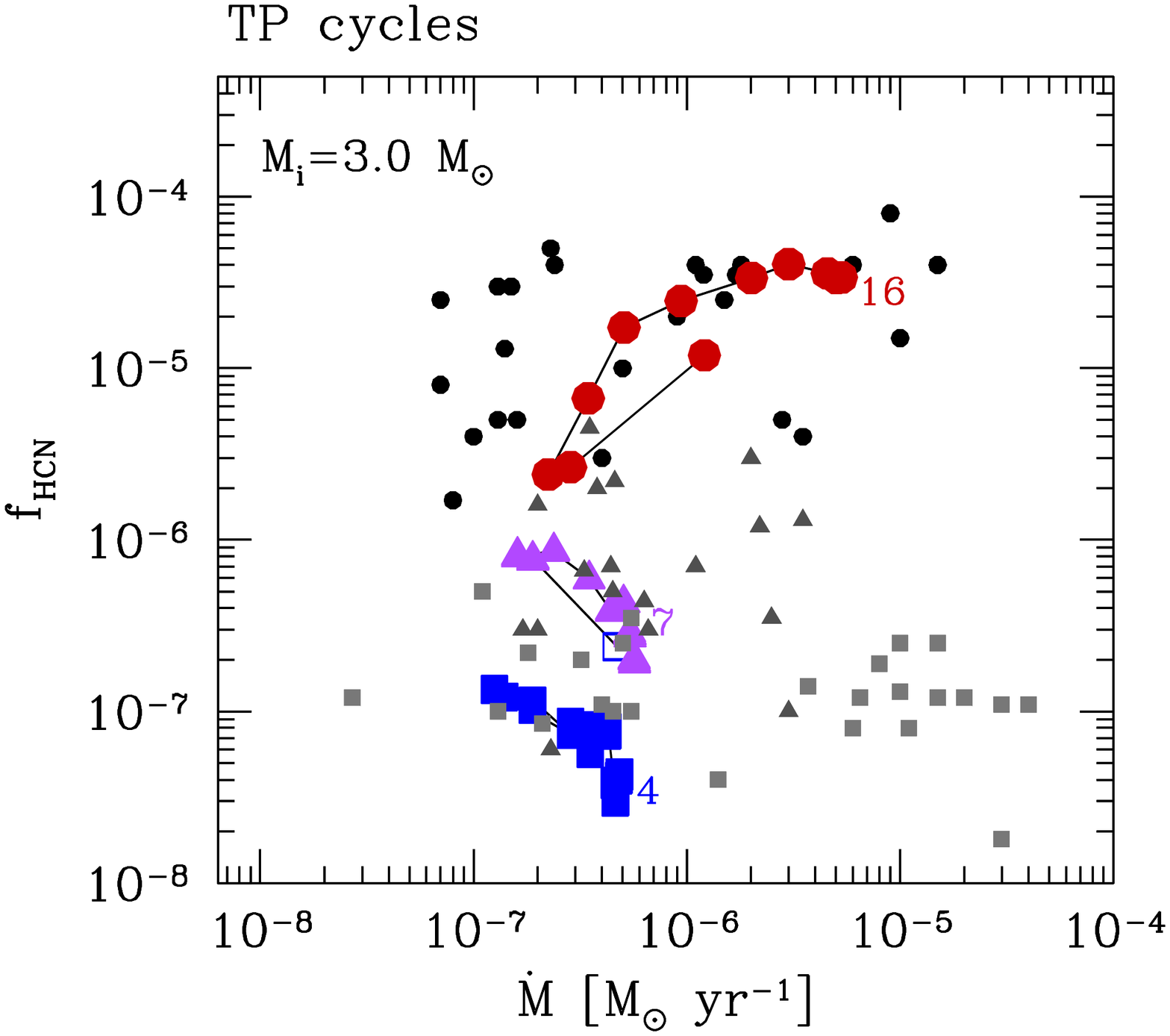}}
\end{minipage}
\hfill
\begin{minipage}{0.32\textwidth}
\resizebox{\hsize}{!}{\includegraphics{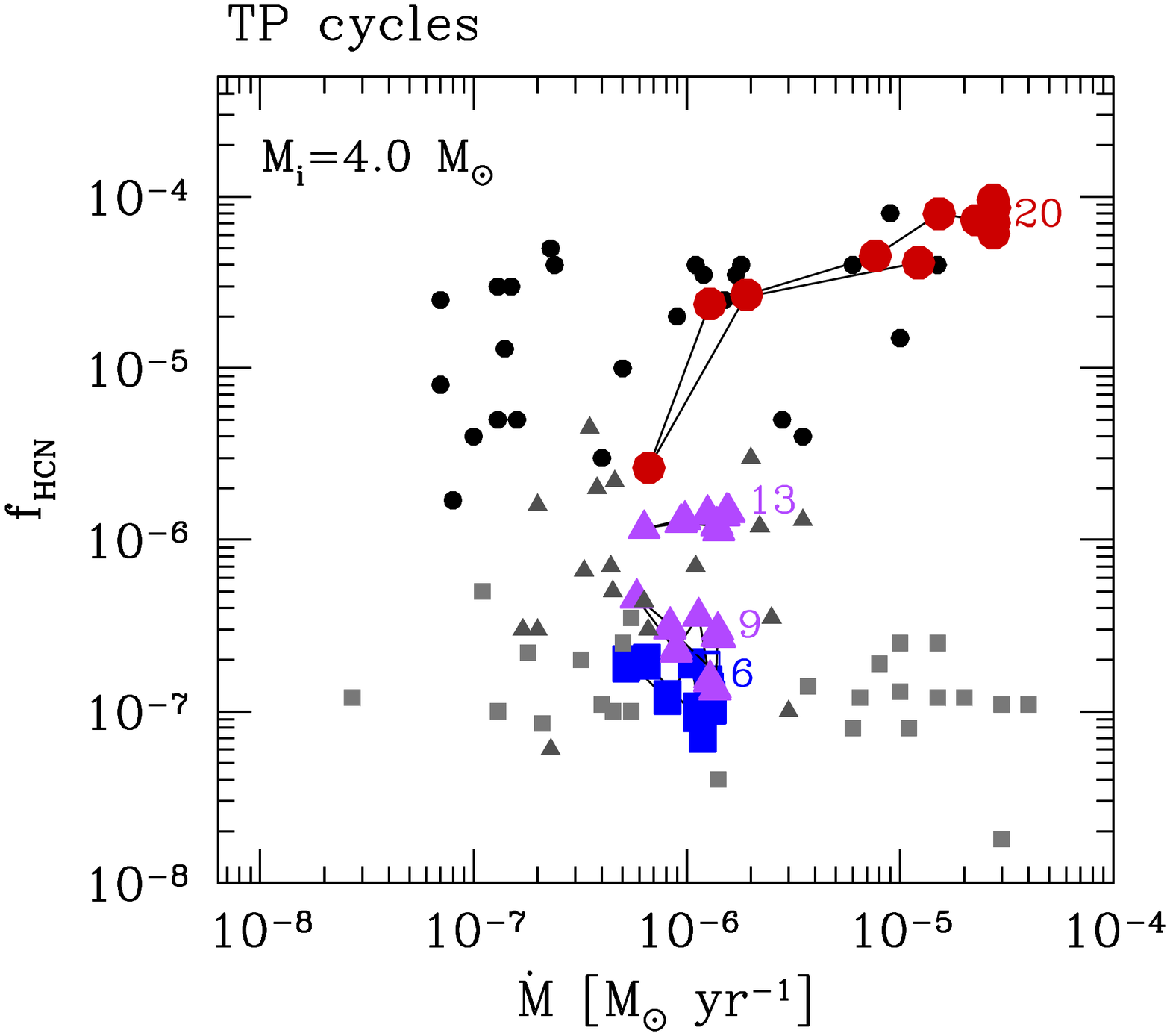}}
\end{minipage}
\caption{Predicted abundances $f_{\rm HCN}$ in the inner CSE ($3 \le r/R \le  5$)
as a function of the mass-loss rate during the entire TP-AGB evolution of six
models with  initial mass 
$M_{\rm i}=1.6\,{\rm M}_{\odot}, 1.8\,{\rm M}_{\odot},\,
2.2\,{\rm M}_{\odot}, 2.6\,{\rm M}_{\odot}, 3.0\,{\rm M}_{\odot}, 4.0\,{\rm M}_{\odot}$, 
 and metallicity 
$Z_{\rm i} =0.017$. The numbers quoted in the plots refer to the selected thermal pulses. 
From top to bottom the panels are organized as follows.
{\em Odd rows}: For each value of $M_{\rm i}$
the sequence of empty symbols correspond to the pre-flash
luminosity maximum stage (pulse cycle phase $\phi=1.0$ in Fig.~\ref{fig_pulse}) over
the whole TP-AGB evolution. Color coding and symbols depend on the photospheric
C/O ratio and are the same as in Fig.~\ref{fig_ice}.
Filled symbols and numbers are used to mark selected thermal pulses.
{\em Even rows}: Evolution of the HCN abundance along the selected
thermal pulse cycles, identified by the corresponding number.
Filled symbols are drawn for ten discrete phases,  $0 \le \phi_{\rm tpc} \le 1$,
in steps of $\Delta \phi_{\rm tpc}=0.1$}
\label{fig_chempulse}
\end{figure*}

Other test calculations were carried out for a few selected thermal pulses of TP-AGB models
that are expected to undergo the transitions to the chemical types M$\rightarrow$S$\rightarrow$C,
following the increase of the photospheric C/O ratio caused by the third dredge-up events.
The results are shown in Fig.~\ref{fig_chempulse} for six choices of the initial stellar
mass, $M_{\rm i} =1.6, 1.8, 2.2, 2.6, 3.0, 4.0\,{\rm M}_{\odot}$.

The variations of the HCN concentrations over a thermal pulse cycle
contribute to  the observed scatter
of the data belonging to the three chemical classes M, S, and C.  
It is also worth noticing that the evolution of 
the HCN abundances  
covers a relatively wide interval of mass-loss rates, while it 
remains well inside the concentration boundaries  
of the three chemical types drawn by the measured data.

For instance, over the $19^{\rm th}$ pulse cycle experienced by the 
$M_{\rm i}=2.2\, {\rm M}_{\odot}$ TP-AGB model, the corresponding 
HCN track performs a wide excursion in terms of 
the mass-loss rate -- ranging from 
$\dot M\approx 10^{-7}\, {\rm M}_{\odot}\,{\rm yr}^{-1}$ to 
$\dot M\approx 2 \times 10^{-5}\, {\rm M}_{\odot}\,{\rm yr}^{-1}$ --,
and the predicted abundances are 
in excellent agreement with the observed location of the C-star data.

\section{Towards a 
consistent chemo-dynamic picture of the inner CSE } 
\label{sect_calibration}
The present work has, for the first time, coupled the evolution of a star during the TP-AGB
phase to the non-equilibrium molecular chemistry that is expected to be active in the inner 
pulsation-shocked zone of its circumstellar envelope. 

To model the chemo-dynamic properties of the shocked gas in the region that
extends from the stellar photosphere to a few stellar radii, we adopted 
the main scheme developed by \citet{WillacyCherchneff_98}, but introducing 
a few important changes. The purpose is to provide an approach that is able
to handle more general cases (i.e. with all stellar parameters that vary along a TP-AGB track) 
and to assure a higher level of self-consistency. In this respect, we performed the following steps.
 
\begin{itemize}
\item First, we relax the common 
assumption of a constant shock parameter $\gamma_{\rm shock}$, which 
was assigned a value of 0.89 in several past studies. 
Since $\gamma_{\rm shock}$
actually fixes the extended scale height for the pre-shock density profile 
in the shocked region, its value is now determined with the aid 
of a suitable continuity condition for the density at the dust condensation radius.
In practice, the boundary condition involves finding the zero of a  function
in which all stellar parameters are involved. In this way $\gamma_{\rm shock}$
is naturally predicted to vary during the TP-AGB evolution.

\item Second, once $\gamma_{\rm shock}$ is known, 
the maximum shock amplitude $\Delta\upsilon_0$ 
follows straightforwardly from the definition 
$\Delta\upsilon_0 = \gamma_{\rm shock} \times \upsilon_{\rm e}$.
In past studies, $\Delta\upsilon_0$ was
often chosen independently from  $\gamma_{\rm shock}$.
We find that $\Delta\upsilon_0$ anti-correlates
with the pulsation period, in close agreement with detailed models of 
pulsation-shocked atmospheres for AGB stars.
Moreover, our predictions for the pre-shock radial profiles of the gas density  and
temperature compare nicely with the time-averaged spatial gradients derived by state-of-the-art
dynamic atmosphere models for pulsating, mass-losing AGB stars.  

\item Third,  we take the remaining two shock parameters, $r_{\rm s,0}$ and 
$\gamma_{\rm ad}^{\rm eff}$,
as free quantities to be calibrated on the base of the HCN abundances   
measured in the CSEs of a populous sample of AGB stars of types M, S, and C.
Following the several tests made with different choices of the shock parameters, 
we eventually obtain a
chemistry-based calibration of our dynamic model. 
The calibrated parameters are used to compute a large number
of chemo-dynamic integrations of the inner CSEs, suitably tailored to 
an extended grid of TP-AGB evolutionary models. These provide
the initial conditions to the CSE models 
in terms of key stellar properties: $M$, $L$, $T_{\rm eff}$,
$P$, $\dot M$, $\rho_{\rm phot}$, and surface 
chemical composition, including the C/O ratio and the equilibrium
molecular chemistry at the photosphere.
In this way a large variety of stellar conditions can be investigated.
\end{itemize}

\begin{figure}
\resizebox{\hsize}{!}{\includegraphics{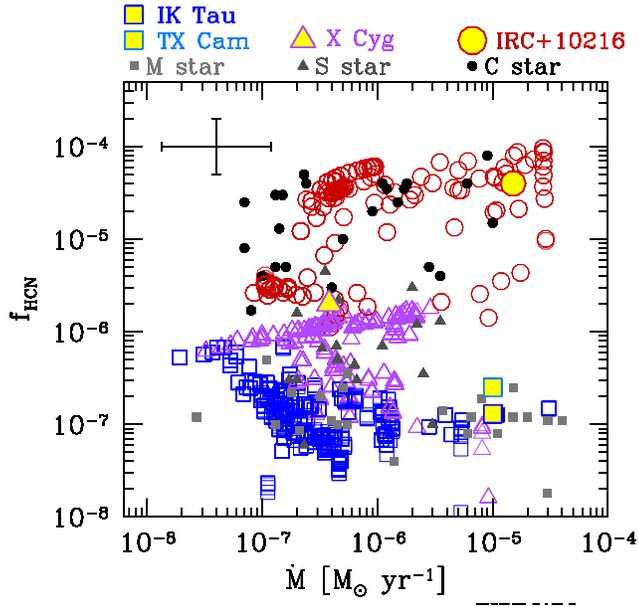}}
\caption{Comparison between observed HCN/H$_2$ and predictions from our best-fit
shock chemistry models applied to several TP-AGB tracks with initial masses in the 
range $1\,{\rm M}_{\odot} \la M_{\rm i} \la 5\,{\rm M}_{\odot}$. 
All dynamic calculations for the inner CSE (with radial extension $1 \le r/R \le 5$)
are carried out with $r_ {s,0}=1 \, R$ and $\gamma_{\rm ad}^{\rm eff}=1.01$.
We warn that the density of the theoretical points on the diagram is not
indicative of detection probability, since we simply plot models from different tracks
without simulating any synthetic sample of TP-AGB stars.
Predicted abundances refer to the distance range 
$3 \le r/R \le 5$ where the HCN chemistry  is frozen out.}
\label{fig_hcn_allmod}
\end{figure}

The calibration procedure is successful and yields a nice  
agreement with the observed HCN abundances.  Figure~\ref{fig_hcn_allmod} presents a 
a comprehensive plot that includes many predicted values of 
the HCN concentrations in the inner CSEs, which
correspond to models with different stellar masses and evolutionary stages during the TP-AGB.
Our models recover  the observed variation of the
data as a function of the chemical type.  

From the chemo-dynamic calibration we derive important implications, which are 
summarised below.
\subsection{Physical properties of the inner shocks}
\begin{itemize}
\item In order to have suitably high densities needed to account for the 
observed HCN data, the shock formation radius should be located
in the innermost zones of the CSE. The best results are obtained assuming $r_{s,0}=R$,
which suggests that the shock waves caused by pulsation should emerge very close to the stellar 
photosphere.  This finding is supported by observational 
and theoretical independent studies.
\item 
In the inner and denser zones of the CSEs,  
the shocks should exhibit a dominant isothermal character. It means that during 
the post-shock relaxation phase the gas temperature cools primarily by emission of radiation.
In our formalism this condition corresponds to assuming low values for the effective
adiabatic index, specifically $\gamma_{\rm ad}^{\rm eff} \simeq 1$.
This indication is also in line with other independent findings obtained on empirical and
theoretical grounds (e.g., see Fig.~\ref{fig_rho2t2}).
\end{itemize}
\subsection{Chemistry routes to the formation of HCN and CN molecules} 
\begin{itemize}
\item The instantaneous chemical equilibrium (ICE) assumption is clearly violated in the 
shocked regions and is definitely inadequate to explain the measured HCN concentrations.
In particular, in the cases of  M and S stars the ICE predictions underestimate 
the abundances by several order of magnitudes. In this respect we confirm, on a larger base
of stellar parameters, the indications from past chemical studies.
\item Our analysis suggests that the HCN abundance is critically linked to the 
H$_2$-H chemistry during the post-shock relaxation phases, rather than  
to the ``very fast chemistry'' in the thin cooling layer as suggested by 
\citet{Cherchneff_06}.

\end{itemize}
\subsection{Dependence of HCN concentrations on main stellar parameters}   
\begin{itemize}
\item The calibrated models successfully account for  the observed range of HCN abundances and measured mass-loss rates, 
extending up to both the minimum and maximum HCN values. The agreement supports
the soundness of the underlying TP-AGB models in terms of C/O enrichment, evolution of effective temperature,
pulsation period, stellar mass and radius, as well as to the appropriateness of the temperature, density, and gas velocity profiles 
across the pulsation-shocked region.  
\item We account for the variation of the HCN abundances as a function 
of the chemical type
neatly indicated by the observations of \citet{Schoier_etal13},
with the mean  values that increase along the sequence M-S-C. In this respect we do not confirm
the almost independence of HCN from the photospheric C/O ratio
suggested by \citet{Cherchneff_06}.
Interestingly, the models with $0.5 \la\,$C/O$\,\la 1.0$ are found to cover 
a broader range HCN concentrations, a feature that the measured data for 
S stars also show.
\item We investigate the effects  on the non-equilibrium chemistry due  
to the variations of the stellar structure driven by occurrence of thermal pulses.
We find that the HCN concentrations may change significantly over a thermal-pulse cycle,
contributing to the observed dispersion of the measured HCN data  as  function
of the mass-loss rates. 
\end{itemize}
\subsection{Caveats and planned improvements}

We are aware of various limitations of our model, and are currently
working on several improvements that will be addressed in follow-up works. 
These involve the assumption that fluid elements
move on closed trajectories, the chemical effects of
dust formation, and in general the chemical network we are employing in this work.
Here we shortly discuss these limitations, and why we believe that
they can be safely neglected in the framework of this paper.

The chemical network used in this paper was originally developed  
by \citet{Cherchneff_12}  in order to model the C star IRC+10216. 
While it includes many reactions about O-bearing species, it is however 
possible that the network omits some reactions that are
important for the chemistry of O-rich (M) stars. We are currently checking this aspect,
and a revised/expanded network will be presented in a future paper.
Here, it suffices to explore qualitatively the potential consequences
upon the results of the present study.

We start from the basic consideration that HCN is a C-bearing
species. Observations show that HCN abundances increase, on average, 
moving along the spectral sequence M-S-C (see Fig.~\ref{fig_hcndata}),
which indicates that the formation HCN is strongly related with the amount
of C that is not locked in CO molecules. In C stars its  abundance
can reach values of the order of $f_{\rm HCN}\sim 10^{-5}-10^{-4}$, 
whereas in M stars HCN can only involve a
residual (tiny) amount of C, and its concentration is typically much
lower, $f_{\rm HCN}\sim 10^{-7}$.

Therefore, our naive expectation is that the addition of chemical
species/reactions involving O-bearing species (e.g., a channel for the
formation of water through ionic reactions) to the chemical network we
used would not affect our results for C stars at all. Instead, in M
(and, to a lesser extent, S) stars, such addition might have an
indirect effect: if a larger fraction of O atoms end up in O-bearing
species (e.g. water), the abundance of CO should decrease a bit
compared to the value we calculated in our models. In turn, this
should increase the amount of C that is available for other C-bearing
species, such as HCN. Were this the case, reproducing the observed distribution
of HCN abundances would require to change the dynamical parameters,
$\gamma_{\rm ad}^{\rm eff}$, $\gamma_{\rm shock}$, and $r_{\rm s,0}$.
However, we  expect such changes to be relatively
small, otherwise we would lose the nice consistency of the dynamical picture obtained
in this work, in particular with the respect to the inner  formation and the 
isothermal character of the shocks.

The model described in this paper does not account for the effects of
dust formation on the chemical budget: in fact, the metals that
condense into dust grains are no more available for gas-phase
chemistry (thus, at $r>R_{\rm cond}$ we expect a significant depletion
of Si and O in oxygen-rich stars, and of C in carbon-rich stars);
furthermore, the very presence of dust might catalyze some reactions
\citep[e.g. H$_2$ formation][and references therein]{Grassi_etal14}.

However, the HCN chemistry should be essentially unaffected by these
inconsistencies. The reason is that in the majority of the cases we
consider (i.e. for most of the TP-AGB evolution), dust formation takes
place outside the region where the HCN abundance is changing. In fact,
we generally have $R_{\rm cond}\gtrsim 2\,R$ for C stars, and $R_{\rm
cond}\sim 3-4\, R$ for M and S stars (see e.g. Fig. 8), whereas the HCN
abundance freezes out  at $\lesssim 1.5\,R$ in C stars, and at
$\lesssim 2.5\,R$ in M and S stars (see, e.g., Fig.~\ref{fig_hcn_rad}). Thus, we believe
that dust formation will not impact on the final abundances of HCN.

A further concern is that the newly condensed dust is likely to accelerate the stellar wind, 
thus taking the Lagrangian trajectories of fluid elements further away from
 our idealized assumption of strict periodicity with zero mass loss. 
However, first we note that through distances $r \la 4-5 R$ the acceleration remains quite
moderate, as can be seen in Fig. 8: for example, at $r=4R$ typical
wind velocities are $\lesssim 4$ km/s, corresponding to a relative
displacement $\Delta r / r \lesssim 0.03 (P/1{\rm yr}) (r/10^{14} {\rm cm})^{-1}$.
More importantly, for $r > R_{\rm cond}$ fluid motions
take place already in the freeze-out region, so that inaccuracies in the description of the 
Lagrangian trajectories are not expected to produce significant effects 
on the HCN abundances.

As for future developments of this work  we wish  
to revise  the chemistry network, to extend the analysis to other  molecular species 
(e.g., H$_2$O, NH$_3$, SiO, SiS), and to 
explore the non-equilibrium chemistry  of AGB CSEs at different metallicities.
We aim at implementing  dust growth in the shocked chemistry of CSEs, 
using the schemes already tested in recent studies \citep{Nanni_etal13, Nanni_etal14}. 
We also plan to improve the prescriptions for the
dynamics of the inner winds of TP-AGB stars, taking advantage of the
results from  state-of-the-art atmosphere models in which shocks, dust condensation,
and  frequency-dependent radiative transfer are simultaneously included
\citep{Bladh_etal15, Eriksson_etal14}.

\section*{Acknowledgments}
This research is supported by the ERC Consolidator Grant funding scheme 
({\em project STARKEY}, G.A. n. 615604). We also acknowledge support from   
 the University of Padova, ({\em Progetto di Ateneo 2012}, ID: CPDA125588/12).

\end{document}